\documentclass{jfm}
\usepackage{graphicx}
\usepackage{epstopdf, epsfig}
\usepackage{subcaption}
\usepackage{scalerel,stackengine}
\usepackage{upgreek}
\usepackage{color}
\newcommand{\firstrev}[1]{{\textcolor{black}{#1}}}
\newcommand{\secondrev}[1]{{\textcolor{black}{#1}}}
\newcommand{\thirdrev}[1]{{\textcolor{black}{#1}}}
\newcommand{\thirdrevtwo}[1]{{\textcolor{black}{#1}}}
\stackMath
\newcommand\reallywidetilde[1]{%
	\savestack{\tmpbox}{\stretchto{%
			\scaleto{%
				\scalerel*[\widthof{\ensuremath{#1}}]{\kern.1pt\mathchar"0365\kern.1pt}%
				{\rule{0ex}{\textheight}}
			}{\textheight}%
		}{2.4ex}}%
	\stackon[-6.9pt]{#1}{\tmpbox}%
}
\newcommand\Sc{\mbox{\textit{Sc}}}  
\newcommand{\norm}[1]{\left\vert\left\vert#1\right\vert\right\vert}

\shorttitle{Reynolds number dependence of Richtmyer--Meshkov instability}
\shortauthor{M. Groom and B. Thornber}

\title{Reynolds number dependence of turbulence induced by the Richtmyer--Meshkov instability using direct numerical simulations}

\author{M. Groom\aff{1}
  \corresp{\email{michael.groom@sydney.edu.au}},
 \and B. Thornber\aff{1}}

\affiliation{\aff{1}School of Aerospace, Mechanical and Mechatronic Engineering, University of Sydney, Sydney, NSW 2006, Australia}

\begin{document}
	

\maketitle

\begin{abstract}
This paper investigates the Reynolds number dependence of a turbulent mixing layer evolving from the Richtmyer-Meshkov instability using a series of direct numerical simulations of a well-defined narrowband initial condition for a range of different Reynolds numbers. The growth rate exponent $\theta$ of the integral width and mixed mass is shown to marginally depend on the initial Reynolds number $\Rey_0$, as does the minimum value of the molecular mixing fraction $\Theta$. The decay rates of turbulent kinetic energy and its dissipation rate are shown to decrease with increasing $\Rey_0$, while the spatial distribution of these quantities is biased towards the spike side of the layer. The normalised dissipation rate $C_\epsilon$ and scalar dissipation rate $C_\chi$ are calculated and are observed to be approaching a high Reynolds number limit. By fitting an appropriate functional form, the asymptotic value of these two quantities is estimated as $C_\epsilon=1.54$ and $C_\chi=0.66$. Finally an evaluation of the mixing transition criterion for unsteady flows is performed, showing that even for the highest $\Rey_0$ case the turbulence in the flow is not yet fully developed. This is despite the observation of a narrow inertial range in the turbulent kinetic energy spectra, with a scaling close to $k^{-3/2}$.
\end{abstract}


\section{Introduction} \label{sec:intro}
This paper is concerned with the effects of Reynolds number on the development of a turbulent mixing layer induced by Richtmyer--Meshkov instability (RMI). RMI occurs when an interface separating two materials of differing densities is accelerated impulsively, usually by an incident shock wave \citep{Richtmyer1960,Meshkov1969}. The instability evolves due to the misalignment of density gradients across the interface and pressure gradients across the shock (typically due to surface perturbations on the interface or a non-uniform/inclined shock wave), which results in a deposition of baroclinic vorticity. This leads to the growth of perturbations on the interface and the development of secondary shear layer instabilities, which drive the transition to a turbulent mixing layer. Unlike the closely related Rayleigh--Taylor instability (RTI), RMI can be induced for both light to heavy and heavy to light configurations. In both cases the initial growth of the interface is linear in time and can be described by analytical expressions. However, as the amplitudes of modes in the perturbation become large with respect to their wavelengths the growth becomes nonlinear, whereby numerical simulation is required to calculate the subsequent evolution of the mixing layer. For a comprehensive and up-to-date review of the literature on RMI, the reader is referred to \citet{Zhou2017a,Zhou2017b}.

The understanding of mixing due to RMI is of great importance in areas such as inertial confinement fusion (ICF) \citep{Lindl2014}, where a spherical capsule containing thermonuclear fuel is imploded using powerful lasers with the aim of compressing the contents to sufficient pressures and temperatures so as \firstrev{to} initiate nuclear fusion. The compression is performed using a series of strong shocks, which trigger hydrodynamic instabilities at the ablation front due to capsule defects and drive asymmetries \citep{Clark2016}. The subsequent mixing of ablator material and fuel that ensues can dilute and cool the hotspot, which reduces the overall efficiency of the implosion. Hence it is important that the mechanism by which this occurs \firstrev{be} well understood. \thirdrevtwo{It has also been shown that the hotspot is very viscous due to the high temperatures involved \citep{Weber2014pre}, with Reynolds numbers in the range of 10-100 and therefore the possibility that ablator material is spread through the hotspot via molecular diffusion. Further evidence for diffusive mixing in the hotspot is given in \citet{Weber2020}, who estimate the Reynolds number of the fill-tube jet that enters the hotspot to be 240 and therefore far lower than the conditions that give rise to fully developed turbulence.} As a contrast to \thirdrevtwo{ICF}, in high speed combustion such as in a scramjet or rotating detonation engine, RMI due to weak shocks improves the mixing of fuel and oxidiser leading to more efficient combustion \citep{Yang2014}. An understanding of mixing due to RMI is also important for many astrophysical phenomena such as supernovae and the dynamics of interstellar media \citep{Zhou2017a}. In all of these applications, quantitative experimental data is difficult to obtain, therefore gaining an understanding of the underlying physics relies to a considerable extent upon the use of numerical simulation. Furthermore, given the broad range of scales involved in these phenomena, as well as the fact that often other physics must be considered such as radiation or chemical/nuclear reactions, it is currently necessary to model the effects of mixing and turbulence to some degree in order to maintain computational tractability. 

This motivates the use of high-fidelity simulation techniques such as large eddy simulation (LES) and direct numerical simulation (DNS) for fundamental problems with the purpose of increasing the understanding of turbulent mixing and guiding the development of reduced-order modelling techniques and sub-grid models. Previous numerical studies of this instability have demonstrated the ability of LES \thirdrev{and implicit LES (ILES)} algorithms to predict mixing at late time due to turbulent stirring in the high Reynolds number limit \citep[see][]{Youngs1994,Hill2006,Thornber2010,Schilling2010,Lombardini2012,Tritschler2014,Oggian2015,Soulard2018}. In the largest such study to date (known as the $\theta$-group collaboration), \citet{Thornber2017} showed that good agreement is obtained for various integral measures such as the mixing layer width, mixedness and total fluctuating kinetic energy across eight independent algorithms. In a follow-up paper, \citet{Thornber2019} computed the transport equation budgets for the mean momentum, mean heavy fluid mass fraction, heavy fluid mass fraction variance, and specific turbulent kinetic energy to provide useful benchmark data for the development of closure models for these quantities. There is still a lack of understanding with regards to the behaviour of the mixing layer during the transitional period between linear growth and fully developed turbulence however. In this regime the use of LES, with either implicit or modelled subgrid terms, is not necessarily well justified and indeed this is where the algorithms in the $\theta$-group collaboration showed the greatest disagreement. In \citet{Groom2019}, the feasibility of performing direct numerical simulations of RMI was assessed for the purpose of investigating this transitional regime. Using the methodology described in that paper, the current work presents a comprehensive study of the Reynolds number dependence of many key quantities of interest in the early time evolution and transition to turbulence of an RMI-induced mixing layer. 

The transition to fully developed turbulence of a turbulent mixing layer evolving from RMI was investigated in shock tube experiments by \citet{Weber2014}, using a broadband initial condition imposed on an interface between helium and argon and either a $M=1.6$ or $M=2.2$ shock Mach number. In that study the authors found an approximate $k^{-5/3}$ inertial range in the scalar variance spectra as well as sufficient separation in the Batchelor and Taylor length scales and final outer-scale Reynolds numbers of $5.7\times10^4$ and $7.2\times10^4$ respectively. This suggests that the turbulence had reached a fully developed state by the latest time considered. \citet{Mohaghar2017} also performed shock tube experiments using nitrogen and carbon dioxide with both single-mode and broadband initial conditions and a $M=1.55$ shock. For both initial conditions the outer-scale Reynolds number was found to be greater than $1\times10^4$ and the ratio of Liepmann--Taylor to inner-viscous length scales to be greater than 1, which is a sufficient criterion for fully developed turbulence in stationary flows \citep{Dimotakis2000}. A scaling of close to $k^{-5/3}$ was also found in the inertial range of the turbulent kinetic energy spectra. In \citet{Mohaghar2019}, results for a second shock Mach number of $M=1.9$ were added and the time-dependent mixing transition criterion of \citet{Zhou2003b} was evaluated, showing that the ratio of diffusion layer to inner-viscous length scales was greater than 1 only after reshock had occurred in the $M=1.55$ case and just prior to reshock in the $M=1.9$ case. \thirdrevtwo{The (stationary) mixing transition criterion was also investigated for a shock-driven gas curtain at three different incident Mach numbers by \citet{Orlicz2015}, who proposed that an outer-scale Reynolds number based on the turbulent kinetic energy rather than the mixing width gives better agreement with the measured Taylor microscales.}

\secondrev{So far the majority of experimental and numerical studies focused on transition to fully developed turbulence due to RMI have explored the effects of Mach number on the temporal development of the flow. For example, \citet{Lombardini2012} investigated the Mach number dependence of transition to fully developed turbulence in RMI by performing large eddy simulations with shock Mach numbers ranging from $M=1.05$ to $M=5$. For these simulations the effects of the unresolved scales of motion were explicitly modelled using the stretched-vortex model of \citet{Misra1997}. A deterministic initial condition was used with a radial power spectrum consisting of a Gaussian profile in wavenumber space. 
\citet{Tritschler2014pre} also examined RMI induced turbulence for a range of different shock Mach numbers, in this case from $M=1.05$ to $M=1.5$, using direct numerical simulations and determined the critical Taylor microscale Reynolds number for fully developed turbulence to be somewhere in the range of $35\le \Rey_{\lambda}\le 80$, substantially lower than previous estimates. A deterministic initial condition was also used, consisting of a dominant single mode perturbation with a multimode perturbation imposed on top of this whose coefficients approximately obey a Gaussian distribution. 
Outside of the effects of compressibilty however, the variation in time-dependent transitional behaviour of the mixing layer is actually due to the variation in Reynolds number, hence it is valuable to explore this parameter space directly as has been done previously for homogeneous turbulence. Direct numerical simulation is the ideal tool for this, as it allows for unparalleled levels of insight into the behaviour of quantities that are typically quite hard to obtain experimentally. This is the main focus of the present study; to explore the Reynolds number dependence of turbulent mixing induced by RMI using direct numerical simulations, with the aim of using the results to infer the behaviour at higher Reynolds numbers.}

A key idea introduced by \citet{Dimotakis2000} to quantify the transition to fully developed turbulence, known as the mixing transition, is to refine the bounds on the second similarity hypothesis of \citet{K41}.  This may be stated as the requirement that
\begin{equation}
\eta\ll l \ll\delta,
\label{eqn:K41}
\end{equation}
for some intermediate scale $l$ in order for the dynamics in the range of scales of size $l$ to be uncoupled from those of the large scales, \secondrev{the largest of which is the outer-scale $\delta$, while also evolving independently of the scales at which viscous effects dominate, characterised by the Kolmogorov scale $\eta$}. By considering the thickness of a laminar vorticity layer growing over spatial extent $\delta$ and using an estimate of $k\eta\approx1/8$ for the beginning of the dissipation range in various high Reynolds number flows \citep{Saddoughi1994}, \citet{Dimotakis2000} refined the criterion given in (\ref{eqn:K41}) to be
\begin{equation}
\eta<\lambda_V<l<\lambda_L<\delta.
\label{eqn:mixing-transtion}
\end{equation}
Here $\lambda_V\approx50\eta$ is referred to as the inner-viscous scale while $\lambda_L=C_{lam}\lambda$ is the Liepmann--Taylor scale, with $C_{lam}\approx5$ a weakly flow-dependent constant \secondrev{and $\lambda$ the Taylor microscale}. An important conclusion of this analysis is that by requiring $\lambda_L/\lambda_V\ge1$, the critical outer-scale Reynolds number for fully developed turbulence must be $\Rey_\delta\gtrsim10^4$, which is in good agreement with the critical values of 1--2$\times 10^4$ observed in experiments. Crucially however, this criterion is only strictly valid for stationary flows. For time-dependent flows, \citet{Zhou2003b} showed that an additional length scale $\lambda_D$ characterising the growth rate of shear-generated vorticity must be considered. The temporal development of such a scale, referred to as the diffusion layer scale, is given by
\begin{equation}
\lambda_D=C_{lam}(\nu t)^{1/2},
\label{eqn:diffusion-scale}
\end{equation}
where $C_{lam}$ is the Liepmann--Taylor growth constant. Following \citet{Zhou2003b}, the lower bound of the energy-containing scales in an unsteady flow is given by the minimum of $\lambda_D$ and $\lambda_L$, therefore the condition for fully developed turbulence becomes
\begin{equation}
\textrm{min}(\lambda_L,\lambda_D)>\lambda_V.
\label{eqn:unsteady-mixing-transition}
\end{equation}

\secondrev{In addition, flows just satisfying the time-dependent mixing transition criterion will not necessarily capture all of the physics of the energy-containing scales that are present at higher Reynolds numbers as there is still some interaction with the dissipation range. \citet{Zhou2007} showed that in order for there to be complete decoupling of the energy-containing and dissipation scales the mode with wavenumber $k_Z=2k_L$, where $k_L$ is the wavenumber of the Liepmann--Taylor scale, must lie within the inertial range. This argument is then used to define the minimum state Reynolds number as the lowest Reynolds number for which the dynamics of the energy-containing scales are completely independent of the dissipation mechanism in the flow and which requires that $k_V=k_Z=2k_L$ (where $k_V$ is the wavenumber of the inner-viscous scale). This definition, along with the definitions for $\lambda_L$ and $\lambda_V$ given previously, is used to determine that the Reynolds number of the minimum state should be $Re^*=1.6\times10^5$, roughly an order of magnitude higher than the criterion of \citet{Dimotakis2000}. At this point the energy-containing scales may be considered to evolve completely independent of the specific value of the Reynolds number.}

One aspect of the simulations presented here that make them particularly challenging, at least from the point of view of achieving a sustained level of turbulence, is the fact that the Reynolds number decreases with time. This challenge also applies to RMI experiments and is due to the dependence of the growth rate exponent $\theta$ on initial conditions \citep{Thornber2010}. As was illustrated in \citet{Groom2019}, if the layer width grows as $\sim t^\theta$ then the Reynolds number based on this width evolves as $\sim t^{2\theta-1}$. For the class of initial conditions presented here, it is expected that $\theta\le 1/3$ \firstrev{\citep{Elbaz2018}} and hence the Reynolds number decreases with time. This is contrasted with simulations/experiments of the Rayleigh--Taylor instability where the layer width grows as $\sim t^2$ and hence the associated Reynolds number grows as $\sim t^3$, which makes it easier to obtain fully developed turbulence. \secondrev{A similar discussion has also been given previously in \citet{Zhou2019b}.}

The paper is organised as follows. In \S\ref{sec:comp}, an overview of the governing equations and numerical methods employed to solve these equations is given, as well as a description of the computational setup. \S\ref{sec:results} details statistics of the velocity and scalar fields as well as the evolution of key length scales and Reynolds numbers. These are used to evaluate the mixing transition criterion for unsteady flows and assess how close the turbulence in the flow is to becoming fully developed. Finally, \S\ref{sec:conclusion} gives a conclusion of the main findings, as well as the direction of future work on this problem.

\section{Computational approach}
\label{sec:comp}
\subsection{Governing equations}
\label{subsec:equations}
The computations presented here solve the three-dimensional, compressible,  multicomponent Navier--Stokes equations, which govern the behaviour of mixtures of miscible gases. These equations can be written in strong conservation form as follows:
\begin{subeqnarray}
	\frac{\p \rho}{\p t}+\boldsymbol{\nabla}\bcdot(\rho \boldsymbol{u}) & = & 0,\\
	\frac{\p \rho \boldsymbol{u}}{\p  t}+\boldsymbol{\nabla}\bcdot(\rho \boldsymbol{u}\boldsymbol{u}^t+p\boldsymbol{\delta}) & = & -\boldsymbol{\nabla\cdot\sigma}, \\
	\frac{\p E}{\p t}+\boldsymbol{\nabla}\bcdot\big[(E+p)\boldsymbol{u}\big] & = & -\boldsymbol{\nabla}\bcdot(\boldsymbol{\sigma\cdot u}+\boldsymbol{q}_c+\boldsymbol{q}_d), \\
    \frac{\p \rho Y_n}{\p t}+\boldsymbol{\nabla}\bcdot(\rho Y_n\boldsymbol{u}) & = & -\boldsymbol{\nabla}\bcdot(\boldsymbol{J}_n).
    \label{eqn:NS}
\end{subeqnarray}
In (\ref{eqn:NS}), $\rho$ is the mass density, $\boldsymbol{u}=[u,v,w]^t$ is the mass-weighted velocity vector, $p$ is the pressure and $Y_n$ is the mass fraction of species $n=1,\ldots,N$, with $N$ the total number of species. $e=E/\rho=e_i+e_k$ is the total energy per unit mass, where $e_k=\frac{1}{2}\boldsymbol{u\cdot u}$ is the kinetic energy and the internal energy $e_i$ is given by the equation of state. All computations are performed using the ideal gas equation of state,
\begin{equation}
e_i(\rho,p,Y_1,\ldots,Y_N) = \frac{p}{\rho(\overline{\gamma}-1)},
\label{eqn:eos}
\end{equation}
where $\overline{\gamma}$ is the ratio of specific heats of the mixture. The viscous stress tensor $\boldsymbol{\sigma}$ for a Newtonian fluid is 
\begin{equation}
\boldsymbol{\sigma} = -\overline{\mu}\big[\boldsymbol{\nabla u}+(\boldsymbol{\nabla u})^t\big]+\frac{2}{3}\overline{\mu}(\boldsymbol{\nabla\cdot u})\boldsymbol{\delta},
\label{eqn:sigma}
\end{equation}
where $\overline{\mu}$ is the dynamic viscosity of the mixture. Note that in (\ref{eqn:sigma}) the bulk viscosity is assumed to be zero according to Stokes' hypothesis. The conductive heat flux is given by Fourier's law,
\begin{equation}
\boldsymbol{q}_c = -\overline{\kappa}\boldsymbol{\nabla}T,
\label{eqn:heat-flux}
\end{equation}
where $\overline{\kappa}$ is the thermal conductivity of the mixture, and $T$ is the temperature. \thirdrevtwo{The thermal conductivity of species $n$ is calculated using kinetic theory as $\kappa_n=\mu_n\left(\frac{5}{4}\frac{\mathcal{R}}{W_n}+c_{p,n}\right)$, while the thermal conductivity of the mixture is calculated using Wilke's rule}. The enthalpy flux arising from changes in internal energy due to mass diffusion is given by
\begin{equation}
\boldsymbol{q}_d = \sum_{n=1}^{N}h_n\boldsymbol{J}_n,
\label{eqn:enthalpy-flux}
\end{equation}
where $h_n=c_{p,n}T$ is the enthalpy of species $n$ and $c_{p,n}$ the specific heat at constant pressure. The mass diffusion flux $\boldsymbol{J}_n$ for species $n$ is
\begin{equation}
\boldsymbol{J}_n = -\rho D_n \boldsymbol{\nabla}Y_n+Y_n\sum_{n=1}^{N}\rho D_n \boldsymbol{\nabla}Y_n,
\label{eqn:mass-flux}
\end{equation}
which is Fick's law plus a correction velocity to ensure mass conservation when more than two species are present. The effective binary diffusivity $D_n$ for species $n$ is given by
\begin{equation}
D_n=\frac{\overline{\mu}}{\rho \Sc_n},
\label{eqn:diffusivity}
\end{equation}
where $\Sc_n$ is the Schmidt number of species $n$. In all of the simulations presented here, $\overline{\mu}=\mu_1=\mu_2$ and $\overline{\gamma}=\gamma_1=\gamma_2$. Setting $\Sc_1=\Sc_2=1$ therefore gives $D_1=D_2=D\thirdrevtwo{=\nu}$. \thirdrevtwo{Such an approximation is common when performing DNS of canonical problems such as RTI \citep{Cook2001} and related flows.}

\subsection{Numerical method}
\label{subsec:numerics}
The governing equations presented in \S\ref{subsec:equations} are solved using the University of Sydney code FLAMENCO, which employs a method of lines discretisation approach in a structured multiblock framework. Spatial discretisation is performed using a Godunov-type finite-volume method, which is integrated in time via a second order TVD Runge-Kutta method. Spatial reconstruction of the inviscid terms is done using a fifth order MUSCL scheme \citep{Kim2005}, which is augmented by a modification to the reconstruction procedure to ensure the correct scaling of pressure, density and velocity fluctuations in the low Mach number limit \citep{Thornber2008b}. The inviscid flux component is calculated using the HLLC Riemann solver \citep{Toro1994}, while the viscous and diffusive fluxes are calculated using second order central differences. This numerical algorithm has been extensively demonstrated to be an effective approach for solving shock-induced turbulent mixing problems \citep[see][]{Thornber2010,Thornber2016,Walchli2017,Groom2019}.
\subsection{Problem description}
\label{subsec:IC}
\begin{table}
	\begin{center}
		\def~{\hphantom{0}}
		\begin{tabular}{lcc}
			Property & Heavy fluid   & Light fluid \\[3pt]
			\thirdrev{$W_n$  (g/mol)}  & 90 & 30 \\
			\thirdrevtwo{$c_{p,n}$ (J/kg-K)} & 231 & 693 \\
			$\gamma_n$   & 5/3 & 5/3 \\
			$\Pran_n$  & 1.0 & 1.0 \\
			$\Sc_n$   & 1.0 & 1.0 \\
		\end{tabular}
		\caption{The molecular weight $W$, ratio of specific heats $\gamma$ and Prandtl and Schmidt numbers of fluid 1 (heavy) and fluid 2 (light).}
		\label{tab:thermo}
	\end{center}
\end{table}
The initial condition used for all simulations here is identical to that of the $\theta$-group collaboration by \citet{Thornber2017}. Two test cases were utilised in that study, referred to as the standard problem and the quarter-scale problem, which used the same computational domain size but with the initial length scales reduced by a factor of four. This allowed for simulations to be run to much later dimensionless times while still being able to obtain grid converged results for the various integral measures of interest. Since the focus of the present study is on the Reynolds number dependence at relatively early dimensionless times, the starting point for the current setup is the standard test case from \citet{Thornber2017}. This maximises the Reynolds numbers at which grid converged DNS solutions may be obtained while still allowing for the simulations to be run up until the onset of late-time behaviour. A summary of how grid convergence is assessed in direct numerical simulations of this initial condition can be found in appendix \ref{app:B}, while full details are given in \citet{Groom2019}. Using the methodology presented in that study, the results for all simulations given here may be considered to be sufficiently converged and independent of the grid resolution used.

\begin{figure}
	\centering
	\includegraphics[width=0.5\textwidth]{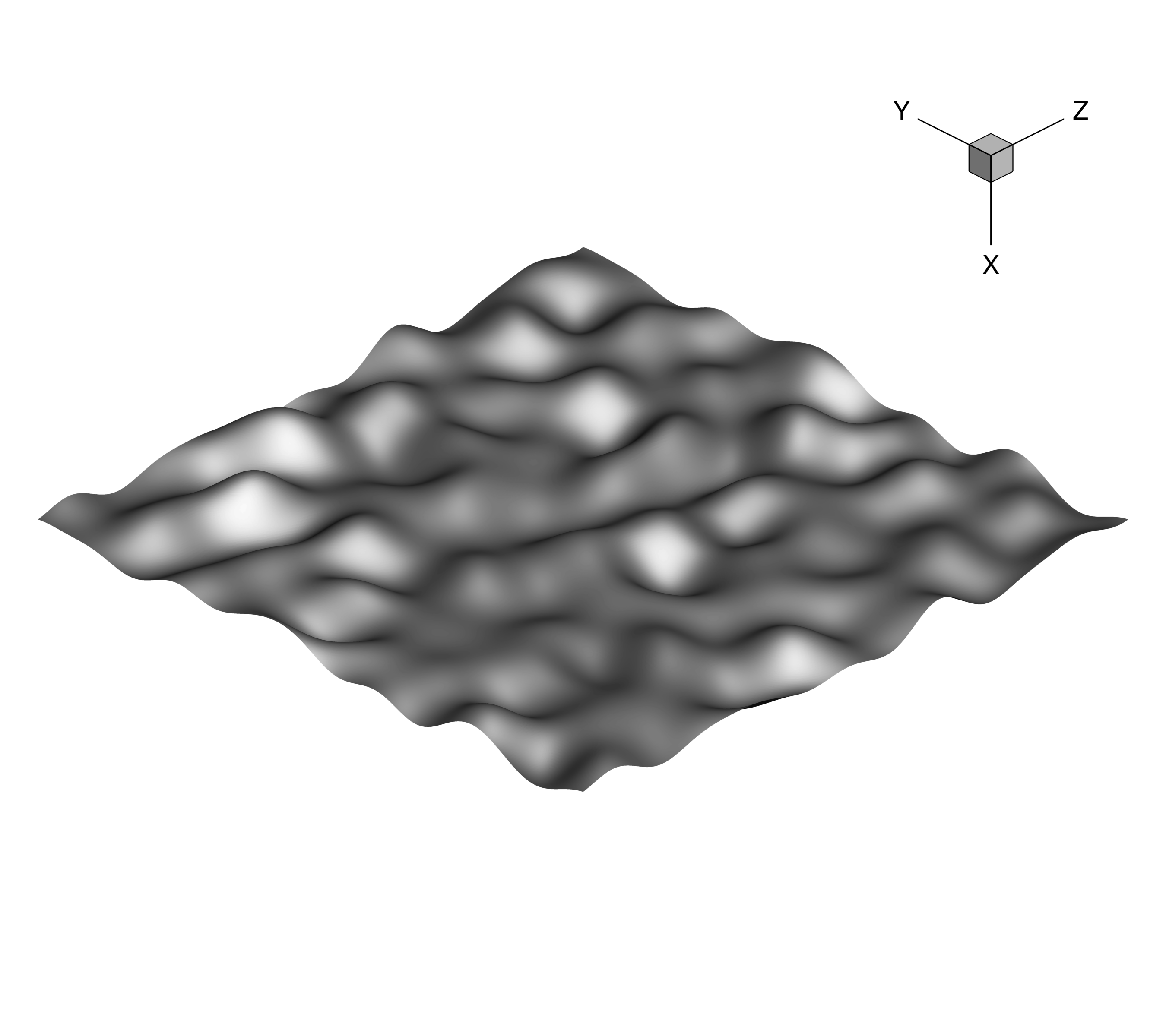}
	\caption{Isosurface of volume fraction $f_1=0.5$ at time $\tau=0$.}
	\label{fig:IC}
\end{figure}

A brief description of the initial condition will now be given. The setup consists of two quiescent gases separated by a material interface and with a shock wave initialised in the heavy gas travelling towards the interface. The material interface is given a surface perturbation, defined in Fourier space as a power spectrum of the form
\begin{equation}
  P(k) = \left\{
\begin{array}{ll}
C, & k_{min}<k<k_{max}, \\
0, & \textrm{otherwise},
\end{array} \right.
\label{eqn:power-spectrum}
\end{equation}
where $k=\sqrt{k_y^2+k_z^2}$ is the radial wave number. The specific perturbation used in this study is a narrowband perturbation \thirdrevtwo{with $k_{min}=4$ and $k_{max}=8$, in other words} containing length scales ranging from $\lambda_{min}=L/8$ to $\lambda_{max}=L/4$ where \thirdrev{$L=2\upi$ m} is the cross section of the computational domain. Setting $C=\lambda_{min}/10$ ensures that all modes are initially growing in the linear regime. The amplitudes and phases of each mode are defined using a set of random numbers that are constant across all grid resolutions and cases, thus allowing for a grid convergence study to be performed for each case. The interface is also initially diffuse for this same reason, with the profile given by an error function with characteristic initial thickness $\updelta=L/32$. The volume fractions $f_1$ and $f_2=1-f_1$ are computed as
\begin{equation}
f_1(x,y,z)=\frac{1}{2}\textrm{erfc}\left\{\frac{\sqrt{\upi}\left[x-S(y,z)\right]}{\updelta}\right\},
\label{eqn:volume-fraction}
\end{equation}
where $S(y,z)=x_0+A(y,z)$, with $A(y,z)$ being the amplitude perturbation satisfying the specified power spectrum and $x_0$ the mean position of the interface. For the purposes of this study it is sufficient to state that $A(y,z)$ is given by
\begin{eqnarray}
A(y,z) = \sum_{m,n=0}^{N}  \big[ & a_{mn}&\cos(mk_0y)\cos(nk_0z)+b_{mn}\cos(mk_0y)\sin(nk_0z) \nonumber\\
+ & c_{mn}&\sin(mk_0y)\cos(nk_0z) + d_{mn}\sin(mk_0y)\sin(nk_0z) \big],
\label{eqn:amplitude}
\end{eqnarray}
where $N=k_{max}L/(2\upi)$, $k_0=2\upi/L$ and $a_{mn}\ldots d_{mn}$ are selected from a Gaussian distribution and scaled such that the overall standard deviation of the perturbation is $0.1\lambda_{min}$. For full details on the derivation of the surface perturbation see \citet{Thornber2010,Thornber2017} and \citet{Groom2020}. A visualisation of the initial perturbation is shown in figure \ref{fig:IC}.

A Cartesian domain of dimensions \thirdrev{$x\times y\times z=2.8\upi\times 2\upi\times 2\upi$ m$^3$} is used for all simulations presented here. Periodic boundary conditions are used in the $y$ and $z$ directions, while in the $x$ direction outflow boundary conditions are imposed very far away from the test section so as to minimise spurious reflections from outgoing waves impacting the flow field. The initial mean positions of the shock wave and the interface are \thirdrev{$x_s=3.0$ m} and \thirdrev{$x_0=3.5$ m} respectively and the initial pressure of both (unshocked) fluids is \thirdrev{$p=1.0\times10^5$ Pa}. The shock Mach number is 1.8439, equivalent to a four-fold pressure increase, the initial densities of the heavy and light fluids are \thirdrev{$\rho_1=3.0$ kg/m$^3$} and \thirdrev{$\rho_2=1.0$ kg/m$^3$} and the post-shock densities are \thirdrev{$\rho_1^+=5.22$ kg/m$^3$} and \thirdrev{$\rho_2^+=1.80$ kg/m$^3$} respectively. This gives a post-shock Atwood number of \firstrev{$At^+=(\rho_2^+-\rho_1^+)/(\rho_2^++\rho_1^+)=0.487$} \thirdrevtwo{(which coincidentally is quite similar to the value of 0.49 used is the gas curtain experiments of \citet{Orlicz2015})}. The variation in density $\rho$ and mass fraction $Y_1$ across the interface is computed using $\rho=\rho_1f_1+\rho_2(1-f_1)$ and $\rho Y_1=\rho_1f_1$ with $f_1$ given by (\ref{eqn:volume-fraction}).

The evolution of the interface is solved in the post-shock frame of reference by applying a factor of \thirdrev{$\Delta u=-291.575$ m/s} to the initial velocities of the shocked and unshocked fluids. In order to be suitable for DNS, the velocity field must be modified so as to include an initial diffusion velocity at the interface \citep{Reckinger2016}. This is performed by considering the incompressible limit of a binary mixture \citep{Livescu2013}, which specifies that
\begin{equation}
\boldsymbol{\nabla\cdot u}=-\boldsymbol{\nabla}\bcdot\left(\frac{D}{\rho}\boldsymbol{\nabla}\rho\right).
\label{eqn:diffusion-velocity}
\end{equation}
To improve the quality of the initial condition, three-point Gaussian quadrature is used in each direction to accurately compute the cell averages required by the finite-volume algorithm. The dynamic viscosity $\mu$ is used to set the initial Reynolds number $\Rey_0$, described in \S\ref{sec:results} below, while all other thermodynamic properties of both fluids are given in table \ref{tab:thermo}.

\section{Results \& discussion}
\label{sec:results}
\subsection{Non-dimensionalisation}
\label{subsec:nondimensional}
\begin{figure}
	\centering
	\includegraphics[width=0.49\textwidth]{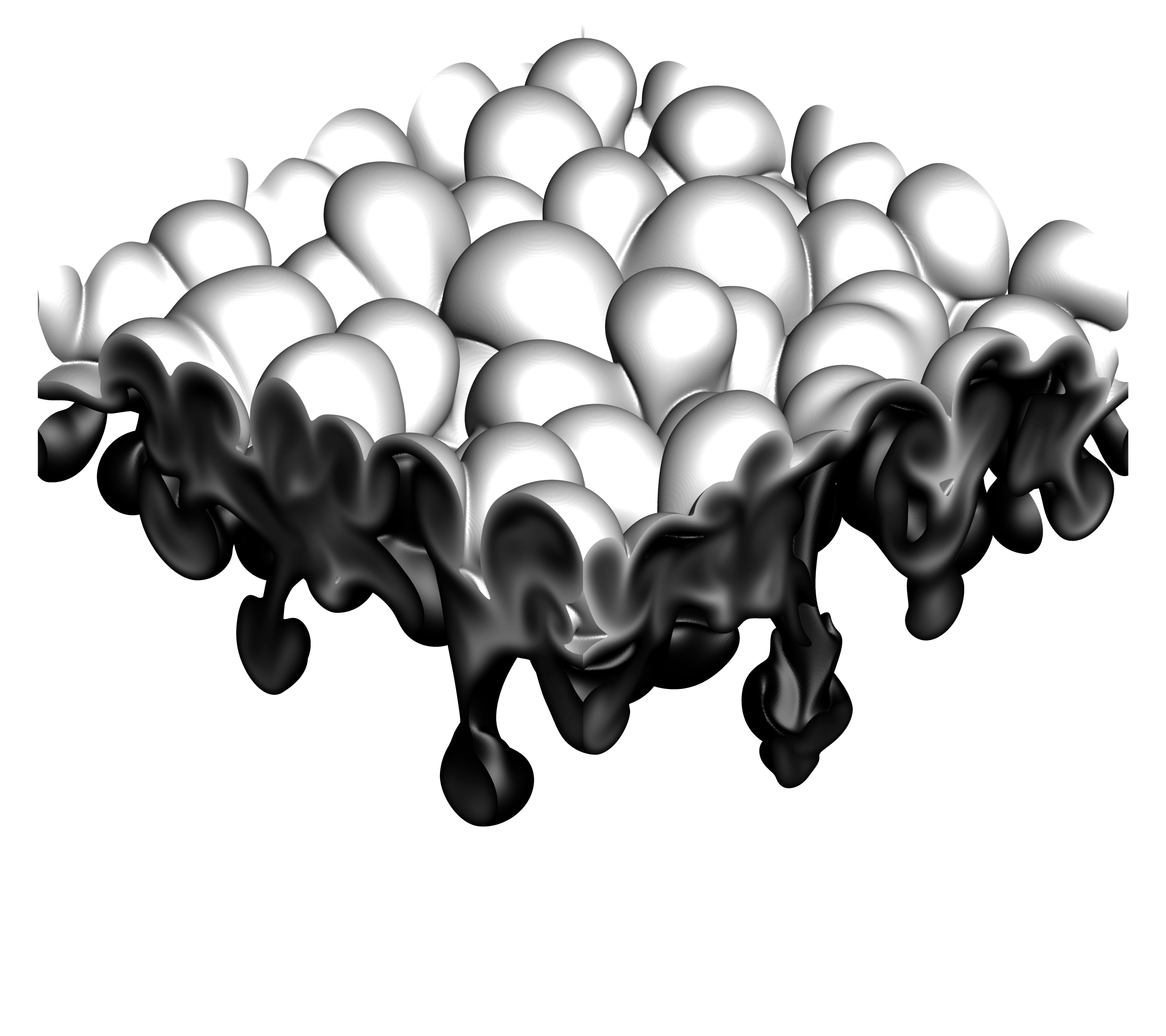}\llap{
		\parbox[b]{0.49\textwidth}{($a$)\\\rule{0ex}{0.40\textwidth}}}
	\includegraphics[width=0.49\textwidth]{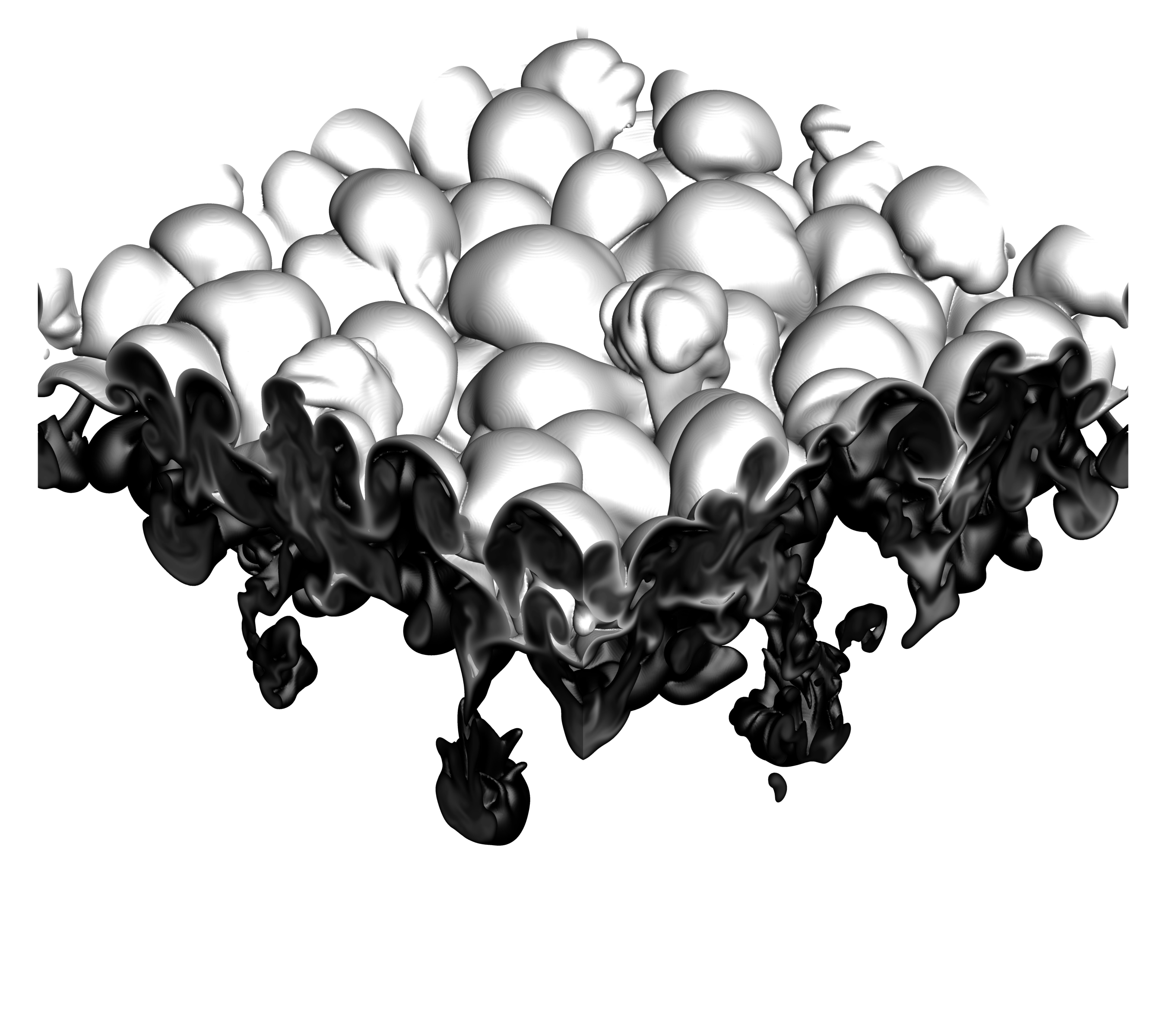}\llap{
		\parbox[b]{0.49\textwidth}{($b$)\\\rule{0ex}{0.40\textwidth}}}
	\caption{Contours of volume fraction $f_1$ for $(a)$ $\Rey_0=174$ and $(b)$ $\Rey_0=697$  at time $\tau=0.94$, bounded by the isosurfaces $f_1=0.1$ (black) and $f_1=0.9$ (white).}
	\label{fig:isosurface}
\end{figure}
All of the quantities presented in the following sections are non-dimensionalised as follows. All velocities are normalised by the initial growth rate of integral width $\dot{W_0}$, given by linear theory. By relating the integral width to the initial variance of the perturbation, \citet{Thornber2017} showed that the estimated initial growth rate is given by
\begin{equation}
\dot{W_0}=0.564\overline{k}At^+\sigma_0^+\Delta u,
\label{eqn:W0dot}
\end{equation}
where $\overline{k}$ is a weighted average wavenumber and $\sigma_0^+$ is the post-shock standard deviation of the perturbation, given by
\begin{subeqnarray}
	\overline{k} & = \frac{\sqrt{\displaystyle\int_0^{\infty}
			k^2P(k)\:\mathrm{d} k}}{\sqrt{\displaystyle\int_0^{\infty}
			P(k)\:\mathrm{d} k}} & , \\
	\sigma_0^+  & = \displaystyle\left(1-\frac{\Delta u}{U_s}\right)\sqrt{\displaystyle\int_0^{\infty}
		P(k)\:\mathrm{d} k} & .
	\label{eqn:kbar}
\end{subeqnarray}

For the current problem, $\overline{k}=\sqrt{7/12}k_{max}$ and the shock velocity is \thirdrev{$U_s=434.61$ m/s}. Following \citet{Youngs2019}, to account for the initial diffuse interface a correction factor $\psi$ is applied to (\ref{eqn:W0dot}) of the form
\begin{equation}
\psi=1+\sqrt{\frac{2}{\upi}}\overline{k}\updelta^+,
\end{equation}
where $\updelta^+=\overline{C}\updelta^-$ is the post-shock characteristic thickness of the interface, $\updelta^-$ is the pre-shock thickness and $\overline{C}=(\rho_1^-+\rho_2^-)/(\rho_1^++\rho_2^+)$ is the mean compression. For the present set of DNS cases, $\updelta^-$ will be slightly larger than the initial characteristic thickness $\updelta_0$ due to diffusion prior to shock arrival. To account for this, $\updelta^-$ is calculated assuming the diffusion occurs purely in $x$-direction, i.e.
\begin{equation}
\updelta^-=\sqrt{4Dt_s+\updelta_0^2},
\end{equation}
where \thirdrev{$t_s=0.0011$ s} is the time taken for the shock to reach the interface and $\updelta_0=\lambda_{min}/(4\sqrt{\upi})$. Therefore the initial growth rate $\dot{W_0}=0.564\overline{k}At^+\sigma_0^+\Delta u/\psi$ ranges from \thirdrev{$9.468$ m/s} to \thirdrev{$9.665$ m/s} for all cases considered here. 

All length scales are non-dimensionalised by \thirdrev{$\overline{\lambda}=2\upi/\overline{k}=1.0283$ m}, while the mean post-shock density \thirdrev{$\overline{\rho^+}=3.51$ kg/m$^3$} is used to non-dimensionalise mass in all relevant quantities. For example, the dimensionless time is defined as $\tau=t\dot{W_0}/\overline{\lambda}$. Based on these reference values, the initial Reynolds number of each case is defined as
\begin{equation}
\Rey_0=\frac{\overline{\rho^+}\dot{W_0}\overline{\lambda}}{\overline{\mu}}.
\end{equation}

Using the initial condition described in \S\ref{subsec:IC}, a series of simulations are performed, each with a different value of $\overline{\mu}$ and hence $\Rey_0$. The values of $\overline{\mu}$ used are \thirdrev{$\overline{\mu}=0.8$, $0.6$, $0.4$, $0.3$, $0.2$, $0.1$ and $0.05$ Pa-s}, which correspond to initial Reynolds numbers $\Rey_0=43$, $57$, $86$, $115$, $174$, $348$ and $697$. \thirdrev{While these viscosities are much higher than would typically occur experimentally, they are equivalent to using much smaller values of $\overline{\lambda}$ to obtain the same Reynolds number due to the various simplifications employed in the governing equations, such as no variation in viscosities with temperature. For a value of $\overline{\mu}=4.25\times10^{-5}$ Pa-s (based on a gas combination of argon and xenon that gives a similar density ratio to the one employed here), the equivalent values of $\overline{\lambda}$ would range from \thirdrev{$1.93\times10^{-4}$ m} to \thirdrev{$3.06\times10^{-3}$ m} respectively.} For each simulation, grid convergence is assessed using the methodology outlined in \citet{Groom2019}. For example, the $\Rey_0=174$, $\Rey_0=348$ and $\Rey_0=697$ cases are found to be suitably converged on grids of $360\times256^2$, $720\times512^2$ and $1440\times1024^2$ cells respectively. All simulations are calculated to a final time of \thirdrev{$t=0.5$ s}, at which point effects due to the finite box size begin to impact the solution \citep{Thornber2016}. An additional simulation with $\Rey_0=1395$ is also performed to a final time of \thirdrev{$t=0.1$ s}, using a domain of size $1.4\upi\times2\upi\times2\upi$ and grids of up to $1440\times2048^2$ cells. \secondrev{The complete set of simulations is summarised in table \ref{tab:simulations}.}

\begin{table}
	\begin{center}
		\def~{\hphantom{0}}
		\begin{tabular}{lcccc}
			$\Rey_0$ & \thirdrev{$\dot{W_0}$ (m/s)} & \thirdrev{Simulation time (s)} & \thirdrev{Domain size (m$^3$)} & Maximum grid resolution \\[3pt]
			43   & 9.468 & 0.5 & $2.8\upi\times 2\upi\times 2\upi$ & $360\times256^2$ \\
			57   & 9.517 & 0.5 & $2.8\upi\times 2\upi\times 2\upi$ & $360\times256^2$ \\
			86   & 9.567 & 0.5 & $2.8\upi\times 2\upi\times 2\upi$ & $360\times256^2$ \\
			115  & 9.593 & 0.5 & $2.8\upi\times 2\upi\times 2\upi$ & $360\times256^2$ \\
			174  & 9.617 & 0.5 & $2.8\upi\times 2\upi\times 2\upi$ & $720\times512^2$ \\
			348  & 9.645 & 0.5 & $2.8\upi\times 2\upi\times 2\upi$ & $720\times512^2$ \\
			697  & 9.659 & 0.5 & $2.8\upi\times 2\upi\times 2\upi$ & $1440\times1024^2$ \\
			1395 & 9.665 & 0.1 & $1.4\upi\times 2\upi\times 2\upi$ & $1440\times2048^2$ \\
		\end{tabular}
		\caption{The initial impulse, total simulation time, domain size and maximum grid resolution employed for each initial Reynolds number.}
		\label{tab:simulations}
	\end{center}
\end{table}

Figure \ref{fig:isosurface} shows visualisations of the solution at $\tau=0.94$ for the $\Rey_0=174$ and $\Rey_0=697$ cases. Bubbles of light fluid can be seen flowing into the heavy fluid on the upper side of the mixing layer, while heavy spikes are penetrating into the light fluid on the lower side. When comparing between the two cases, it can be observed that the effects of Reynolds number are more apparent at the spike side than the bubble side of the mixing layer. Whereas the structure of the bubble front is largely the same between the two cases, there is substantially more fine scale detail in the spikes for the $\Rey_0=697$ case. Thus it can be hypothesised that the transition to fully developed turbulence begins preferentially on the spike side, likely due to the higher velocity and stronger gradients of the spikes feeding the growth of secondary shear layer instabilities at a faster rate. The following sections will explore this transitional behaviour further through an analysis of the variation with Reynolds number in the velocity and scalar fields.

\subsection{Mixing measures \& growth rates}
\label{subsec:mix}
It is reasonably well established that multimode RMI will evolve into a turbulent mixing layer whose width is proportional to $t^\theta$, however there are still differences in the exact value of $\theta$ reported in the literature \citep{Zhou2017a}. \citet{Thornber2010} showed that these discrepancies can be at least partially explained by dependence on initial conditions, and for narrowband perturbations where the instability growth is due to non-linear coupling/backscatter from the energetic modes a value of $\theta=0.26$ was obtained from numerical simulations. This was found to be in good agreement with the experimental measurements of \citet{Dimonte2000} which gave $\theta=0.25\pm0.05$. However, \citet{Thornber2016} showed that the value of $\theta$ is sensitive to the length of dimensionless time a simulation (or experiment) is run for and gave an updated value of $\theta=0.275$. Similarly, in the recent $\theta$-group collaboration using eight independent algorithms \citep{Thornber2017}, a value of $\theta=0.219$ was obtained for the standard narrowband case (the same initial condition considered in the present study) while the quarter-scale version of that case that was run to much later dimensionless time gave $\theta=0.291$. 

\citet{Elbaz2018} gave a theoretical argument that for incompressible and immiscible fluids, the bubble front should reach a self-similar state once at least 3-4 mode coupling generations have occurred, with $\theta_b=1/3$. \citet{Soulard2018} applied an eddy damped quasinormal \thirdrevtwo{Markovian} (EDQNM) closure to RM turbulence in the low Atwood number limit and also obtained \thirdrev{$\theta_b=1/3$} for narrowband perturbations with a constant initial power spectrum. This is quite close to the results of \citet{Reese2018}, who found $\theta=0.34\pm0.01$ in vertical shock tube experiments (after adjusting the concentration field to remove large-scale structures from the mixing layer). Experiments in air and sulphur hexafluoride conducted by \citet{Prasad2000} examining late-time behaviour \thirdrev{for a nominally single-mode perturbation} found $0.26\le\theta\le0.33$, roughly spanning the range of different values from simulations of narrowband perturbations.
Recently, \citet{Youngs2019} modified a Buoyancy-Drag model based on results from the $\theta$-group study to account for initial conditions. This analysis also provided a new method for estimating the asymptotic value of $\theta$ at late-time and found that $0.32\le\theta\le0.36$, in excellent agreement with the theoretical and experimental results mentioned above.

\begin{figure}
	\centering
	\bottominset{\includegraphics[width=0.2\textwidth]{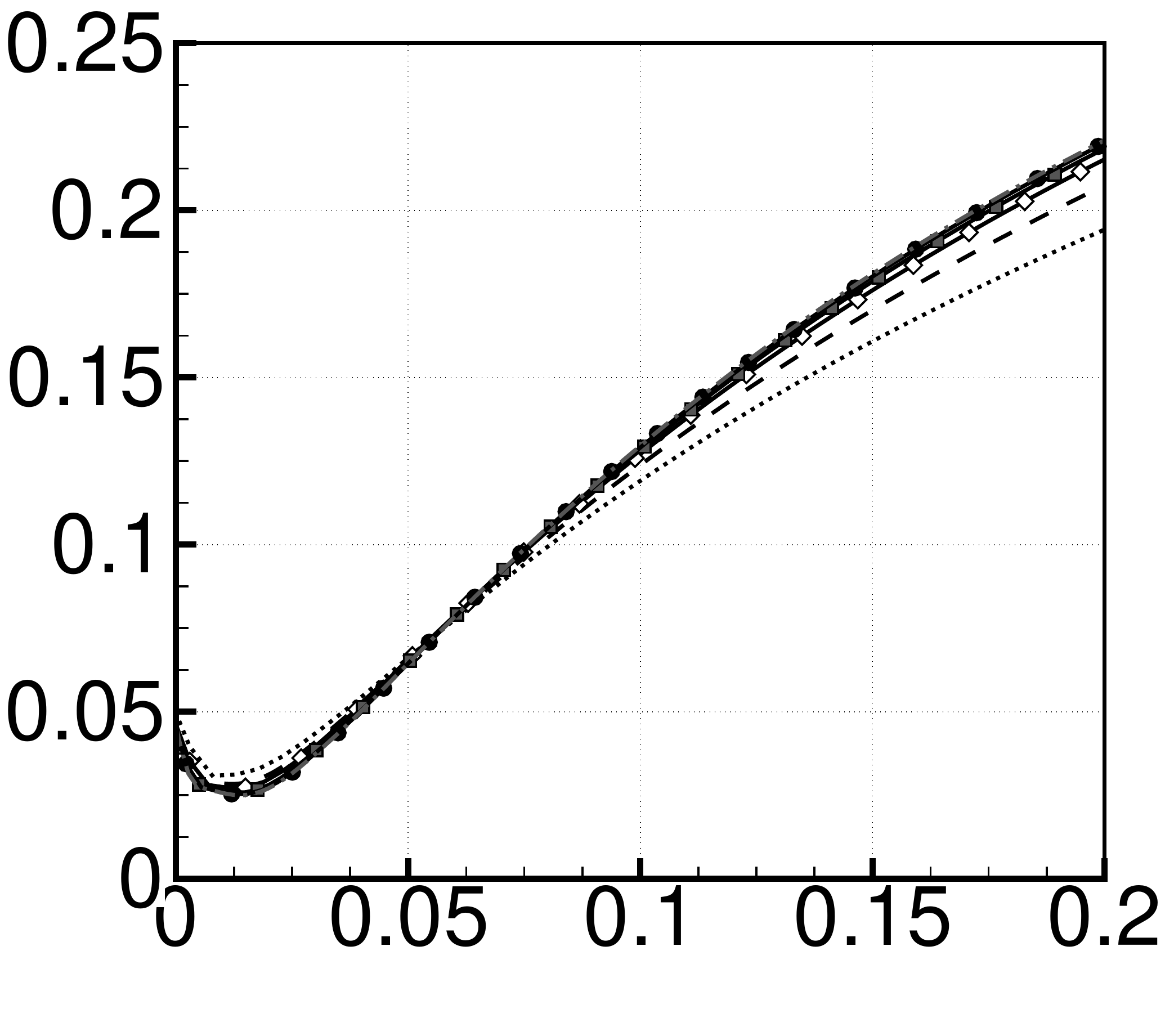}}{\includegraphics[width=0.49\textwidth]{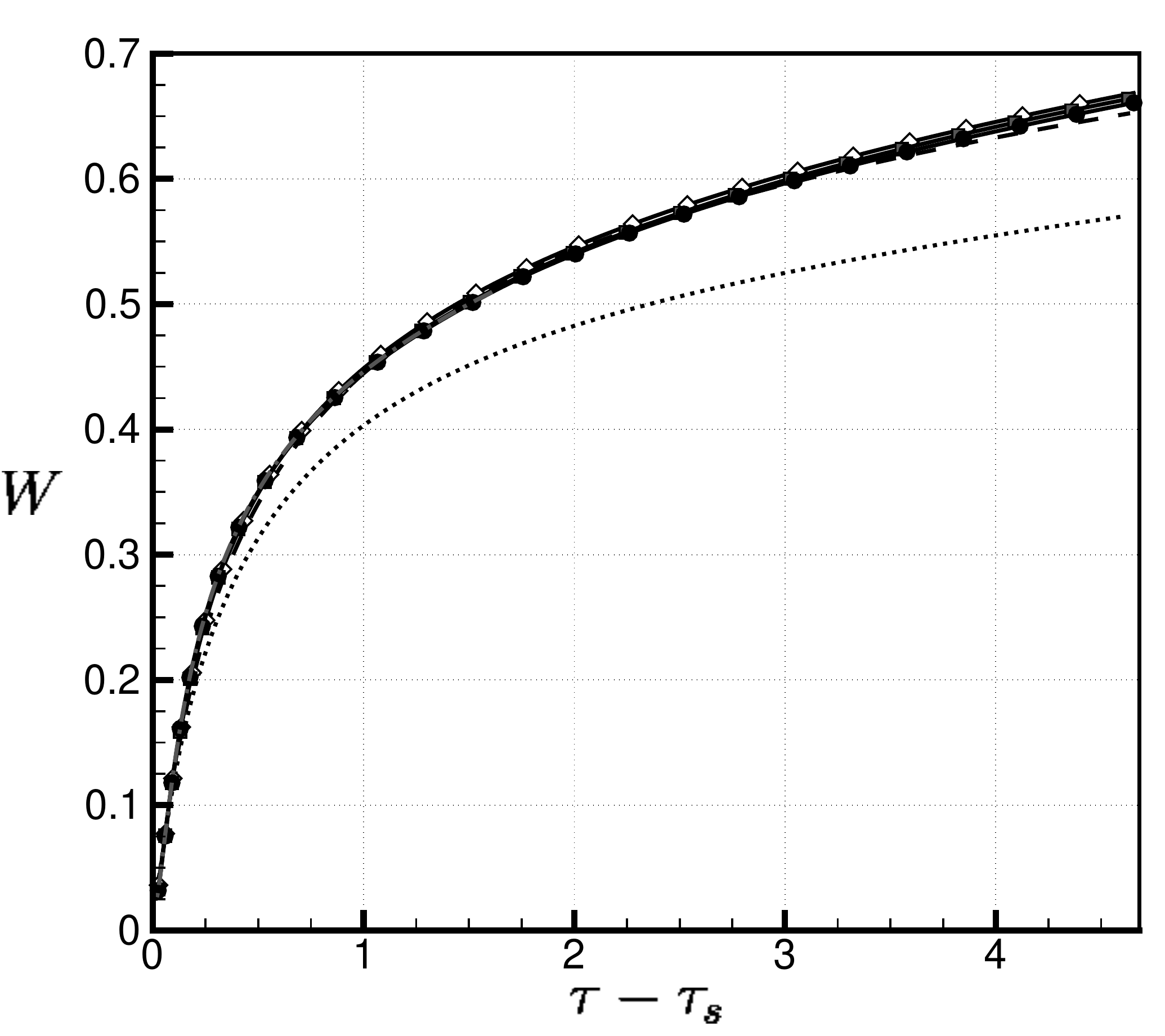}}{25pt}{10pt}\llap{
		\parbox[b]{0.495\textwidth}{($a$)\\\rule{0ex}{0.40\textwidth}}}
	\includegraphics[width=0.49\textwidth]{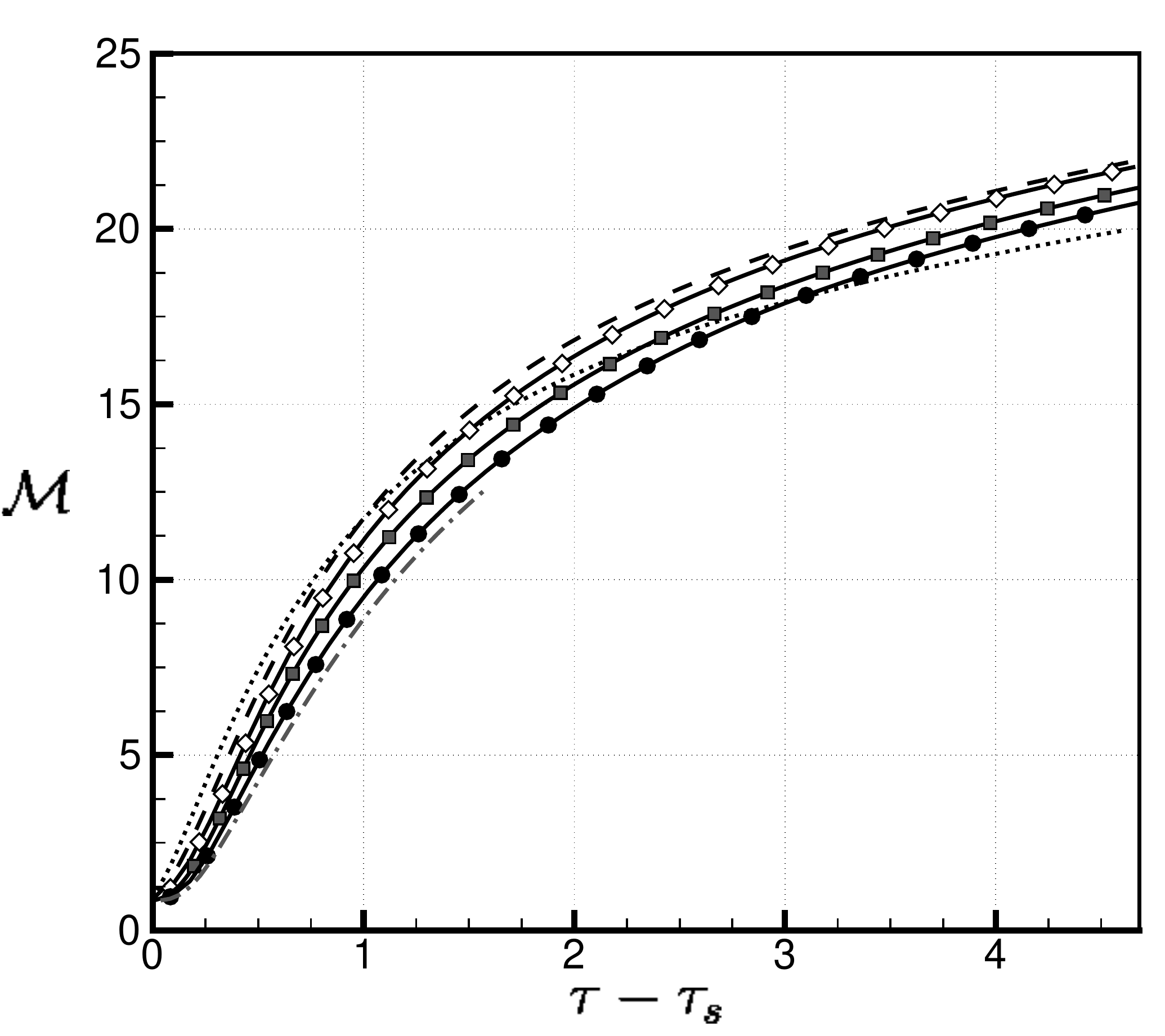}\llap{
		\parbox[b]{0.49\textwidth}{($b$)\\\rule{0ex}{0.40\textwidth}}}
	\caption{Temporal evolution of $(a)$ integral width and $(b)$ mixed mass. Shown are data for $\Rey_0=43$ (dotted black lines), $\Rey_0=86$ (dashed black lines), $\Rey_0=174$ \firstrev{(white diamonds)}, $\Rey_0=348$ \firstrev{(grey squares)}, $\Rey_0=697$ (black circles) and $\Rey_0=1395$ \firstrev{(dash-dot grey lines)}.}
	\label{fig:W-M}
\end{figure}

One area on which little data has been published is the effects of Reynolds number on $\theta$, which can be discerned using data taken from the present set of DNS results. A caveat must first be made; the results presented here are for comparatively early dimensionless times and should not be interpreted as representative of any late-time self-similar state. A commonly used quantity for estimating $\theta$ is the integral width, given by
\begin{equation}
W=\int \langle f_1\rangle\langle f_2\rangle\:\mathrm{d} x,
\label{eqn:integral-width}
\end{equation}
where $\langle \ldots \rangle$ denotes a plane average over the statistically homogeneous directions (in this case $y$ and $z$). Another quantity that may be considered to be a more direct measure of the mixing layer evolution is the mixed mass \citep{Zhou2016a}, which is given by
\begin{equation}
\mathcal{M}=\int \langle \rho Y_1Y_2\rangle\:\mathrm{d} x.
\label{eqn:imixed-mass}
\end{equation}
An important feature of the mixed mass is that it is a conserved quantity. Figure \ref{fig:W-M} shows the evolution in time of $W$ and $\mathcal{M}$, with both quantities exhibiting a non-trivial variation with Reynolds number. At the latest time considered, $W$ is smallest for the $\Rey_0=43$ case and largest for the $\Rey_0=174$ case. This ordering can be explained by the variation that occurs in dissipation of kinetic energy due to viscous action and dissipation due to turbulence as the Reynolds number is increased. For low Reynolds numbers, such as $\Rey_0=43$ and $\Rey_0=57$ (not shown), the growth in $W$ is damped by viscous dissipation, in other words the largest scales are evolving under the influence of viscosity. \thirdrev{Significant Schmidt number effects are also expected at sufficiently low Reynolds numbers, since in the limit of $\Rey_0\rightarrow0$ (and with $\Sc=1$) the growth in $W$ becomes dominated by the diffusion velocity (and hence grows as $\sim t^{1/2}$).} For high Reynolds numbers $W$ grows independently of viscous effects, and the growth rate is instead damped by turbulent dissipation. This is beginning to occur in the two highest $\Rey_0$ cases, where comparisons with the ILES data from \citet{Thornber2017} show that $W$ is tending towards the high Reynolds number limit \citep{Groom2019}. The $\Rey_0=174$ case is representative of an intermediate regime where damping due to viscous dissipation has reduced but the amount of turbulence in the flow is still relatively low, thus the damping on the growth of $W$ is lowest.

\def\stackalignment{r}

A different variation with Reynolds number is observed for $\mathcal{M}$, where at the latest time considered the $\Rey_0=43$ case has the lowest amount of mixed mass, followed by the $\Rey_0=697$ case, while the $\Rey_0=86$ case has the highest amount of mixed mass. At early times $\mathcal{M}$ decreases with increasing $\Rey_0$, which can be explained in terms of increasing levels of molecular diffusion leading to greater mixing. As the simulations progress however, the amount of mixed mass in the $\Rey_0=43$ case is eventually overtaken by that in the $\Rey_0=86$ case, which in turn is overtaken by the $\Rey_0=115$ case (not shown). This is most likely due to a combination of two factors that influence the rate at which molecular mixing occurs; the steepness of gradients across the interface and the interfacial surface area. As mixing progresses in the lowest $\Rey_0$ cases, the gradients across the interface (which control the rate of molecular diffusion) are reduced and hence the mixing rate slows. When combined with the fact that there is less interfacial surface area (i.e. the area across which molecular diffusion can occur) due to inhibition of turbulence, this explains why the amount of mixed mass in these cases is eventually overtaken by that in the higher Reynolds number cases. Indeed, this trend \thirdrev{is} expected to continue if the simulations were run to later times, where eventually the highest $\Rey_0$ case would obtain the highest amount of mixed mass.

Using nonlinear regression to fit a function of the form $W=\beta(\tau-\tau_0)^\theta$ allows the exponent $\theta$ to be obtained for each case, with the fit performed from \firstrev{$\tau-\tau_s=2.2$} to \firstrev{$\tau-\tau_s=4.6$}. \thirdrev{This fitting window is chosen based on the period over which the instantaneous value of $\theta$ obtained from a buoyancy-drag model is constant \citep{Groom2020}}. In order of ascending $\Rey_0$, the calculated values are \thirdrev{$\theta=0.172\pm1.78\times10^{-4}$, $\theta=0.163\pm2.58\times10^{-5}$, $\theta=0.178\pm6.99\times10^{-6}$, $\theta=0.197\pm1.40\times10^{-5}$, $\theta=0.215\pm4.74\times10^{-5}$, $\theta=0.214\pm5.60\times10^{-5}$ and $\theta=0.214\pm1.85\times10^{-4}$}. These values should be compared to the value of $\theta=0.219$ that was obtained from ILES simulations of the same initial condition \citep{Thornber2017}. \thirdrev{Note that the error bounds are merely a measure of how well the assumed functional form can explain the variation in the data, they are not indicative of the uncertainty in the data itself (which would require multiple realisations to be run in order to estimate).} There is a clear trend of increasing values of $\theta$ with increasing $\Rey_0$ at low Reynolds numbers, although the variation is only 25\% at most, \thirdrevtwo{while at the highest Reynolds numbers considered $\theta$ is becoming independent of $\Rey_0$. There is still a clear dependence on initial conditions over this range of dimensionless times however, since $\theta<1/3$ indicates that the growth in $W$ is not yet self-similar \citep{Elbaz2018}.} The same procedure is also preformed for the mixed mass $\mathcal{M}$, for which the corresponding values of $\theta$ are \thirdrev{$\theta=0.189\pm3.40\times10^{-5}$, $\theta=0.186\pm1.00\times10^{-4}$, $\theta=0.195\pm1.42\times10^{-4}$, $\theta=0.198\pm1.48\times10^{-4}$, $\theta=0.204\pm1.82\times10^{-4}$, $\theta=0.219\pm2.54\times10^{-4}$ and $\theta=0.214\pm2.58\times10^{-4}$}. These values are not substantially different from those calculated using $W$, except at the lowest Reynolds numbers considered. At even lower Reynolds numbers than those in the present study (and $\Sc=1$), it is likely that the calculated values of $\theta$ using $W$ and $\mathcal{M}$ would begin to differ more substantially.

\begin{figure}
	\centering
	\includegraphics[width=0.49\textwidth]{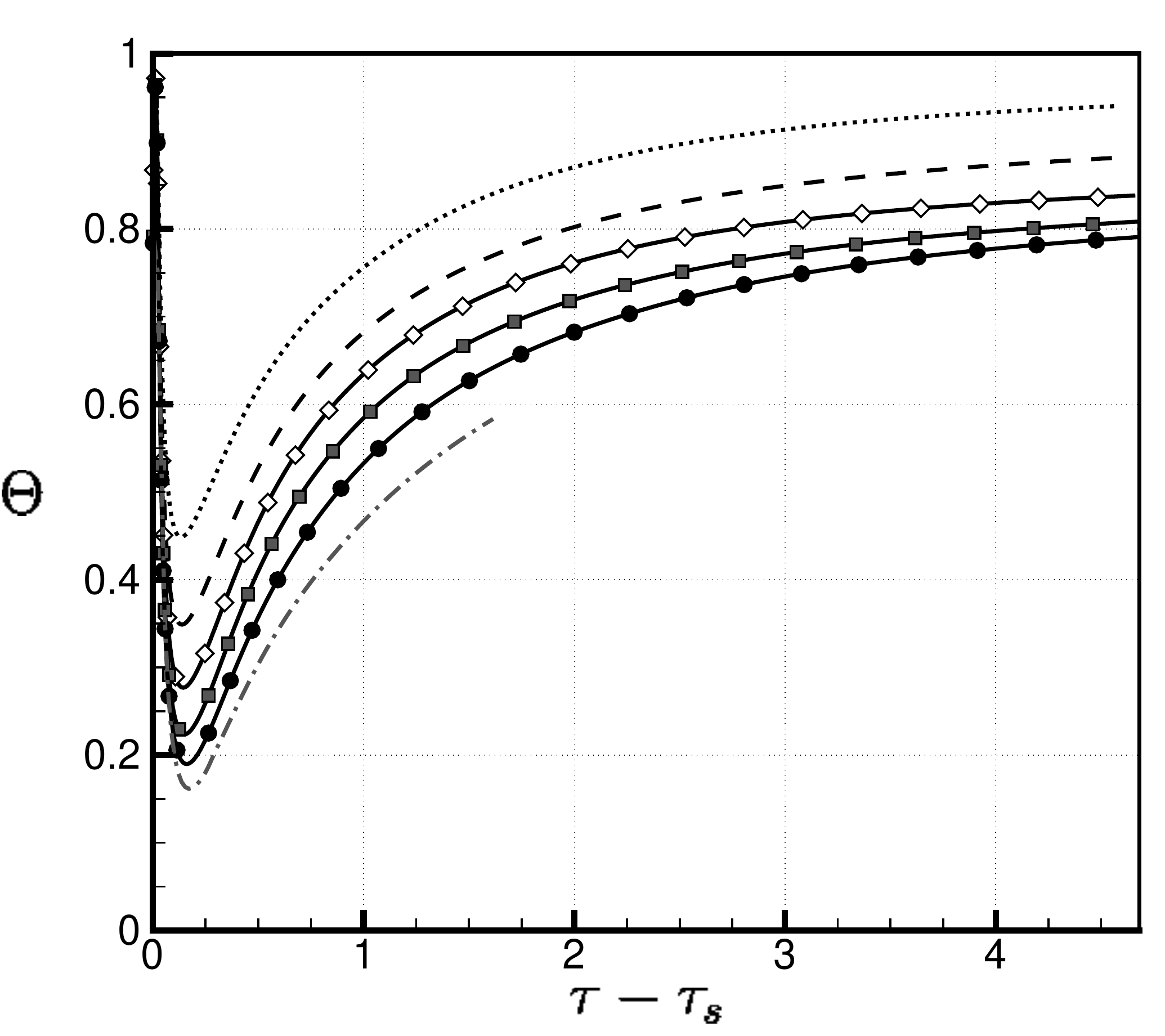}\llap{
		\parbox[b]{0.49\textwidth}{($a$)\\\rule{0ex}{0.40\textwidth}}}
	\includegraphics[width=0.49\textwidth]{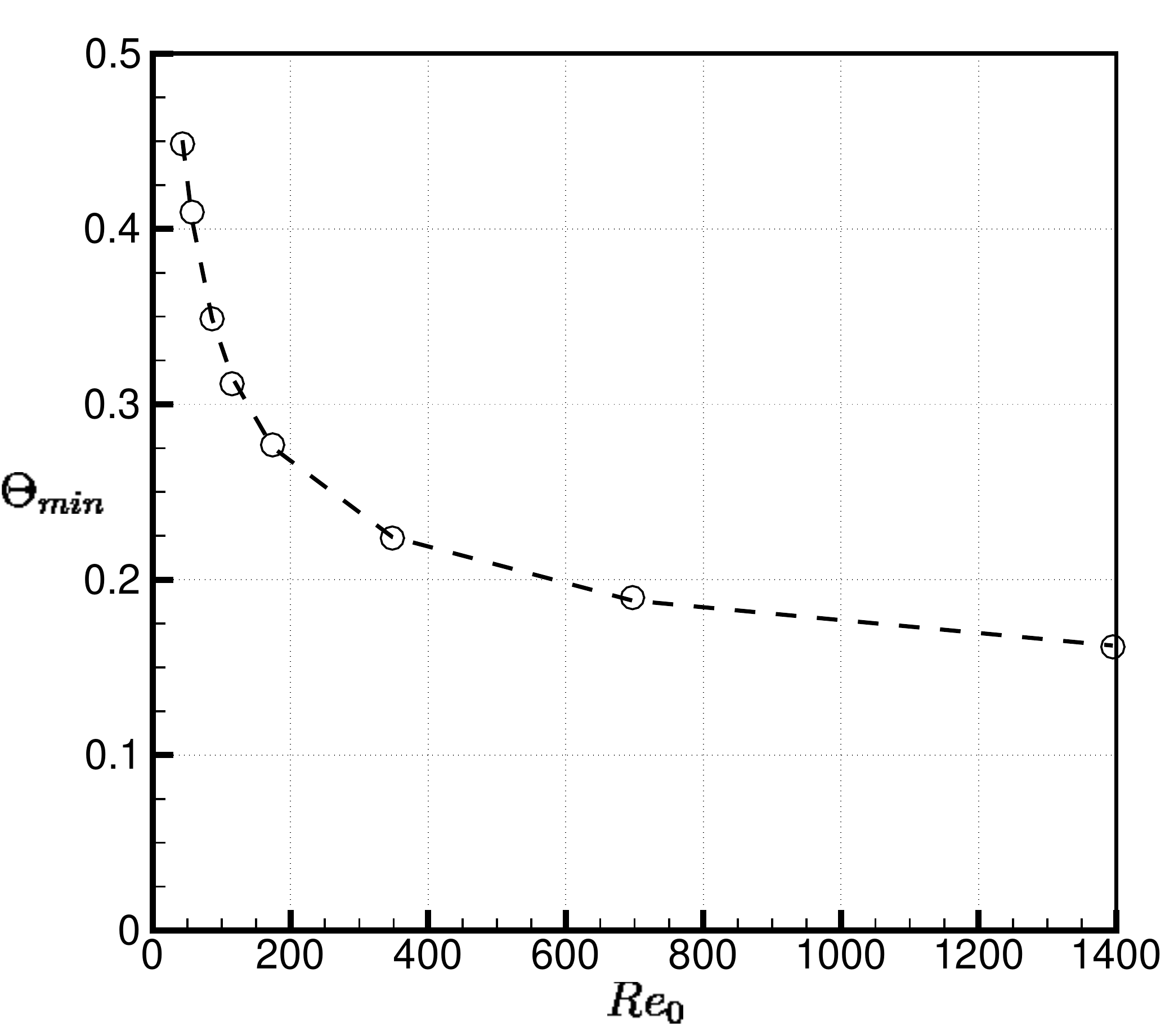}\llap{
	\parbox[b]{0.49\textwidth}{($b$)\\\rule{0ex}{0.40\textwidth}}}
	\caption{$(a)$ Temporal evolution of molecular mixing fraction at the mixing layer centre plane. Shown are data for $\Rey_0=43$ (dotted black lines), $\Rey_0=86$ (dashed black lines), $\Rey_0=174$ \firstrev{(white diamonds)}, $\Rey_0=348$ \firstrev{(grey squares)}, $\Rey_0=697$ (black circles) and $\Rey_0=1395$ \firstrev{(dash-dot grey lines)} . \thirdrevtwo{$(b)$ Minimum value of molecular mixing fraction vs. initial Reynolds number, including curve fits to the data (dashed lines).}}
	\label{fig:Theta-Psi}
\end{figure}

The degree of how effectively the two fluids are mixed may be quantified by the (global) molecular mixing fraction, given by
\begin{equation}
\Theta=\frac{\int\langle f_1f_2\rangle\:\mathrm{d}x}{\int\langle f_1\rangle\langle f	_2\rangle\:\mathrm{d}x}.
\label{eqn:theta}
\end{equation}
$\Theta$ can take values anywhere between 0 and 1, with $\Theta=0$ corresponding to complete heterogeneity and $\Theta=1$ corresponding to complete homogeneity of mixing. A similar measure may also be defined based on the mixed mass, known as the normalised mixed mass $\Psi$ \citep{Zhou2016a}. 
Figure \ref{fig:Theta-Psi} shows the evolution in time of $\Theta$, \thirdrevtwo{which displays} a clear trend \firstrev{toward} a more heterogeneous mixture with increasing Reynolds number. \thirdrevtwo{Note that results for $\Psi$ are not shown as the  behaviour is almost identical to that of $\Theta$}. After the initial compression by the shock, at which point the mixing layer is highly homogeneous, the interface is rapidly stretched by instability growth due to the impulsive acceleration. This stretching of the interface, combined with the increasing amplitude of each mode, leads to a rapid increase in the heterogeneity of the mixing layer. \firstrev{This is soon balanced by the onset of secondary instabilities, as well as (in the low Reynolds number limit) molecular diffusion due to steepening gradients across the interface}, leading to a minimum in $\Theta$ \thirdrevtwo{(and $\Psi$)}. 
\thirdrevtwo{The value of this minimum varies between $0.449$ and $0.161$ as $\Rey_0$ is increased and thus the value, and to a lesser degree the temporal location, is observed to depend on the initial Reynolds number}. There is also evidence that a high Reynolds number limit exists, for example the distance between the $\Rey_0=1395$ and $\Rey_0=697$ minima is less than that between the $\Rey_0=697$ and $\Rey_0=348$ minima. \thirdrevtwo{The variation of the minimum values of $\Theta$ with initial Reynolds number $\Rey_0$ is also shown in figure \ref{fig:Theta-Psi}, along with the curve of best fit to the data}, obtained using nonlinear regression with a functional form of 
\begin{equation}
\thirdrevtwo{f=A+\sqrt{B/(\Rey_0-C)}.}
\end{equation}
The optimal parameters are $A=0.10$, $B=5.34$ and $C=-0.60$ for $\Theta_{min}$. \thirdrevtwo{Although the correct behaviour as $\Rey_0\rightarrow0$ is not captured by the assumed functional form}, the high Reynolds number limit for both quantities may be estimated and is given by the coefficient $A$. 


Beyond the point of minimum mix, $\Theta$ \thirdrevtwo{(and $\Psi$) starts} to increase and by the end of the simulation \thirdrevtwo{is} close to obtaining an asymptotic value, which is also observed to be a function of the initial Reynolds number. A simple Richardson extrapolation of the end time values of $\Theta$ for the $\Rey_0=174$, $\Rey_0=348$ and $\Rey_0=697$ cases gives an estimate of $0.765$ for the high Reynolds number limit. \thirdrevtwo{This is} slightly lower than the value obtained with ILES for this problem \citep{Groom2019}. At much later dimensionless times, $\Theta$ \thirdrevtwo{has} been shown to gradually decay as self-similarity is approached \citep{Thornber2017}, however this phenomenon may only occur for sufficiently high Reynolds number turbulence.

\subsection{Reynolds number effects}
\label{subsec:reynolds}
\subsubsection{Velocity field}
\begin{figure}
	\centering
	\includegraphics[width=0.49\textwidth]{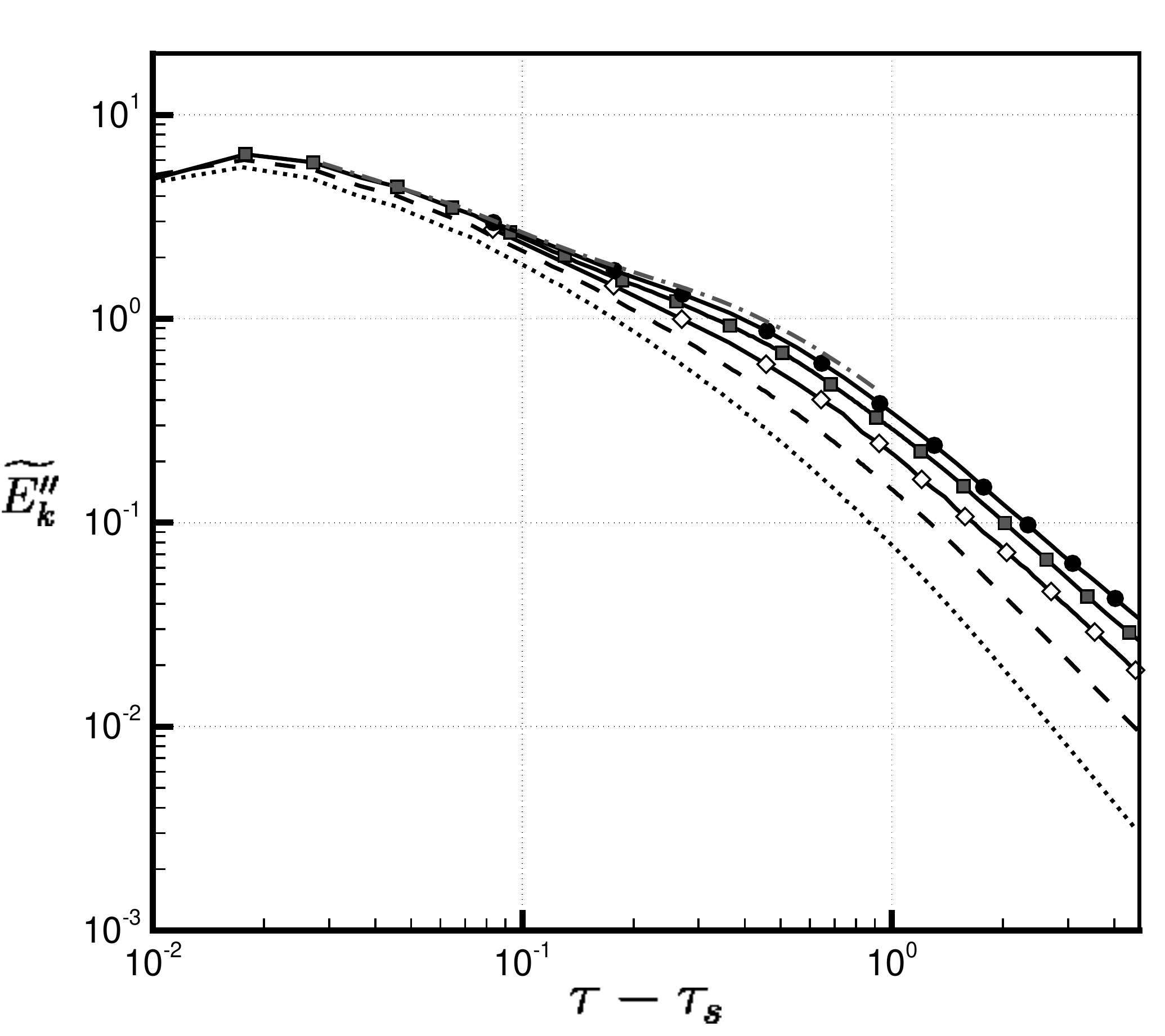}\llap{
		\parbox[b]{0.49\textwidth}{($a$)\\\rule{0ex}{0.40\textwidth}}}
	\includegraphics[width=0.49\textwidth]{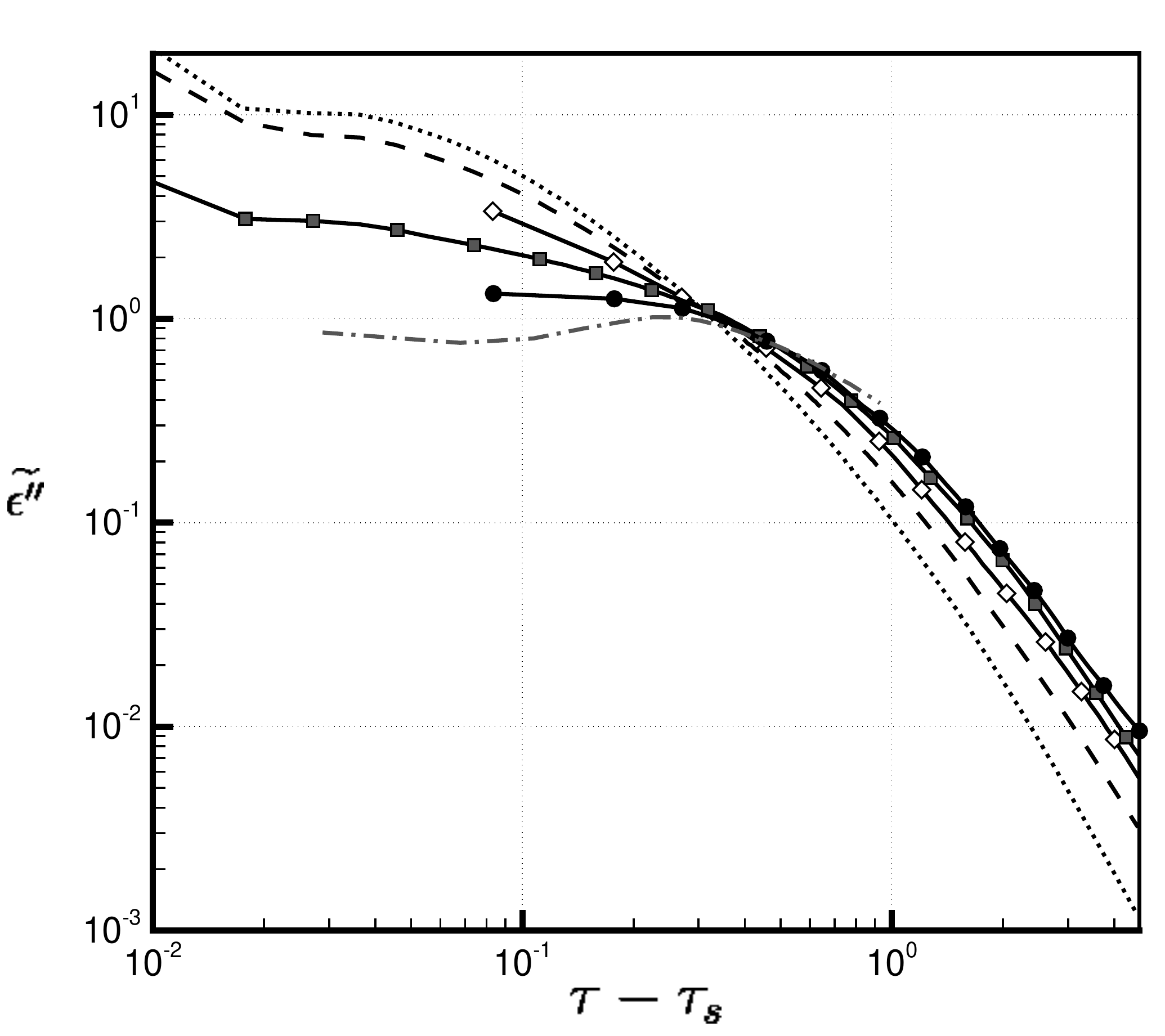}\llap{
		\parbox[b]{0.49\textwidth}{($b$)\\\rule{0ex}{0.40\textwidth}}}
	\caption{Temporal evolution of $(a)$ turbulent kinetic energy and $(b)$ dissipation rate at the mixing layer centre plane. Shown are data for $\Rey_0=43$ (dotted black lines), $\Rey_0=86$ (dashed black lines), $\Rey_0=174$ \firstrev{(white diamonds)}, $\Rey_0=348$ \firstrev{(grey squares)}, $\Rey_0=697$ (black circles) and $\Rey_0=1395$ \firstrev{(dash-dot grey lines)}.}
	\label{fig:TKE-EPS}
\end{figure}

The observed Reynolds number dependence in \S\ref{subsec:mix} motivates a systematic study of Reynolds number effects in both the velocity and scalar fields. The first quantity considered is the turbulent kinetic energy, defined as
\begin{equation}
\widetilde{E_k^{\prime\prime}} = \frac{1}{2}\widetilde{u_i^{\prime\prime}u_i^{\prime\prime}},
\label{eqn:TKE}
\end{equation} 
where $\psi^{\prime\prime}=\psi-\widetilde{\psi}$ indicates a fluctuating quantity and $\widetilde{\psi}=\overline{\rho\psi}/\overline{\rho}$ is a Favre average. A plane average taken over the statistically homogeneous directions \thirdrevtwo{(i.e. $y$-$z$ planes)} is used to calculate the ensemble average $\overline{\psi}$ of a quantity $\psi$. The dissipation rate of the Favre-averaged turbulent kinetic energy is given by
\begin{equation}
\widetilde{\epsilon^{\prime\prime}}=\frac{2\bar{\mu}}{\bar{\rho}}\left(\widetilde{s_{ij}^{\prime\prime}s_{ij}^{\prime\prime}}-\frac{1}{3}\widetilde{\theta^{\prime\prime2}}\right),
\label{eqn:eps}
\end{equation}
where $s_{ij}^{\prime\prime}=\frac{1}{2}\left(\frac{\p u_i^{\prime\prime}}{\p x_j}+\frac{\p u_j^{\prime\prime}}{\p x_i}\right)$ is the fluctuating strain rate tensor and $\theta^{\prime\prime}=\frac{\p u_l^{\prime\prime}}{\p x_l}$ \citep{Chassaing2002}. Figure \ref{fig:TKE-EPS} shows the evolution in time of $\widetilde{E_k^{\prime\prime}}$ and $\widetilde{\epsilon^{\prime\prime}}$ at the mixing layer centre plane $x_c$, which is defined as the $x$ position of equal mixed volumes \citep{Walchli2017}, given by
\begin{equation}
\int_{-\infty}^{x_c}\langle f_2\rangle\:\mathrm{d}x=\int_{x_c}^{\infty}\langle f_1\rangle\:\mathrm{d}x.
\label{eqn:xc}
\end{equation}
Both quantities exhibit a decay in time as the kinetic energy initially deposited by the shock wave is converted into internal energy by irreversible processes. The initial amount of turbulent kinetic energy is also essentially the same for all cases, with only very small differences observed due to slightly different values of $\updelta^-$. The dissipation rate is initially highest for the $\Rey_0=43$ case and lowest for the $\Rey_0=1395$ case, and at all times considered there is more turbulent kinetic energy in the flow for increasing $\Rey_0$. The turbulent kinetic energy is also monotonically decreasing in time for all cases, as is the dissipation rate in all cases except for the $\Rey_0=1395$ case, which exhibits a maximum at time \firstrev{$\tau-\tau_s=0.224$}. At late times the dissipation rate increases with increasing $\Rey_0$ due to the presence of more turbulent structures in the mixing layer. Beyond the initial transient stage, the turbulent kinetic energy decays as $\widetilde{E_k^{\prime\prime}}\sim t^{-n}$. Using linear regression, the decay rate $n$ is found to be $2.20$, $2.00$ $1.83$, $1.76$, $1.68$, $1.59$ and $1.51$ in order of lowest to highest initial Reynolds number (excluding $\Rey_0=1395$). \thirdrev{All of these decay rates are steeper than the $t^{-10/7}$ or $t^{-6/5}$ decay typical of homogeneous turbulence with a Batchelor or Saffman spectrum.} Similarly, the dissipation rate is expected to decay as $\widetilde{\epsilon^{\prime\prime}}\sim t^{-(n+1)}$ with the actual decay rates found to be $3.24$, $2.97$, $2.80$, $2.73$, $2.66$, $2.67$ and $2.40$, in good agreement with the measured values of $n$. In \citet{Groom2019}, the total fluctuating kinetic energy of the mixing layer was found to scale as $\sim t^{-1.41}$ for the $\Rey_0=348$ case. Following the dimensional arguments given in \citet{Thornber2010}, the total fluctuating kinetic energy is proportional to the mixing layer width multiplied by the mean kinetic energy and should therefore scale as $t^\theta t^{-n}=t^{\theta-n}$. For the $\Rey_0=348$ case, this gives a value of ${\theta-n}=-1.38$ which is reasonably close, while the decay rates for the other cases may be related to their bulk values in a similar manner.

\begin{figure}
	\centering
	\includegraphics[width=0.33\textwidth]{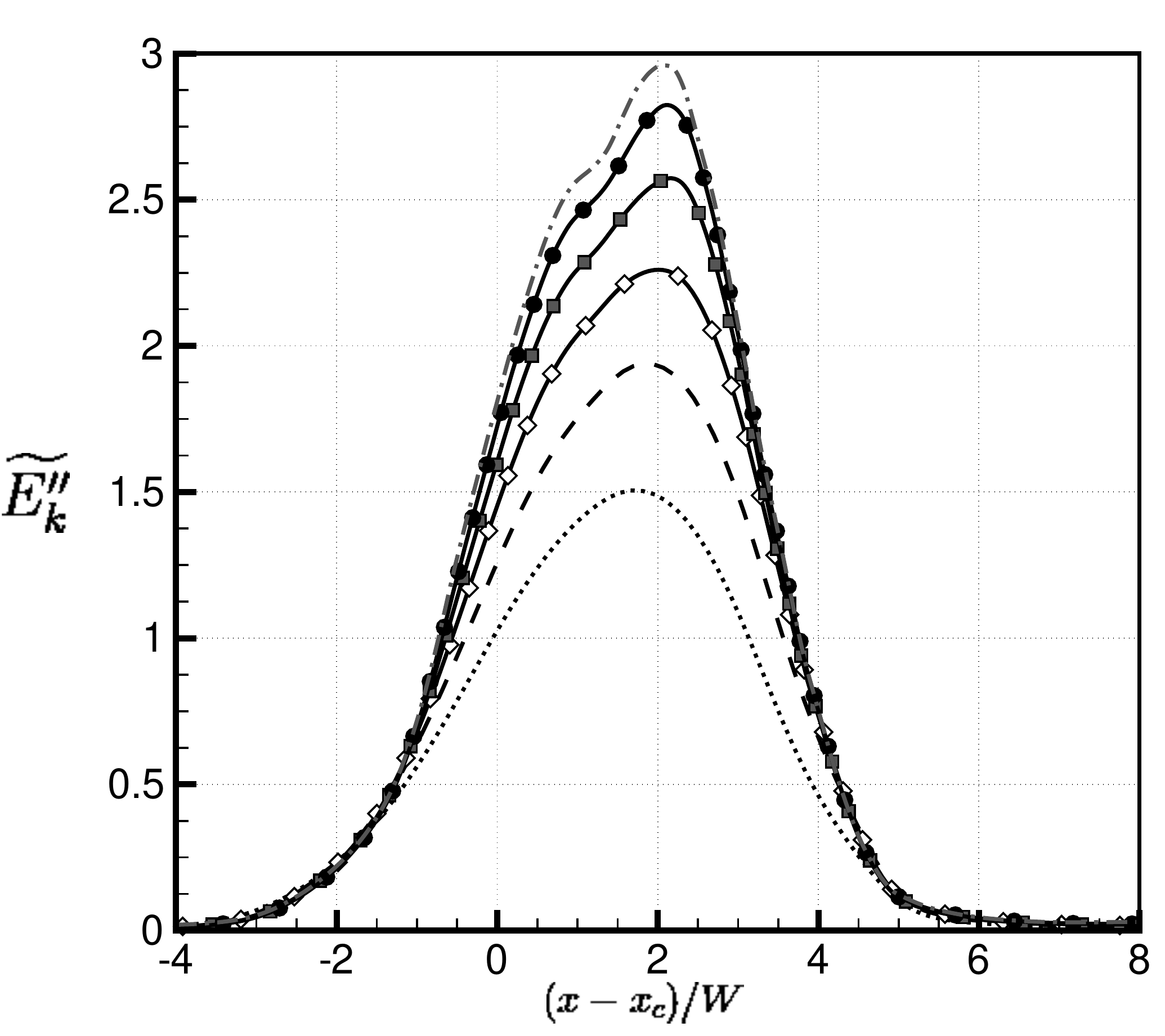}\llap{\parbox[b]{0.33\textwidth}{($a$)\\\rule{0ex}{0.267\textwidth}}}
	\includegraphics[width=0.33\textwidth]{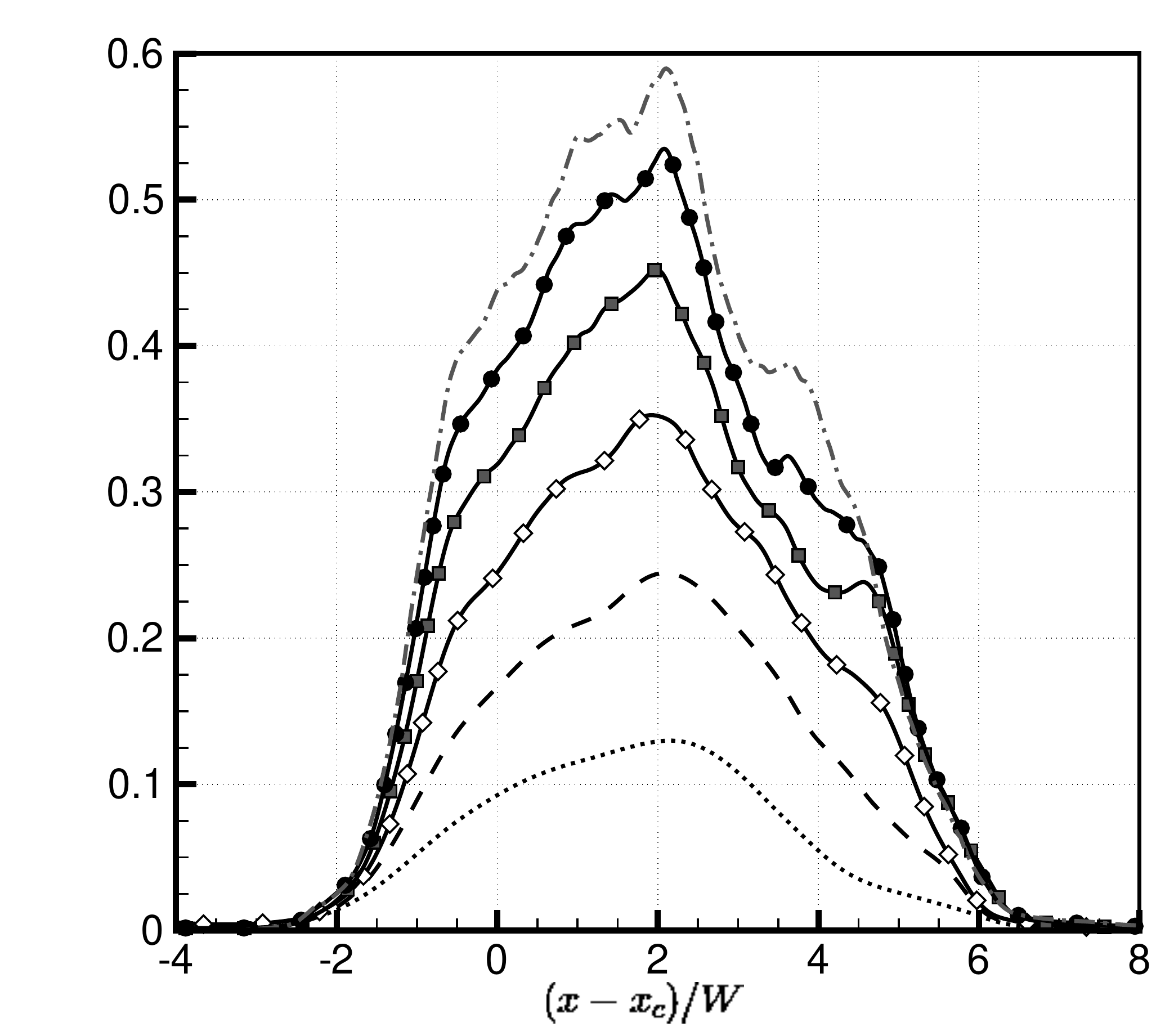}\llap{\parbox[b]{0.33\textwidth}{($b$)\\\rule{0ex}{0.267\textwidth}}}
	\includegraphics[width=0.33\textwidth]{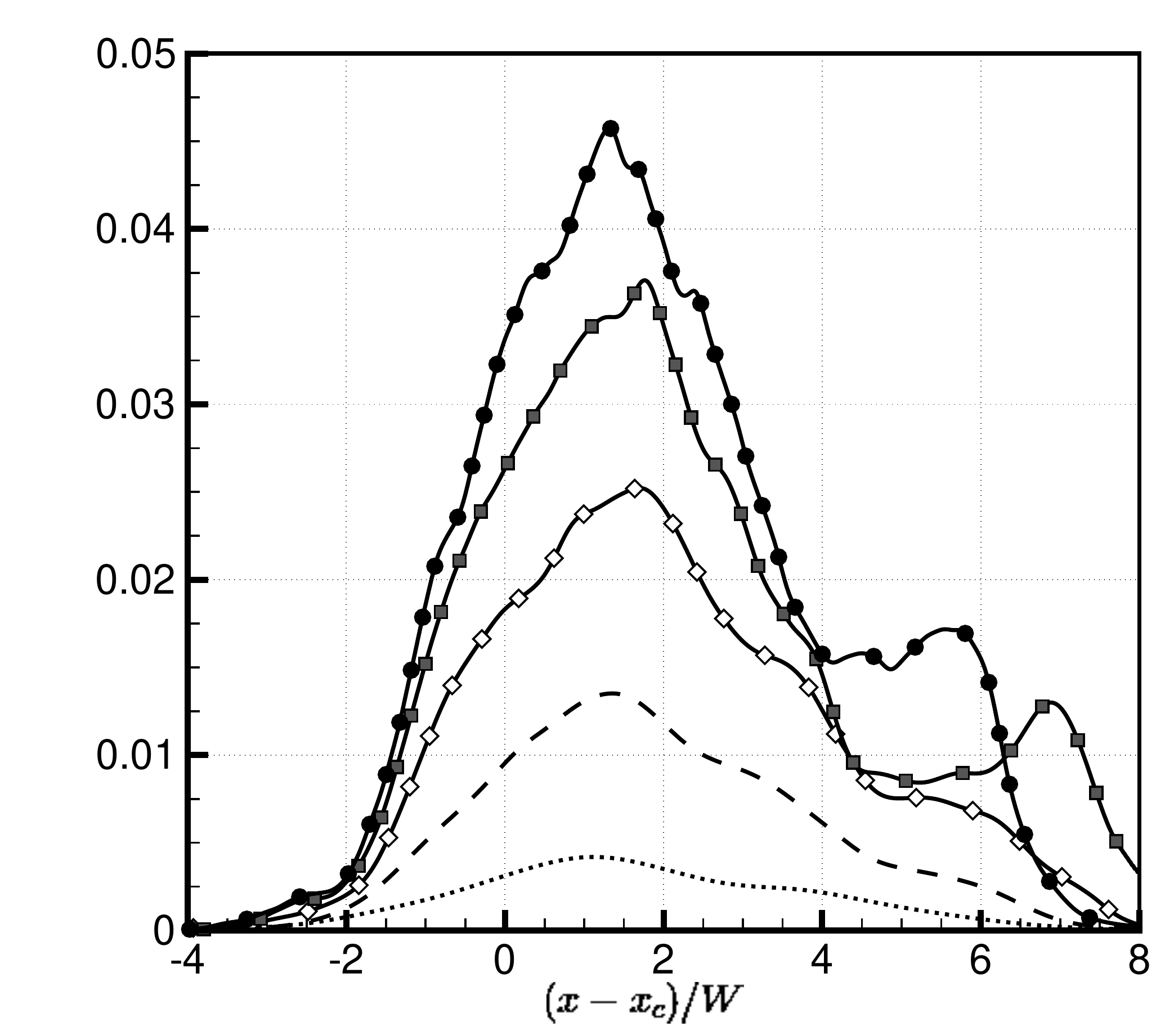}\llap{\parbox[b]{0.333\textwidth}{($c$)\\\rule{0ex}{0.267\textwidth}}}
	\caption{Spatial distribution of turbulent kinetic energy in the $x$-direction for times ($a$) $\tau=0.187$, ($b$) $\tau=0.939$ and ($c$) $\tau=4.70$. Shown are data for $\Rey_0=43$ (dotted black lines), $\Rey_0=86$ (dashed black lines), $\Rey_0=174$ \firstrev{(white diamonds)}, $\Rey_0=348$ \firstrev{(grey squares)}, $\Rey_0=697$ (black circles) and $\Rey_0=1395$ \firstrev{(dash-dot grey lines)}.}
	\label{fig:TKE-X}
\end{figure}

\begin{figure}
	\centering
	\includegraphics[width=0.33\textwidth]{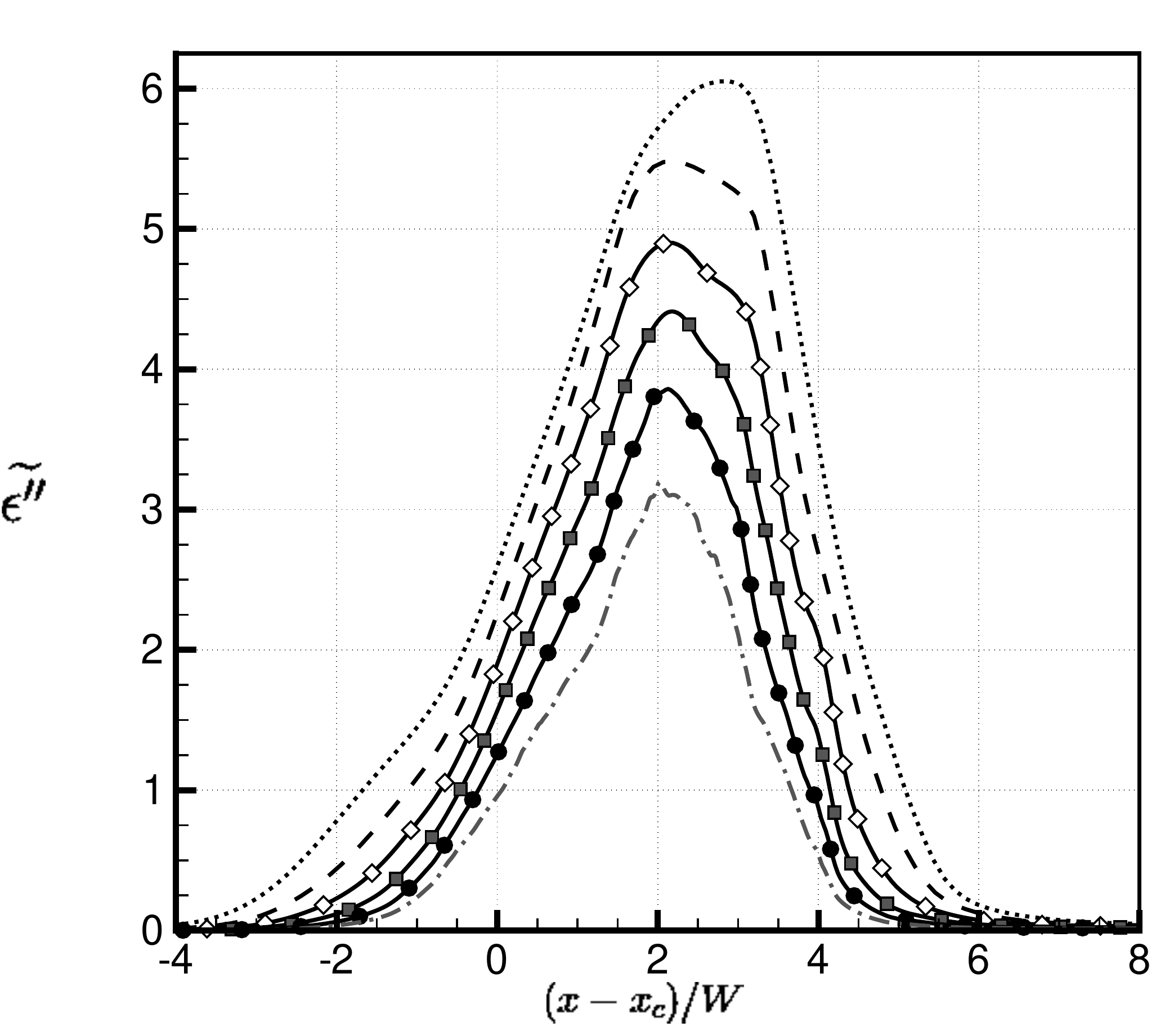}\llap{
		\parbox[b]{0.33\textwidth}{($a$)\\\rule{0ex}{0.267\textwidth}}}
	\includegraphics[width=0.33\textwidth]{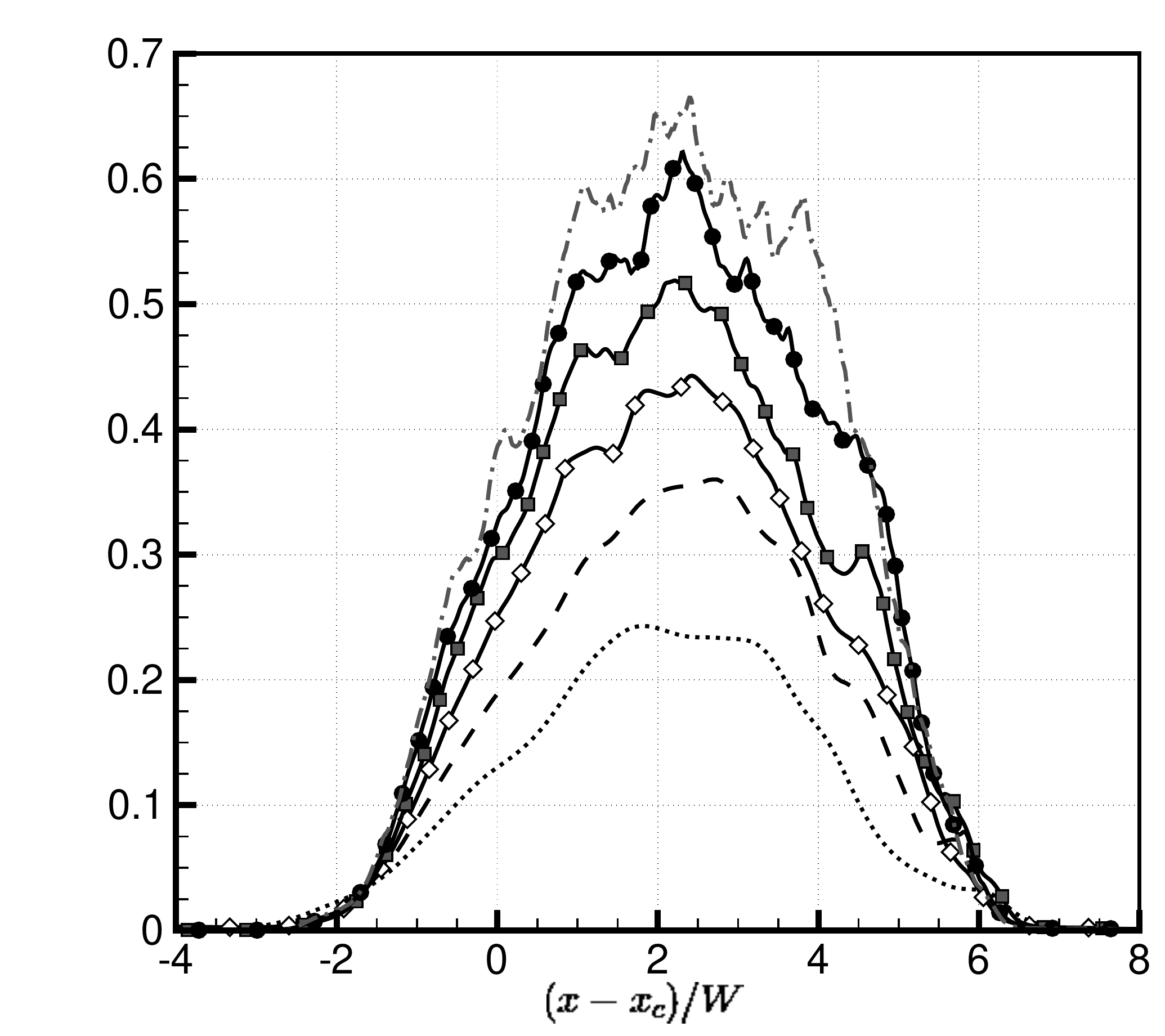}\llap{
		\parbox[b]{0.33\textwidth}{($b$)\\\rule{0ex}{0.267\textwidth}}}
	\includegraphics[width=0.33\textwidth]{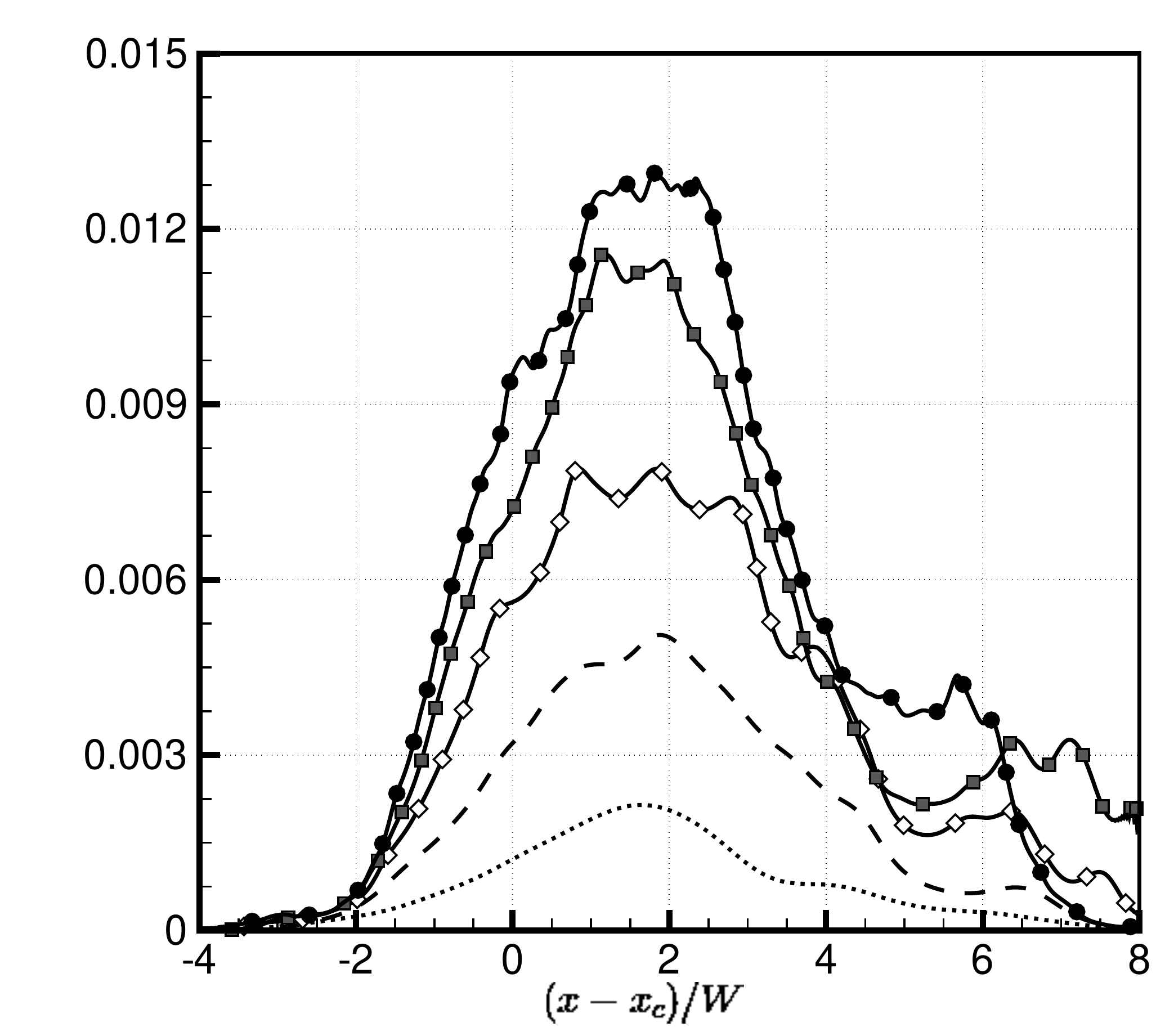}\llap{
		\parbox[b]{0.333\textwidth}{($c$)\\\rule{0ex}{0.267\textwidth}}}
	\caption{Spatial distribution of dissipation rate in the $x$-direction for times ($a$) $\tau=0.187$, ($b$) $\tau=0.939$ and ($c$) $\tau=4.70$. Shown are data for $\Rey_0=43$ (dotted black lines), $\Rey_0=86$ (dashed black lines), $\Rey_0=174$ \firstrev{(white diamonds)}, $\Rey_0=348$ \firstrev{(grey squares)}, $\Rey_0=697$ (black circles) and $\Rey_0=1395$ \firstrev{(dash-dot grey lines)}.}
	\label{fig:EPS-X}
\end{figure}

Figures \ref{fig:TKE-X} and \ref{fig:EPS-X} show the spatial distribution across the layer of $\widetilde{E_k^{\prime\prime}}$ and $\widetilde{\epsilon^{\prime\prime}}$ at three different points in time. The $x$-coordinate is normalised by the integral width $W$ and is centred about $x_c$. To give some context to the figures, the locations at which $\langle f_1 \rangle=0.99$ and $\langle f_1 \rangle=0.01$ (i.e. the 99\% bubble and spike heights) range between $(x-x_c)/W=-2.8$ to $-3.0$ and $(x-x_c)/W=4.9$ to $5.1$ respectively throughout the simulation. At the earliest time shown, the turbulent kinetic energy profile is biased towards the spike side of the layer, with the peak occurring at a distance of about two integral widths from the layer centre in all cases. This is also observed for the dissipation rate. As time progresses, the profiles become more symmetric about $x_c$, although there is a persistent bias towards the spike side, indicating that more turbulent fluctuations are occurring there. The difference between the profiles of the highest and lowest $\Rey_0$ cases also increases throughout the simulation, particularly at the very fringes of the spike side of the layer. Indeed, at the latest time considered, there is a substantially higher amount of turbulent kinetic energy (as well as a larger dissipation rate) in this region for the $\Rey_0=348$ and $\Rey_0=697$ cases than in any of the lower Reynolds number cases. This suggests that there are spikes penetrating deep into the light fluid at these Reynolds numbers, and which break down more rapidly at lower Reynolds numbers. In fact, the spikes in the $\Rey_0=348$ case actually penetrate further. This is because of a lower amount of turbulent dissipation inhibiting their growth, as was previously mentioned in \S \ref{subsec:mix} for the integral width. A similar phenomenon can also be observed in \citet{Thornber2017} for the ILES codes with greater numerical dissipation. 


\begin{figure}
	\centering
	\includegraphics[width=0.33\textwidth]{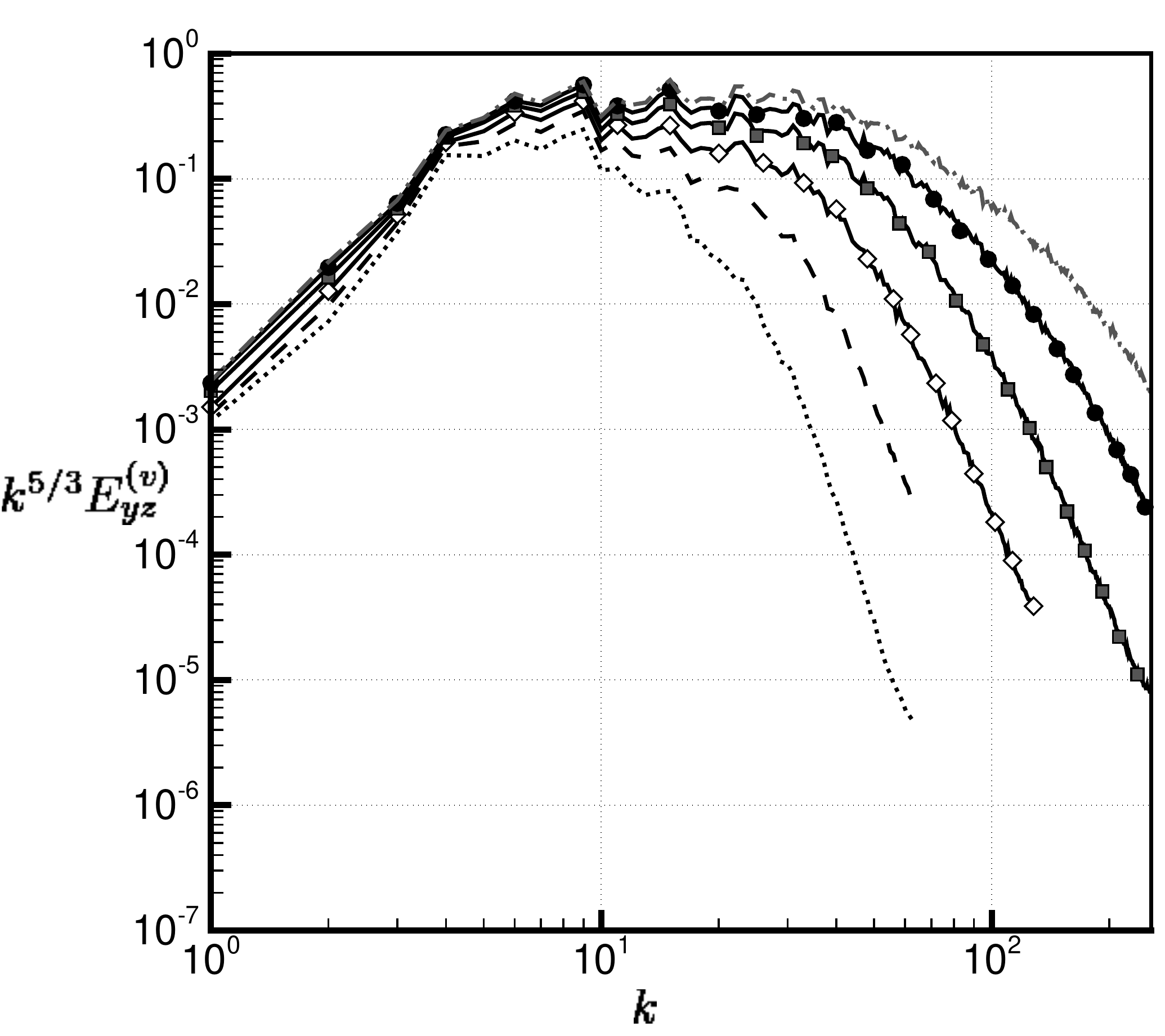}\llap{
		\parbox[b]{0.33\textwidth}{($a$)\\\rule{0ex}{0.267\textwidth}}}
	\includegraphics[width=0.33\textwidth]{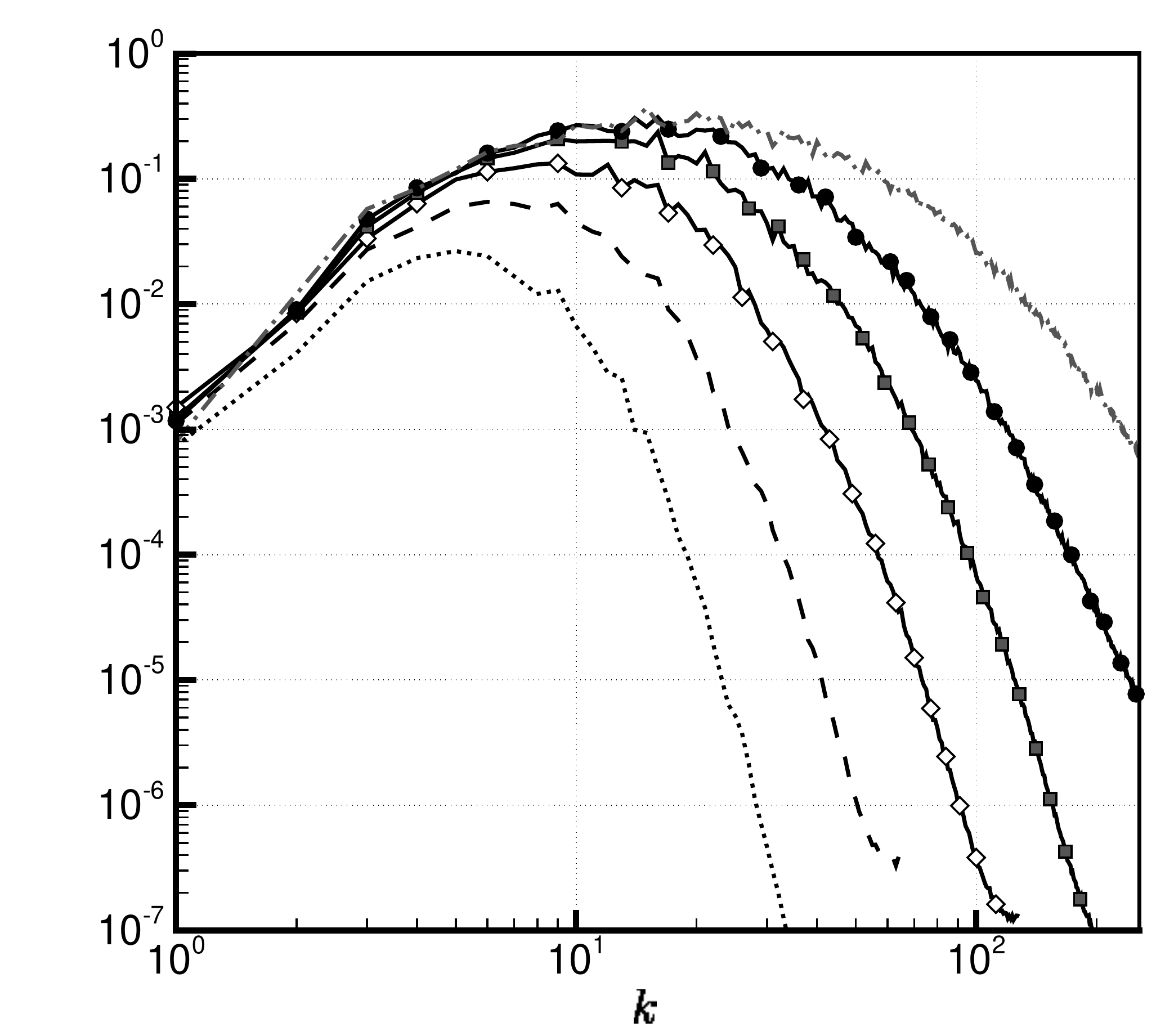}\llap{
		\parbox[b]{0.33\textwidth}{($b$)\\\rule{0ex}{0.267\textwidth}}}
	\includegraphics[width=0.33\textwidth]{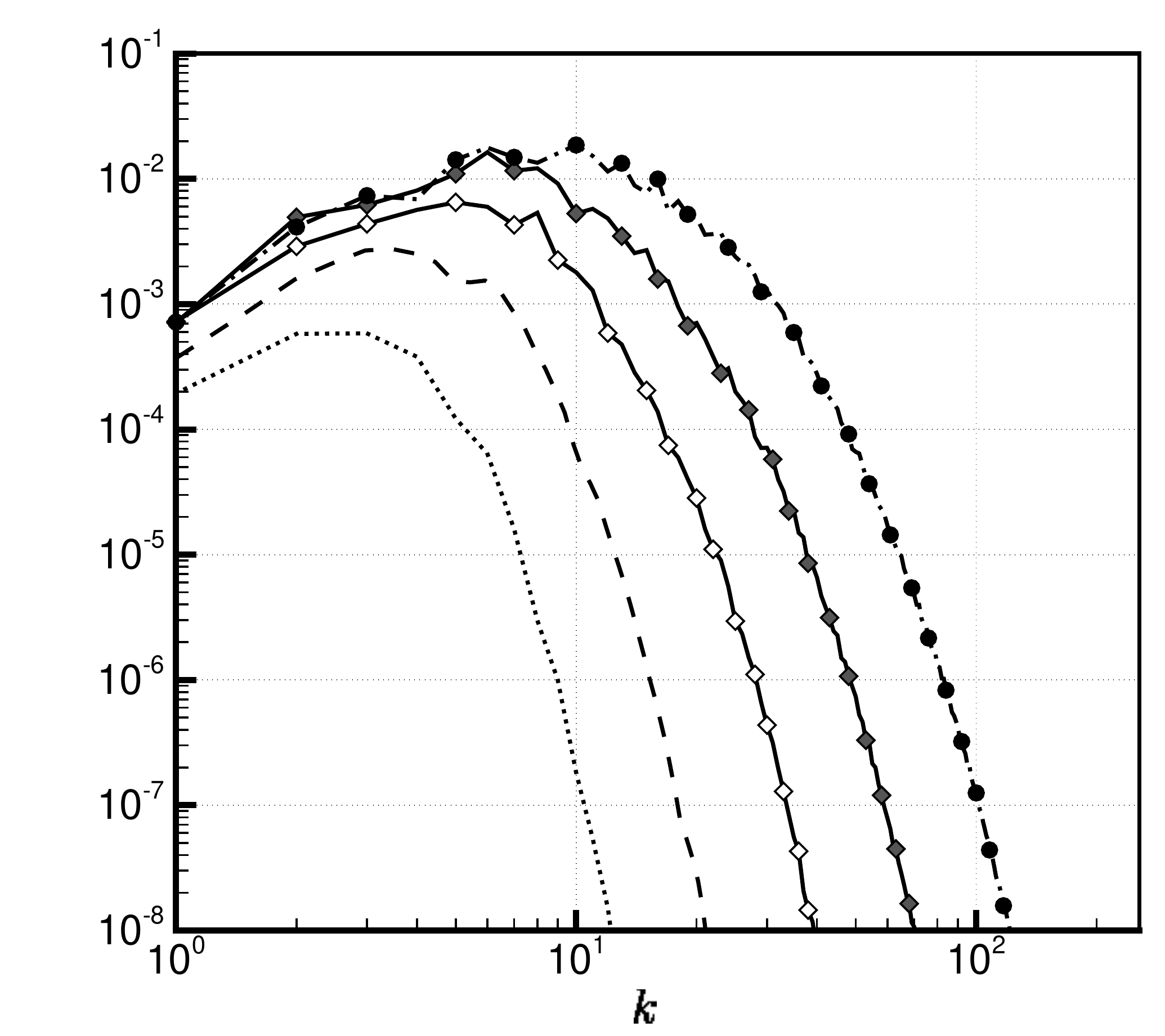}\llap{
		\parbox[b]{0.33\textwidth}{($c$)\\\rule{0ex}{0.267\textwidth}}}
	\caption{\thirdrev{Compensated} power spectra of the average transverse turbulent kinetic energy per unit volume taken at the mixing layer centre plane for times ($a$) $\tau=0.187$, ($b$) $\tau=0.939$ and ($c$) $\tau=4.70$. Shown are data for $\Rey_0=43$ (dotted black lines), $\Rey_0=86$ (dashed black lines), $\Rey_0=174$ \firstrev{(white diamonds)}, $\Rey_0=348$ \firstrev{(grey squares)}, $\Rey_0=697$ (black circles) and $\Rey_0=1395$ \firstrev{(dash-dot grey lines)}.}
	\label{fig:KEV-K}
\end{figure}

\begin{figure}
	\centering
	\includegraphics[width=0.33\textwidth]{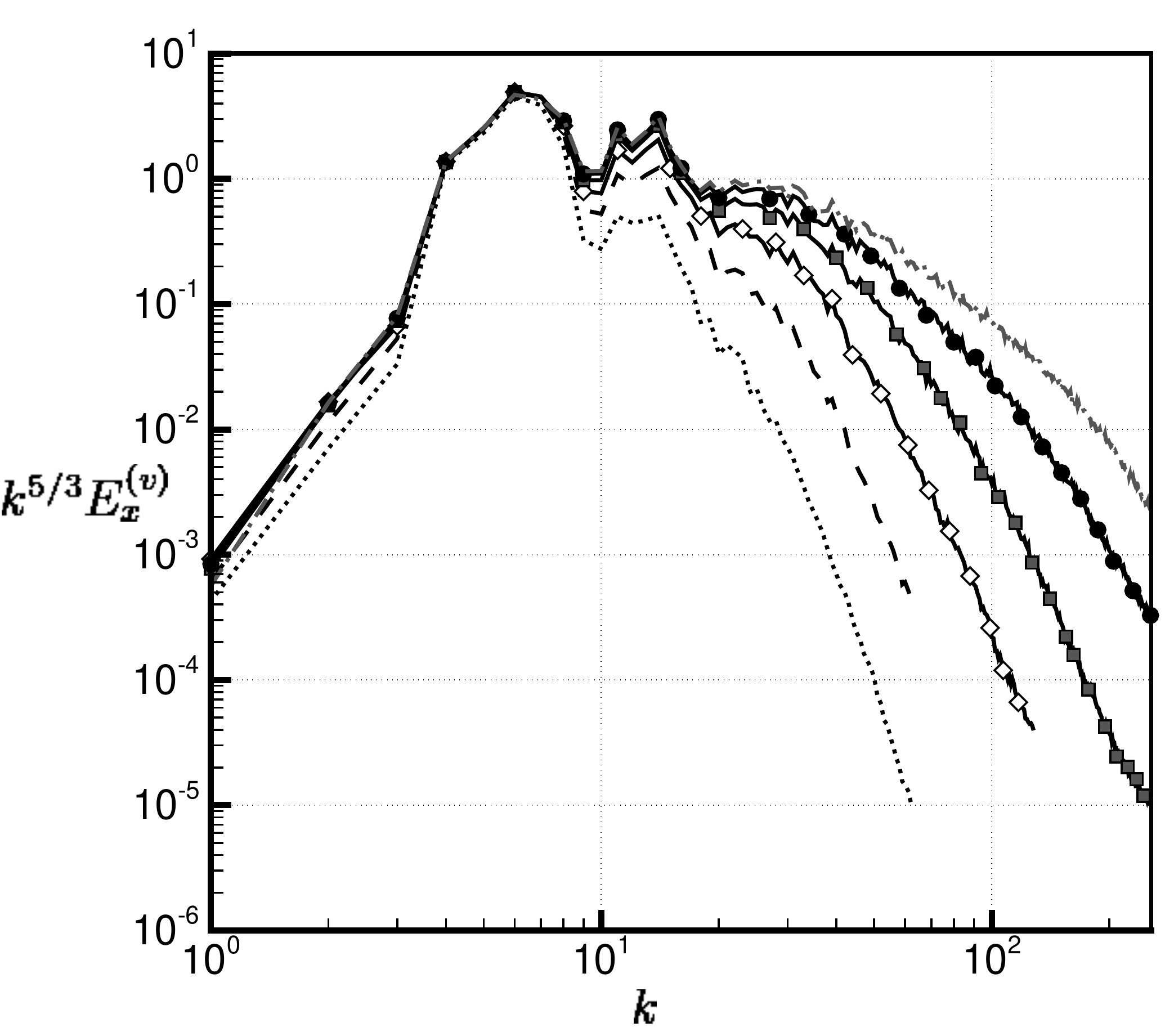}\llap{
		\parbox[b]{0.33\textwidth}{($a$)\\\rule{0ex}{0.267\textwidth}}}
	\includegraphics[width=0.33\textwidth]{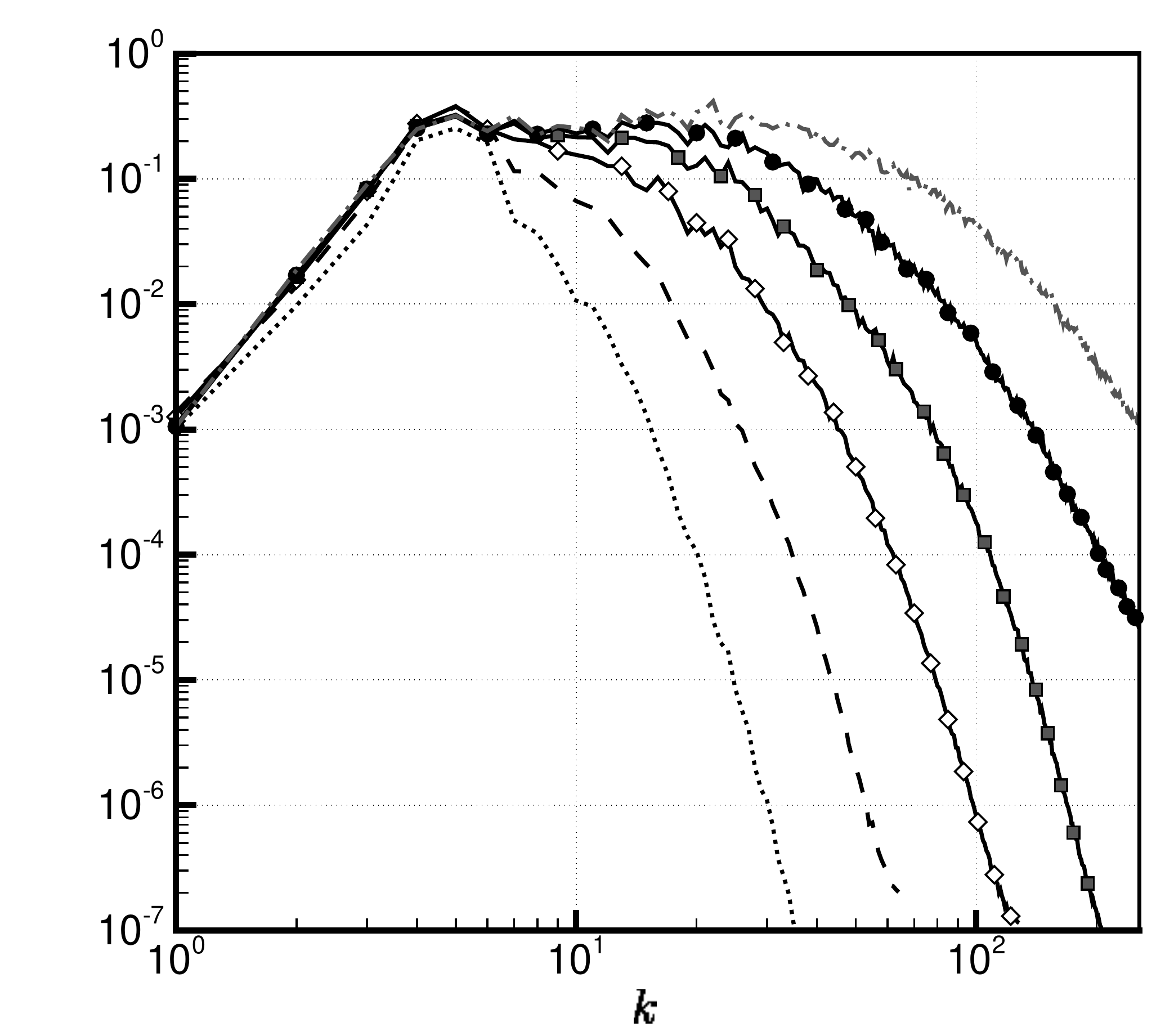}\llap{
		\parbox[b]{0.33\textwidth}{($b$)\\\rule{0ex}{0.267\textwidth}}}
	\includegraphics[width=0.33\textwidth]{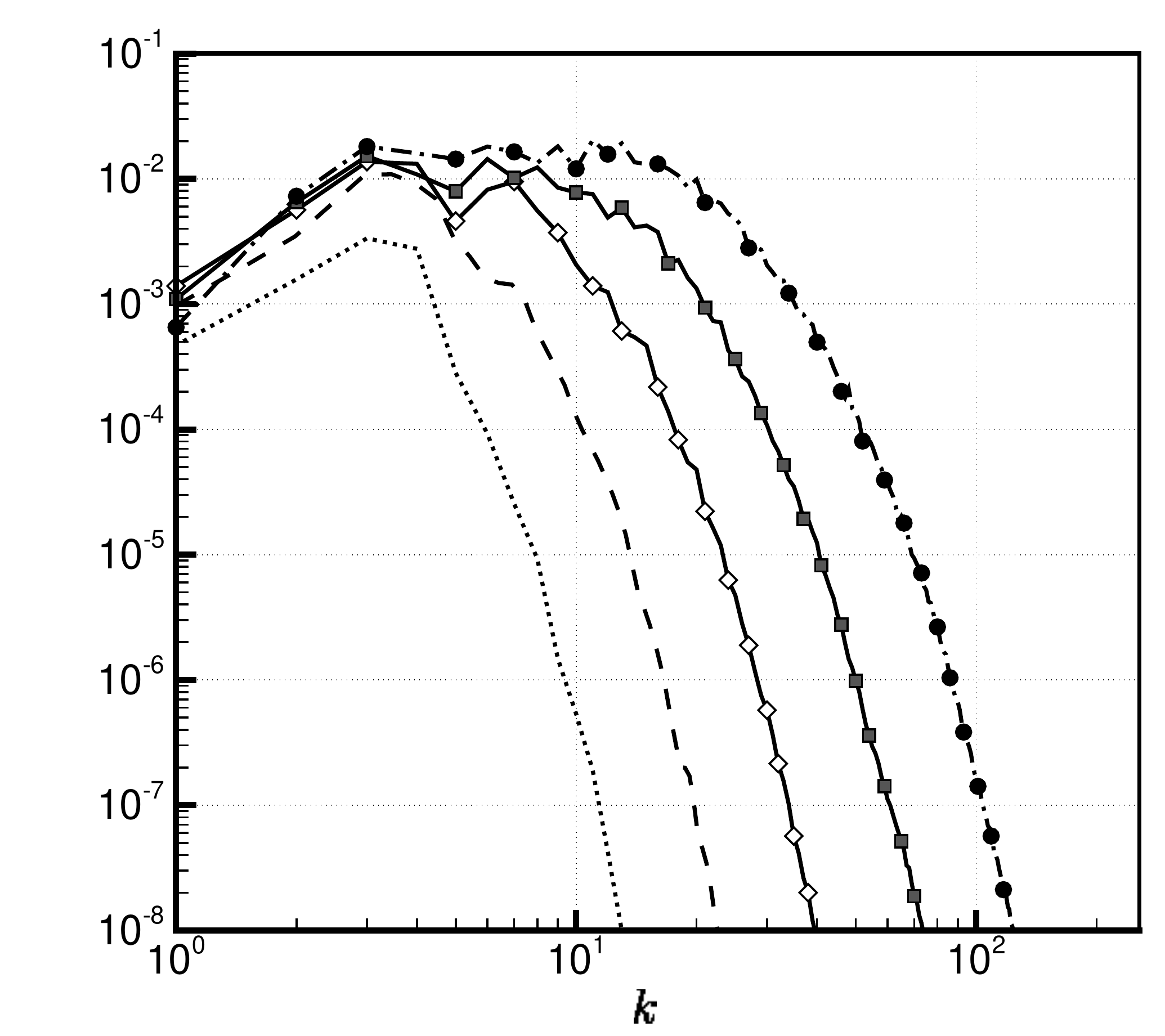}\llap{
		\parbox[b]{0.33\textwidth}{($c$)\\\rule{0ex}{0.267\textwidth}}}
	\caption{\thirdrev{Compensated} power spectra of the normal turbulent kinetic energy per unit volume taken at the mixing layer centre plane for times ($a$) $\tau=0.187$, ($b$) $\tau=0.939$ and ($c$) $\tau=4.70$. Shown are data for $\Rey_0=43$ (dotted black lines), $\Rey_0=86$ (dashed black lines), $\Rey_0=174$ \firstrev{(white diamonds)}, $\Rey_0=348$ \firstrev{(grey squares)}, $\Rey_0=697$ (black circles) and $\Rey_0=1395$ \firstrev{(dash-dot grey lines)}.}
	\label{fig:KEX-K}
\end{figure}

The distribution of turbulent kinetic energy is also examined in spectral space. Radial power spectra for each component of the turbulent kinetic energy per unit volume are calculated at the mixing layer centre plane as
\begin{equation}
E_{i}^{(v)}(k) = \widehat{\psi_i}^\dagger\widehat{\psi_i},
\label{eqn:E2D}
\end{equation}
where $\psi_i=\sqrt{\rho}u_i^{\prime\prime}$, $k=\sqrt{k_y+k_z}$ is the radial wavenumber in the $y$-$z$ plane at $x=x_c$, $\widehat{(\ldots)}$ denotes the 2D Fourier transform taken over this plane and $\widehat{(\ldots)}^\dagger$ is the complex conjugate of this transform \citep{Cook2002}. As isotropy is expected in the homogeneous directions, the spectra $E_{y}^{(v)}$ and $E_{z}^{(v)}$ are averaged to give a single transverse spectrum $E_{yz}^{(v)}$. This spectrum \thirdrev{(compensated by the Kolmogorov $k^{5/3}$ scaling)} is shown in figure \ref{fig:KEV-K} for each case at three different times, while the \thirdrev{compensated} spectra of the normal component $E_{x}^{(v)}$ is shown in figure \ref{fig:KEX-K}. There is an extremely similar distribution of energy at the large scales across all cases at time $\tau=0.187$, particularly for the normal component. This is in agreement with the observations made for figure \ref{fig:TKE-EPS} at early time. Substantially more energy is contained at the small scales as $\Rey_0$ is increased, however this represents a small fraction of the total turbulent kinetic energy in the flow. By the end of the simulation there are much greater differences in the energy contained in the large scales between cases, while the differences at the small scales are even greater than at earlier times. 


\begin{figure}
	\centering
	\includegraphics[width=0.33\textwidth]{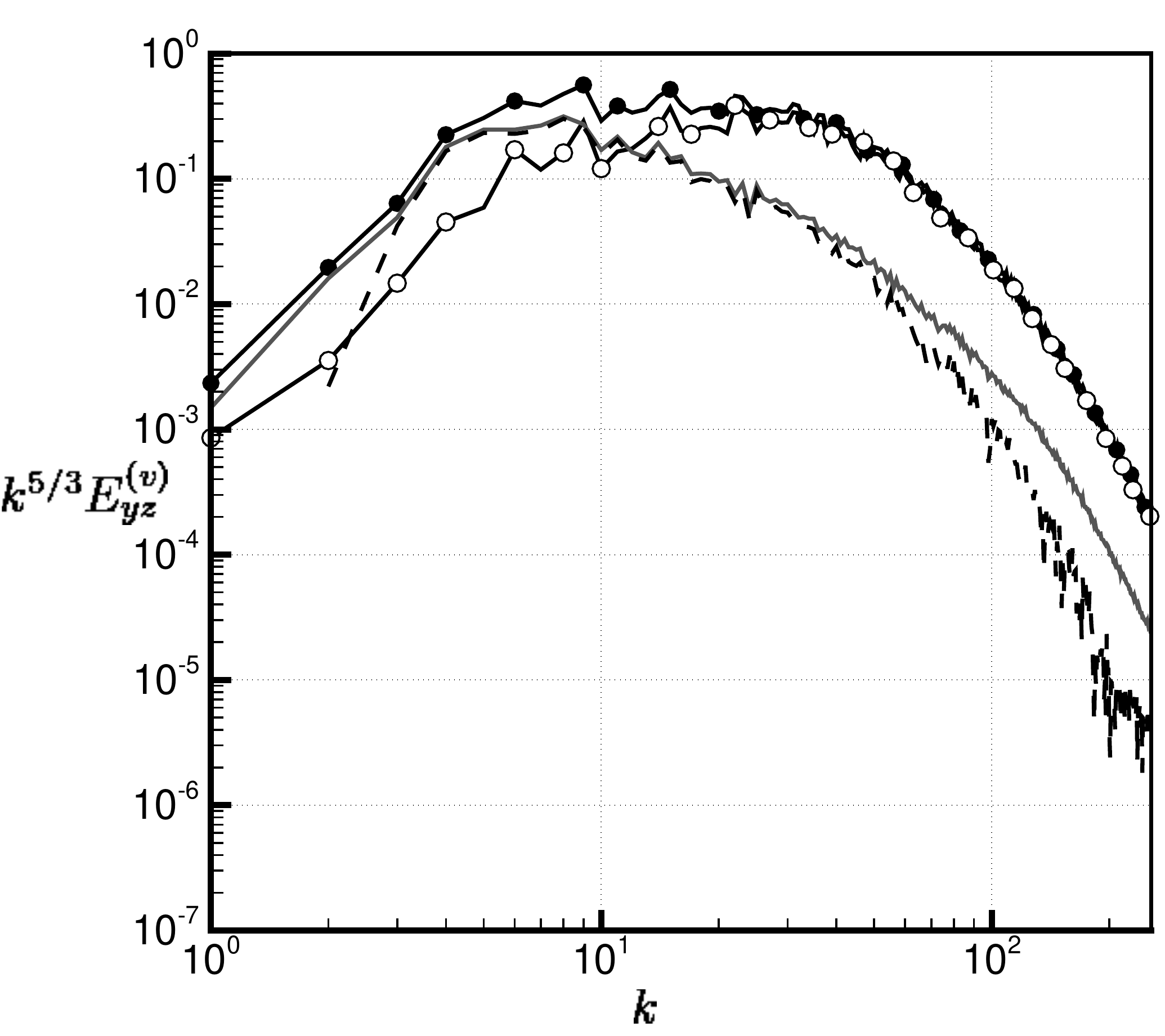}\llap{
		\parbox[b]{0.33\textwidth}{($a$)\\\rule{0ex}{0.267\textwidth}}}
	\includegraphics[width=0.33\textwidth]{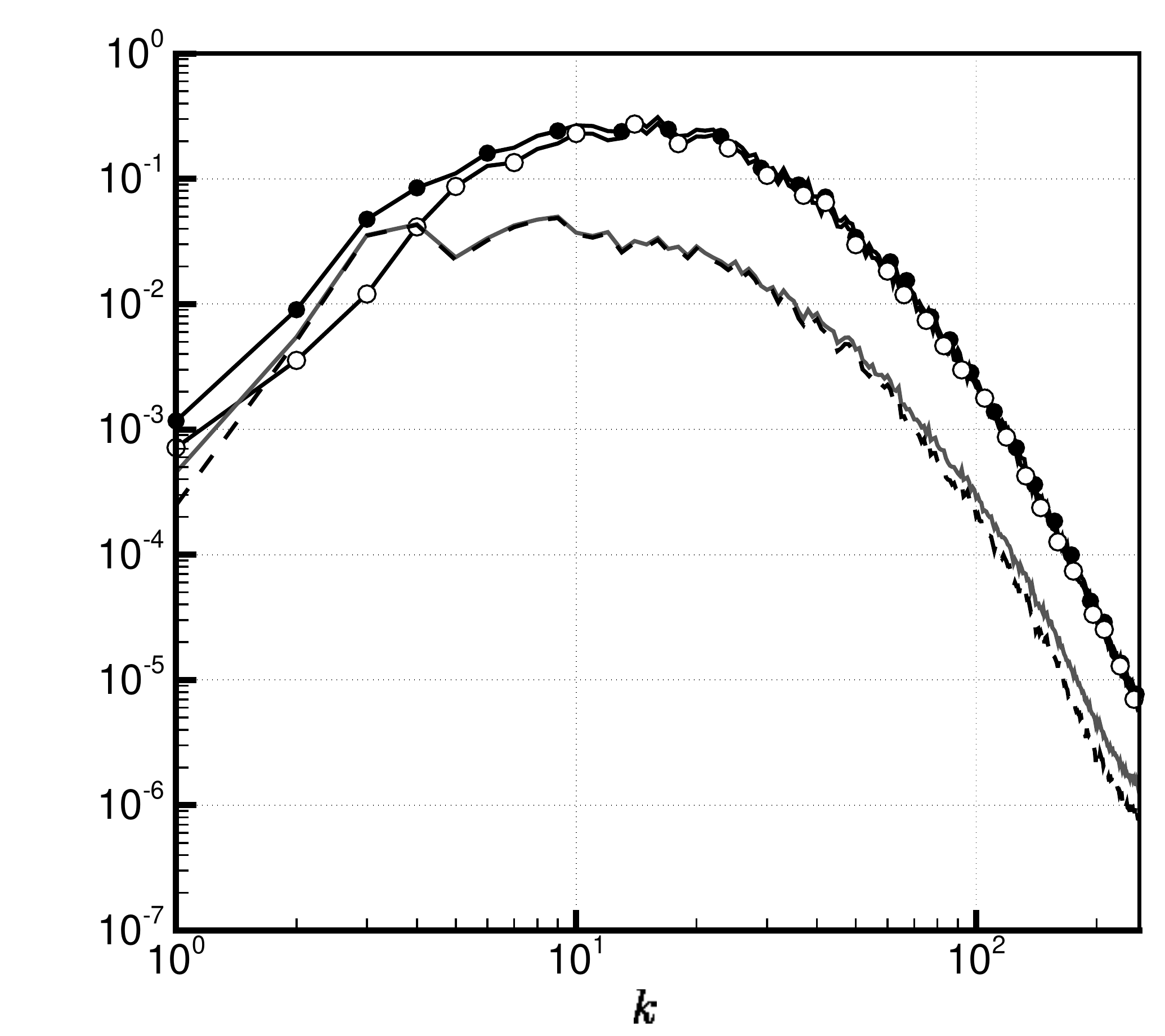}\llap{
		\parbox[b]{0.33\textwidth}{($b$)\\\rule{0ex}{0.267\textwidth}}}
	\includegraphics[width=0.33\textwidth]{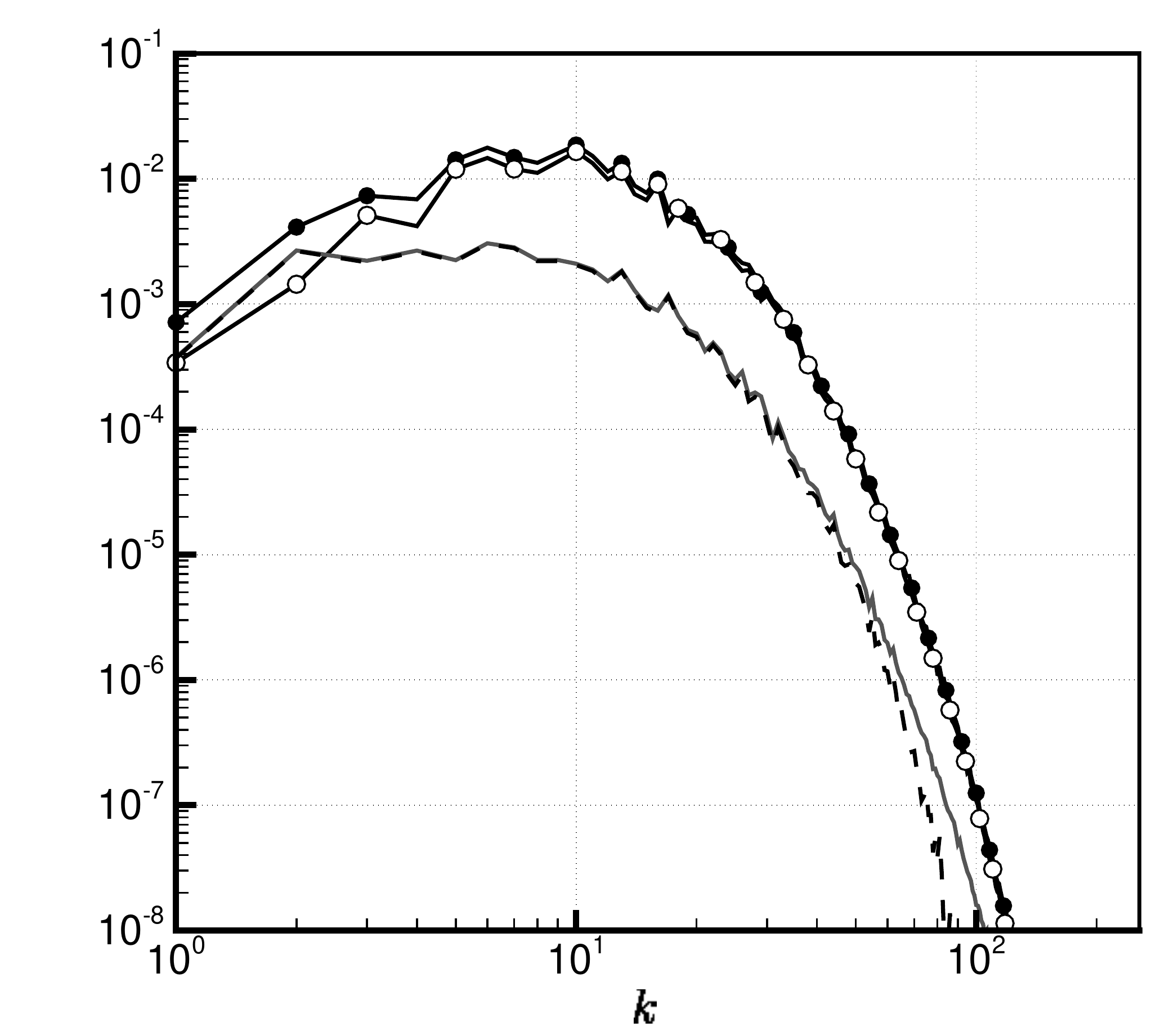}\llap{
		\parbox[b]{0.33\textwidth}{($c$)\\\rule{0ex}{0.267\textwidth}}}
	\caption{Decomposition of the \thirdrev{compensated} average transverse turbulent kinetic energy per unit volume (black circles) in the $\Rey_0=697$ case into solenoidal \firstrev{(white circles)}, total dilatational \firstrev{(solid grey lines)} and compressible \firstrev{(dashed black lines)} components for times ($a$) $\tau=0.187$, ($b$) $\tau=0.939$ and ($c$) $\tau=4.70$.}
	\label{fig:KEV-Re913}
\end{figure}

The \thirdrev{$k^{5/3}E_{yz}^{(v)}$} spectra at time $\tau=0.187$ indicate the presence of a power law scaling of the intermediate wavenumbers for the higher $\Rey_0$ cases, spanning roughly half a decade. The slope is close to a Kolmogorov $k^{-5/3}$ scaling, with the \thirdrev{(uncompensated)} spectra for the $\Rey_0=697$ and $\Rey_0=1395$ cases observed to scale as $k^{-1.78}$ and $k^{-1.65}$ respectively when measured over the range of wavenumbers $8\le k\le 30$. \thirdrev{Similar scalings are also observed in the $k^{5/3}E_{x}^{(v)}$ spectra at times $\tau=0.939$ and $\tau=4.70$.} \citet{Tritschler2014pre} also observed a power law scaling in the turbulent kinetic energy spectra from their DNS at early time (shortly after shock passage) over a similar span of wavenumbers with a scaling close to $k^{-5/3}$. 
%
%
It is hypothesised that there may be a substantial influence from acoustic waves on the spectra at early time. To investigate this, the vector field $\boldsymbol{\psi}_{yz}=\sqrt{\rho}[v^{\prime\prime},w^{\prime\prime}]^t=\sqrt{\rho}\boldsymbol{u}_{yz}^{\prime\prime}$ is decomposed into its solenoidal and dilatational components as
\begin{equation}
\boldsymbol{\psi}_{yz}=\boldsymbol{\nabla}\phi+\boldsymbol{\nabla}\times\xi.
\end{equation}
A further distinction is made between dilatation due to compressibility effects and dilatation due to variable-density mixing in the incompressible limit. This is performed by using the relation in (\ref{eqn:diffusion-velocity}) to calculate the divergence of $\boldsymbol{\psi}_{yz}$ in the incompressible limit as
\begin{equation}
\boldsymbol{\nabla}\bcdot\boldsymbol{\psi}_{yz}=\frac{D}{\sqrt{\rho}}\left(\frac{\boldsymbol{\nabla}\rho\bcdot\boldsymbol{\nabla}\rho}{\rho}-\nabla^2\rho\right)+\frac{1}{2\sqrt{\rho}}\boldsymbol{\nabla}\rho\bcdot\boldsymbol{u}_{yz}^{\prime\prime}=g,
\end{equation}
and further decomposing $\phi$ into $\phi=\zeta+\alpha$ where $\nabla^2\alpha=g$. In spectral space, the Fourier transform of the total dilatational component $\boldsymbol{\nabla}\phi$ is calculated as
\begin{equation}
\mathcal{F}\{\boldsymbol{\nabla}\phi\}=\frac{\boldsymbol{k}\bcdot\widehat{\boldsymbol{\psi}}_{yz}}{|\boldsymbol{k}|^2}\boldsymbol{k},
\end{equation}
which gives the solenoidal component $\mathcal{F}\{\boldsymbol{\nabla}\times\xi\}=\widehat{\boldsymbol{\psi}}_{yz}-\mathcal{F}\{\boldsymbol{\nabla}\phi\}$. The compressible component is calculated as
\begin{equation}
\mathcal{F}\{\boldsymbol{\nabla}\zeta\}=\frac{\boldsymbol{k}\bcdot\widehat{\boldsymbol{\psi}}_{yz}+\mathrm{i}\widehat{g}}{|\boldsymbol{k}|^2}\boldsymbol{k}.
\end{equation}
These components are used to calculate the solenoidal, total dilatational and compressible turbulent kinetic energy, denoted as $E_{yz}^{(s)}$, $E_{yz}^{(d)}$ and $E_{yz}^{(c)}$ respectively. The \thirdrev{compensated} solenoidal, total dilatational and compressible turbulent kinetic energy for the $\Rey_0=697$ case are plotted in figure \ref{fig:KEV-Re913} (results for the $\Rey_0=1395$ case are shown later in figure \ref{fig:KEV-Re1826}), alongside the \thirdrev{compensated} total transverse turbulent kinetic energy \thirdrev{$k^{5/3}E_{yz}^{(v)}$}. At time $\tau=0.187$ it can be seen that there is a significant contribution from the total dilatational component to the overall energy spectrum in the intermediate wavenumber range, which is almost entirely due to compressibility effects. At later times the spectrum is dominated by the solenoidal component at all but the lowest wavenumbers. This shows that care must be taken when interpreting spectra at early times in RMI flows, as there are significant acoustic effects that are present in the energy-containing scales and to a lesser degree the inertial scales too. These effects are the dominant source of dilatation in the flow, except at the highest wavenumbers, and can significantly alter the shape of the spectrum compared to a fully incompressible flow. \thirdrevtwo{In particular, they can lead to the appearance of an inertial range when in fact one does not exist. The procedure detailed in \S\ref{subsec:transition} shows how to give more precise estimates of the scaling of any inertial range that may form in the energy spectra, as well as when it is appropriate to do so.}

\subsubsection{Scalar field}
\label{subsec:scalar}
\begin{figure} 
	\centering
	\includegraphics[width=0.49\textwidth]{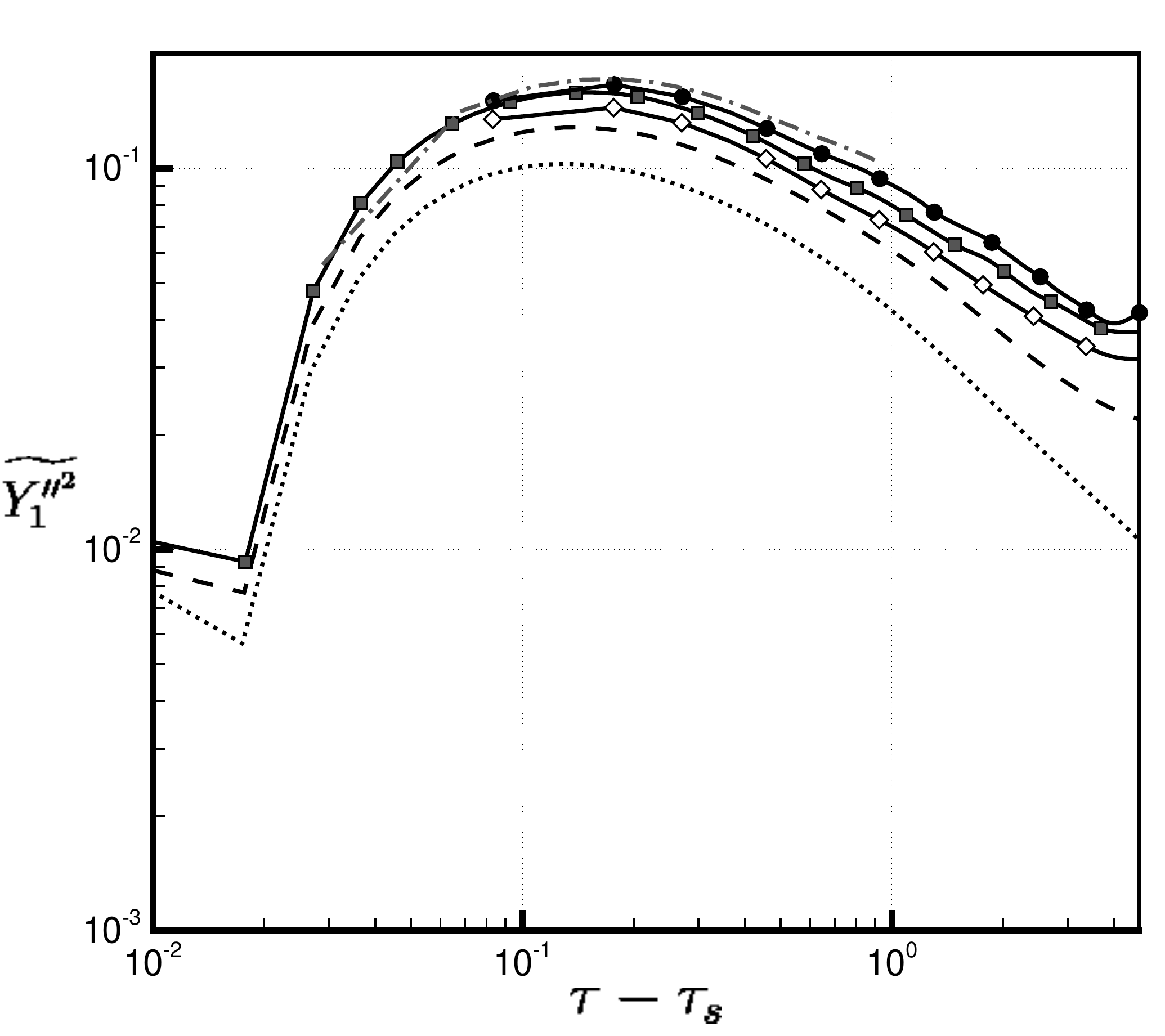}\llap{
		\parbox[b]{0.49\textwidth}{($a$)\\\rule{0ex}{0.40\textwidth}}}
	\includegraphics[width=0.49\textwidth]{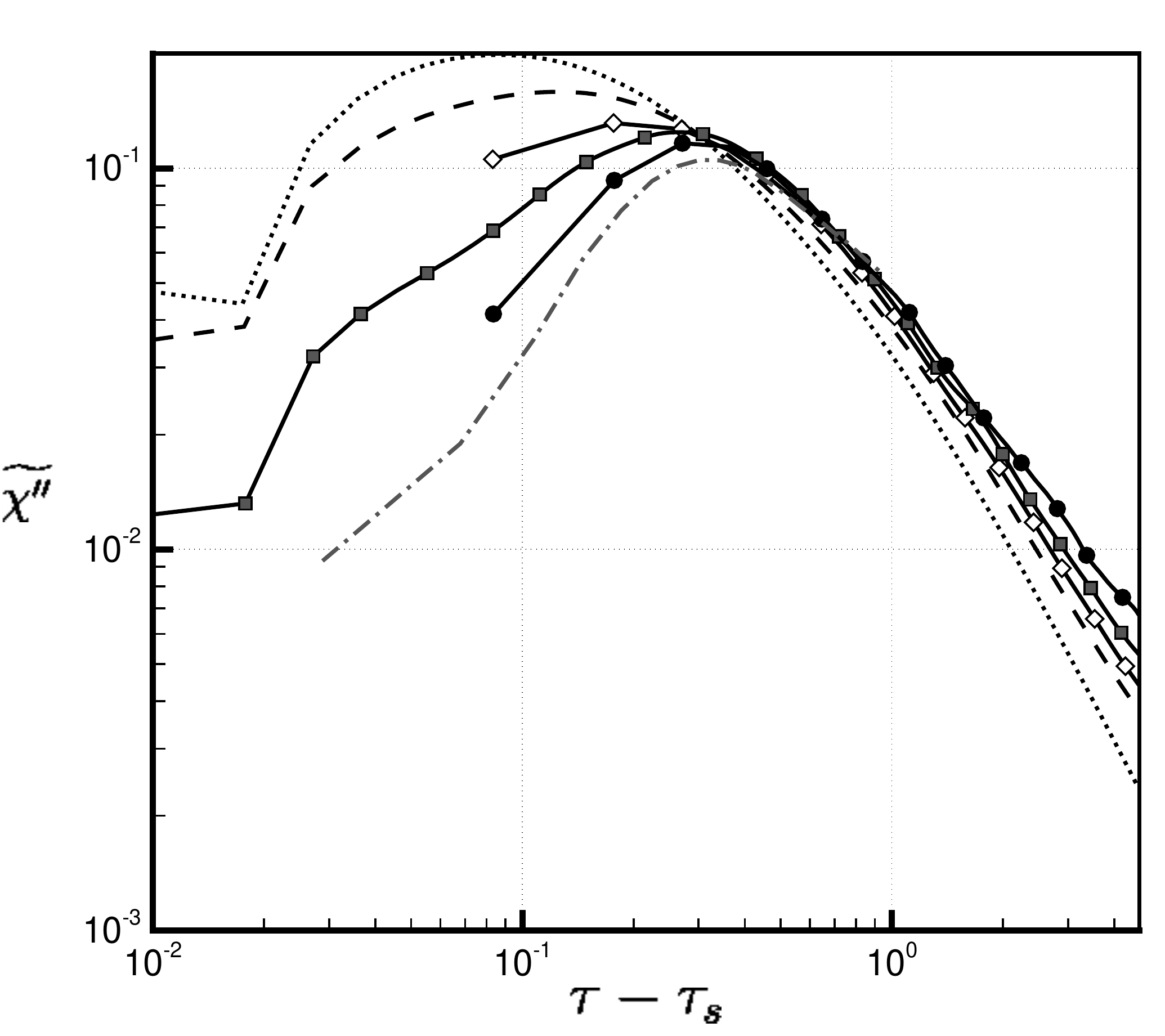}\llap{
		\parbox[b]{0.49\textwidth}{($b$)\\\rule{0ex}{0.40\textwidth}}}
	\caption{Temporal evolution of $(a)$ scalar variance and $(b)$ scalar dissipation rate at the mixing layer centre plane. Shown are data for $\Rey_0=43$ (dotted black lines), $\Rey_0=86$ (dashed black lines), $\Rey_0=174$ \firstrev{(white diamonds)}, $\Rey_0=348$ \firstrev{(grey squares)}, $\Rey_0=697$ (black circles) and $\Rey_0=1395$ \firstrev{(dash-dot grey lines)}.}
	\label{fig:MFV-Chi}
\end{figure}

The variance of the heavy fluid mass fraction is denoted by $\widetilde{Y_1^{\prime\prime^2}}$, with its corresponding dissipation rate given by 
\begin{equation}
\widetilde{\chi^{\prime\prime}} = \overline{D}\reallywidetilde{\left(\frac{\p Y_1^{\prime\prime}}{\p x_j}\right)^2}.
\label{eqn:chi}
\end{equation} 
Figure \ref{fig:MFV-Chi} shows the evolution in time of $\widetilde{Y_1^{\prime\prime^2}}$ and $\widetilde{\chi^{\prime\prime}}$ at the mixing layer centre plane. A maximum in the scalar variance is observed for all cases and occurs at approximately the same time as the minimum in the mixing measures $\Theta$ and $\Psi$ \firstrev{(see figure \ref{fig:Theta-Psi})}. The value of this maximum also increases as $\Rey_0$ is increased, indeed at all points in time a higher value of $\Rey_0$ corresponds to a higher scalar variance. A maximum in the scalar dissipation rate is also observed, however the relation between this maximum and the maximum in scalar variance changes as $\Rey_0$ is varied. For low $\Rey_0$, the maximum in scalar dissipation rate occurs prior to the maximum in scalar variance, with these cases having the greatest scalar dissipation rate at early time. As $\Rey_0$ is increased, the location of the maximum scalar dissipation rate shifts later and later in time so that for the higher $\Rey_0$ cases this maximum occurs later than the maximum scalar variance. In the $\Rey_0=1395$ case, the maximum in $\widetilde{\chi^{\prime\prime}}$ occurs slightly later than the maximum in $\widetilde{\epsilon^{\prime\prime}}$ \firstrev{(see figure \ref{fig:TKE-EPS})}, at a time of \firstrev{$\tau-\tau_s=0.326$}. It also appears that a high Reynolds number limit should exist in the value and location of this maximum, although it is not able to be estimated from the current data as no clear pattern of convergence is present. 

\begin{figure}
	\centering
	\includegraphics[width=0.33\textwidth]{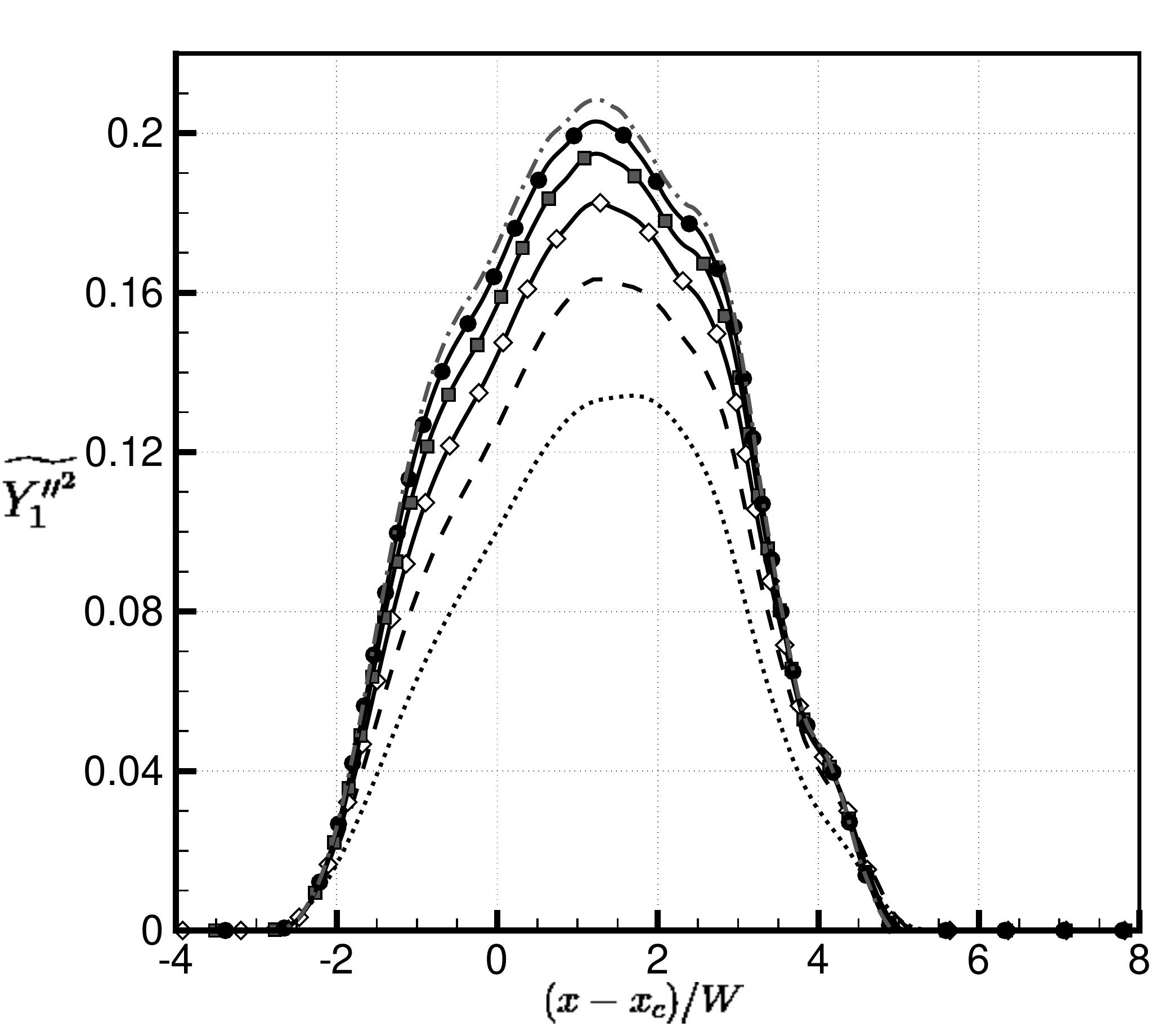}\llap{
		\parbox[b]{0.33\textwidth}{($a$)\\\rule{0ex}{0.267\textwidth}}}
	\includegraphics[width=0.33\textwidth]{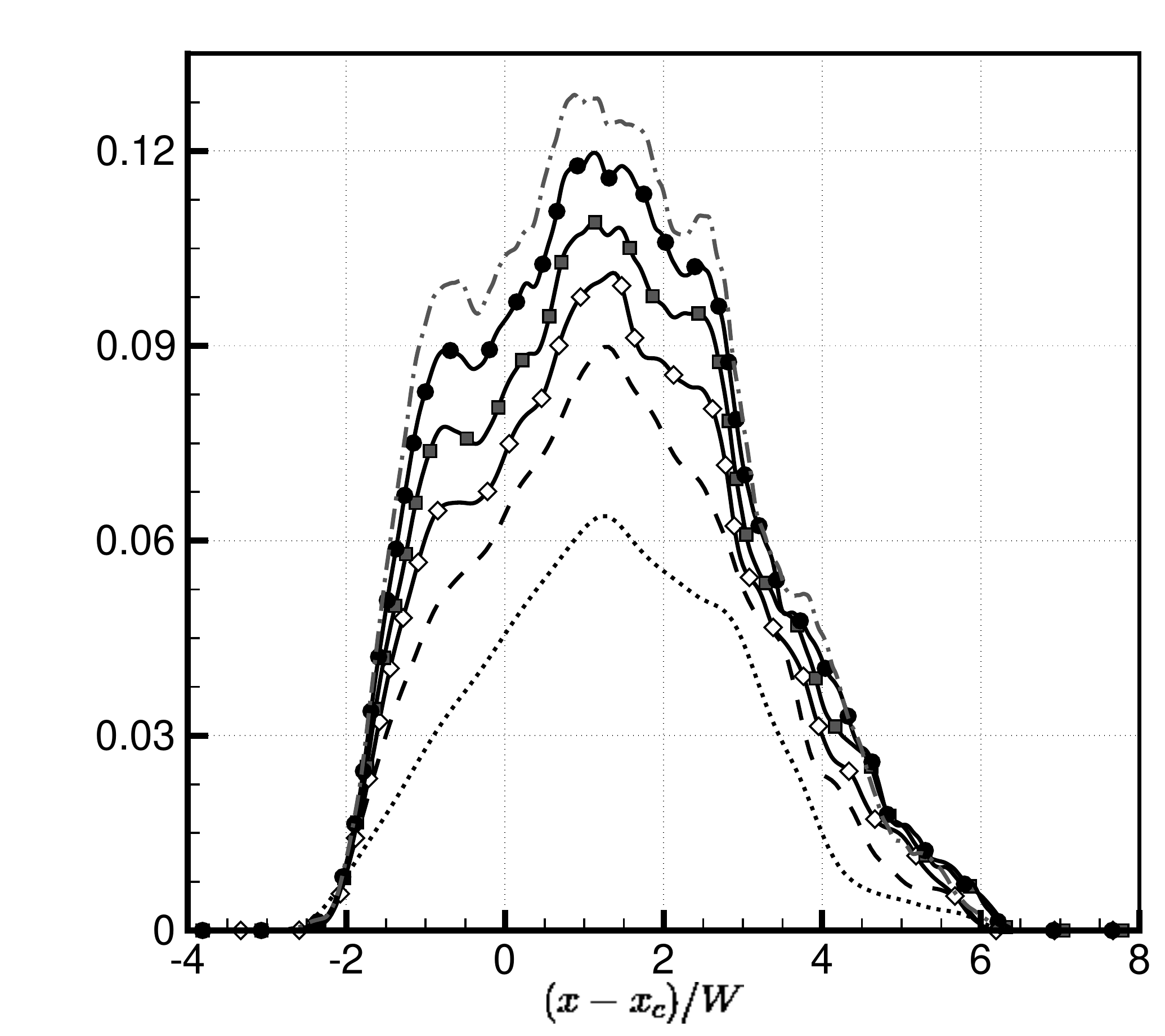}\llap{
		\parbox[b]{0.333\textwidth}{($b$)\\\rule{0ex}{0.267\textwidth}}}
	\includegraphics[width=0.33\textwidth]{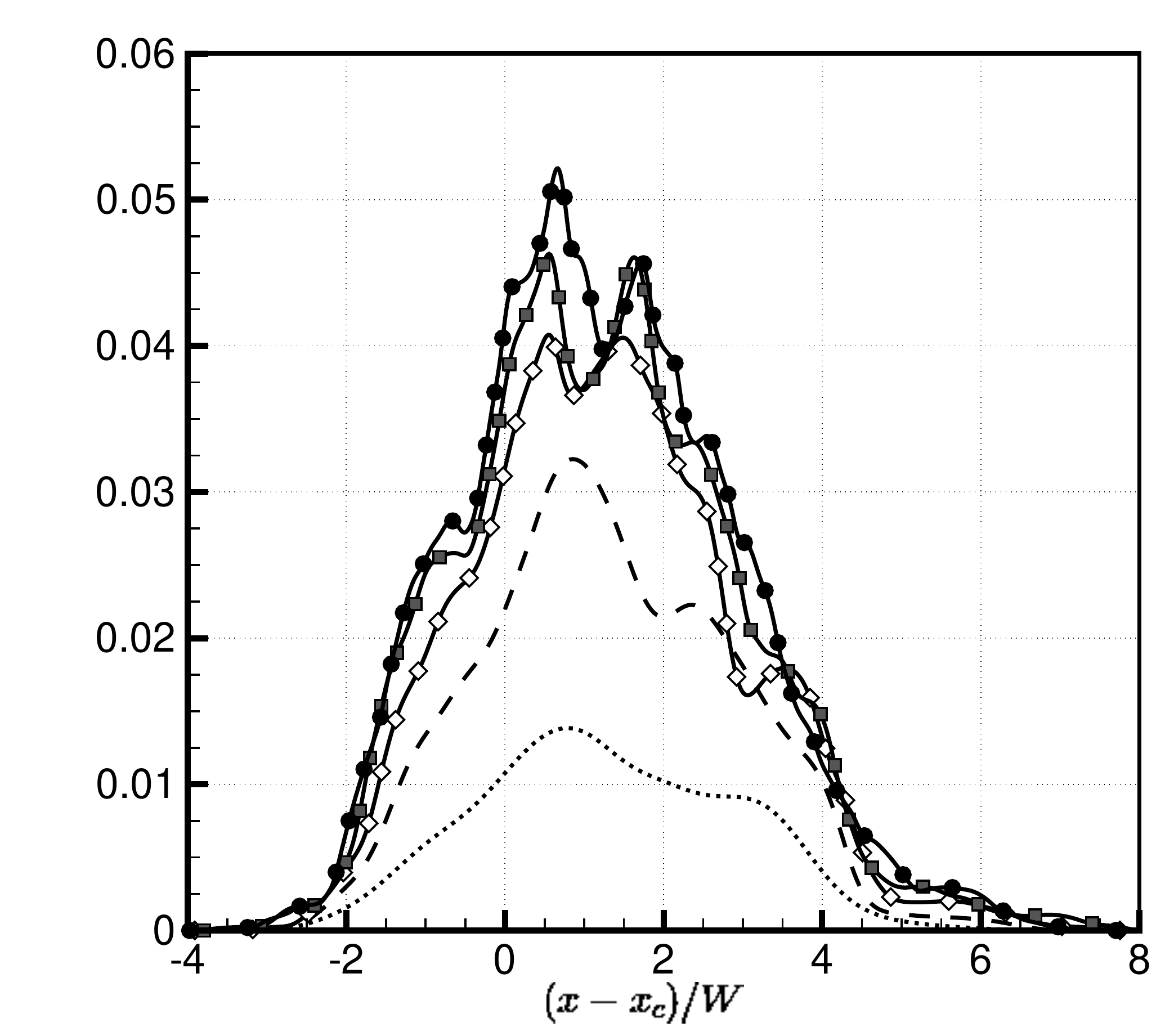}\llap{
		\parbox[b]{0.333\textwidth}{($c$)\\\rule{0ex}{0.267\textwidth}}}
	\caption{Spatial distribution of scalar variance in the $x$-direction for times ($a$) $\tau=0.187$, ($b$) $\tau=0.939$ and ($c$) $\tau=4.70$. Shown are data for $\Rey_0=43$ (dotted black lines), $\Rey_0=86$ (dashed black lines), $\Rey_0=174$ \firstrev{(white diamonds)}, $\Rey_0=348$ \firstrev{(grey squares)}, $\Rey_0=697$ (black circles) and $\Rey_0=1395$ \firstrev{(dash-dot grey lines)}.}
	\label{fig:MFV-X}
\end{figure}

\begin{figure}
	\centering
	\includegraphics[width=0.33\textwidth]{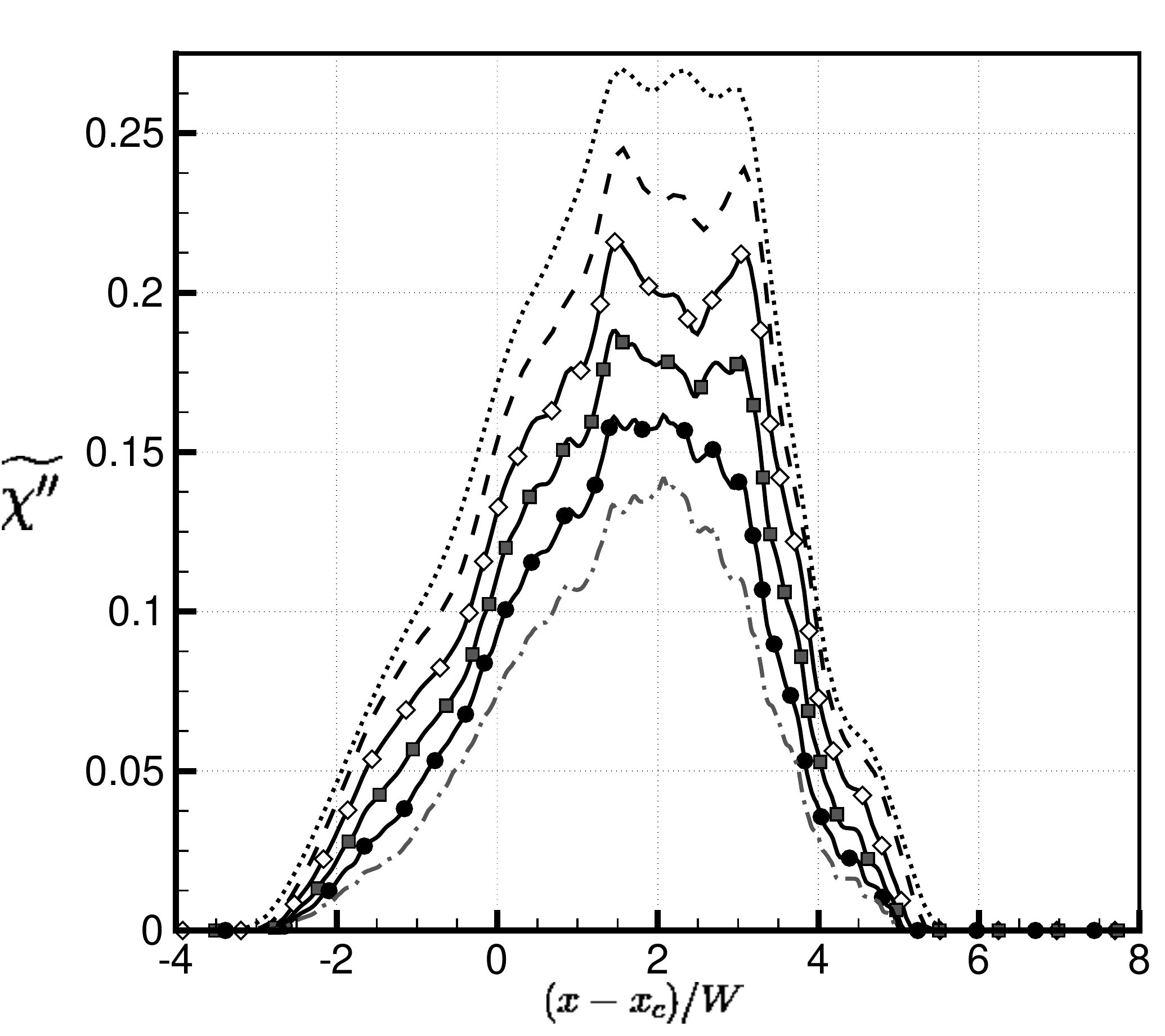}\llap{
		\parbox[b]{0.33\textwidth}{($a$)\\\rule{0ex}{0.267\textwidth}}}
	\includegraphics[width=0.33\textwidth]{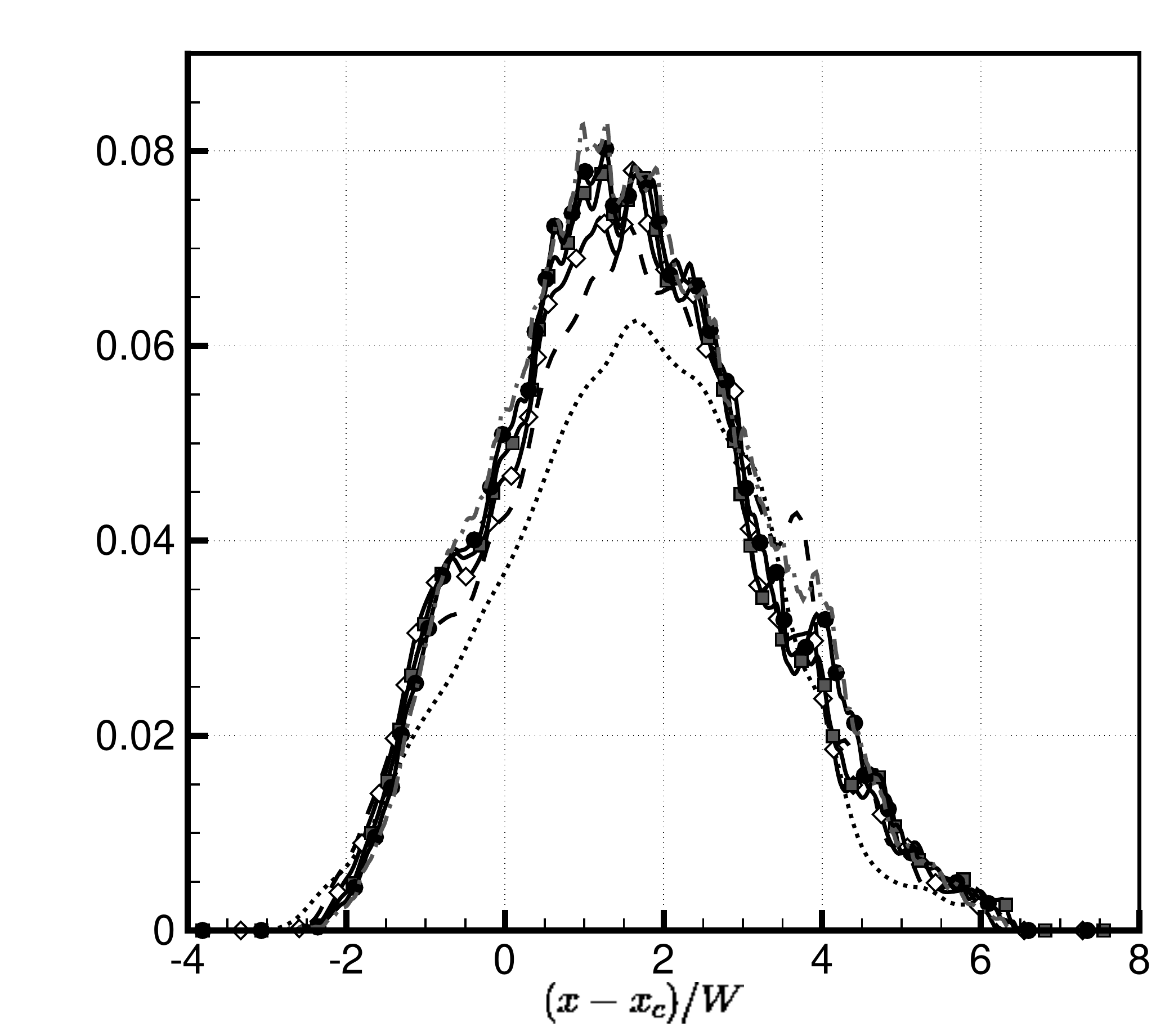}\llap{
		\parbox[b]{0.333\textwidth}{($b$)\\\rule{0ex}{0.267\textwidth}}}
	\includegraphics[width=0.33\textwidth]{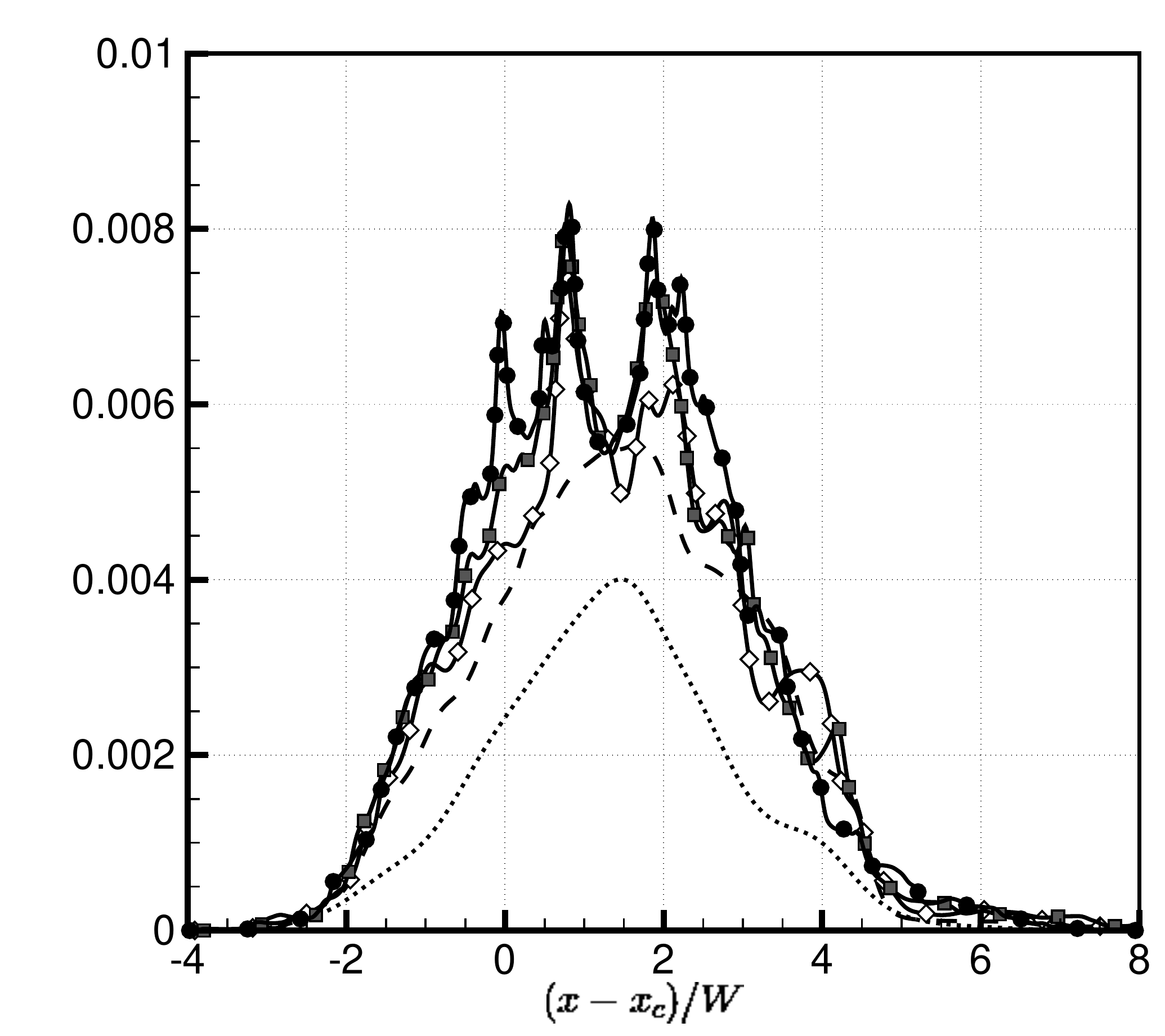}\llap{
		\parbox[b]{0.333\textwidth}{($c$)\\\rule{0ex}{0.267\textwidth}}}
	\caption{Spatial distribution of scalar dissipation rate in the $x$-direction for times ($a$) $\tau=0.187$, ($b$) $\tau=0.939$ and ($c$) $\tau=4.70$. Shown are data for $\Rey_0=43$ (dotted black lines), $\Rey_0=86$ (dashed black lines), $\Rey_0=174$ \firstrev{(white diamonds)}, $\Rey_0=348$ \firstrev{(grey squares)}, $\Rey_0=697$ (black circles) and $\Rey_0=1395$ \firstrev{(dash-dot grey lines)}.}
	\label{fig:Chi-X}
\end{figure}

Figures \ref{fig:MFV-X} and \ref{fig:Chi-X} show the spatial distribution of $\widetilde{Y_1^{\prime\prime^2}}$ and $\widetilde{\chi^{\prime\prime}}$ across the layer. Similar trends to those observed for the spatial distribution of turbulent kinetic energy and dissipation rate are also seen here. The peaks in scalar variance and scalar dissipation rate are also biased towards the spike side of the layer, however they occur closer to the mixing layer centre. There is also less variation in the data, particularly for the scalar dissipation rate at later times.  Very little scalar variance/dissipation rate is located at the fringes of the spike side of the layer at late time. 


\subsubsection{Normalised dissipation rates}
\begin{figure} 
	\centering
	\includegraphics[width=0.49\textwidth]{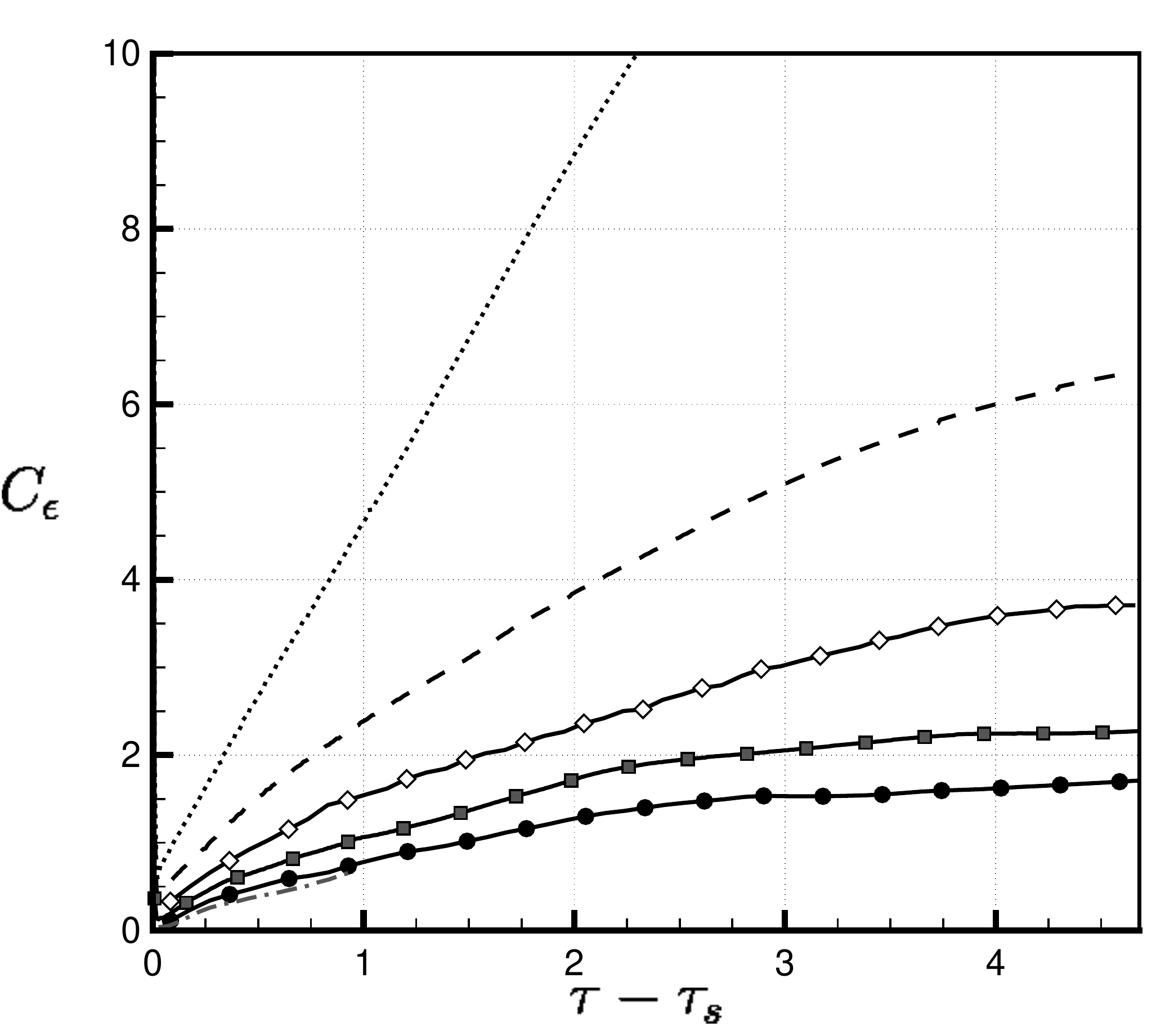}\llap{
		\parbox[b]{0.49\textwidth}{($a$)\\\rule{0ex}{0.40\textwidth}}}
	\includegraphics[width=0.49\textwidth]{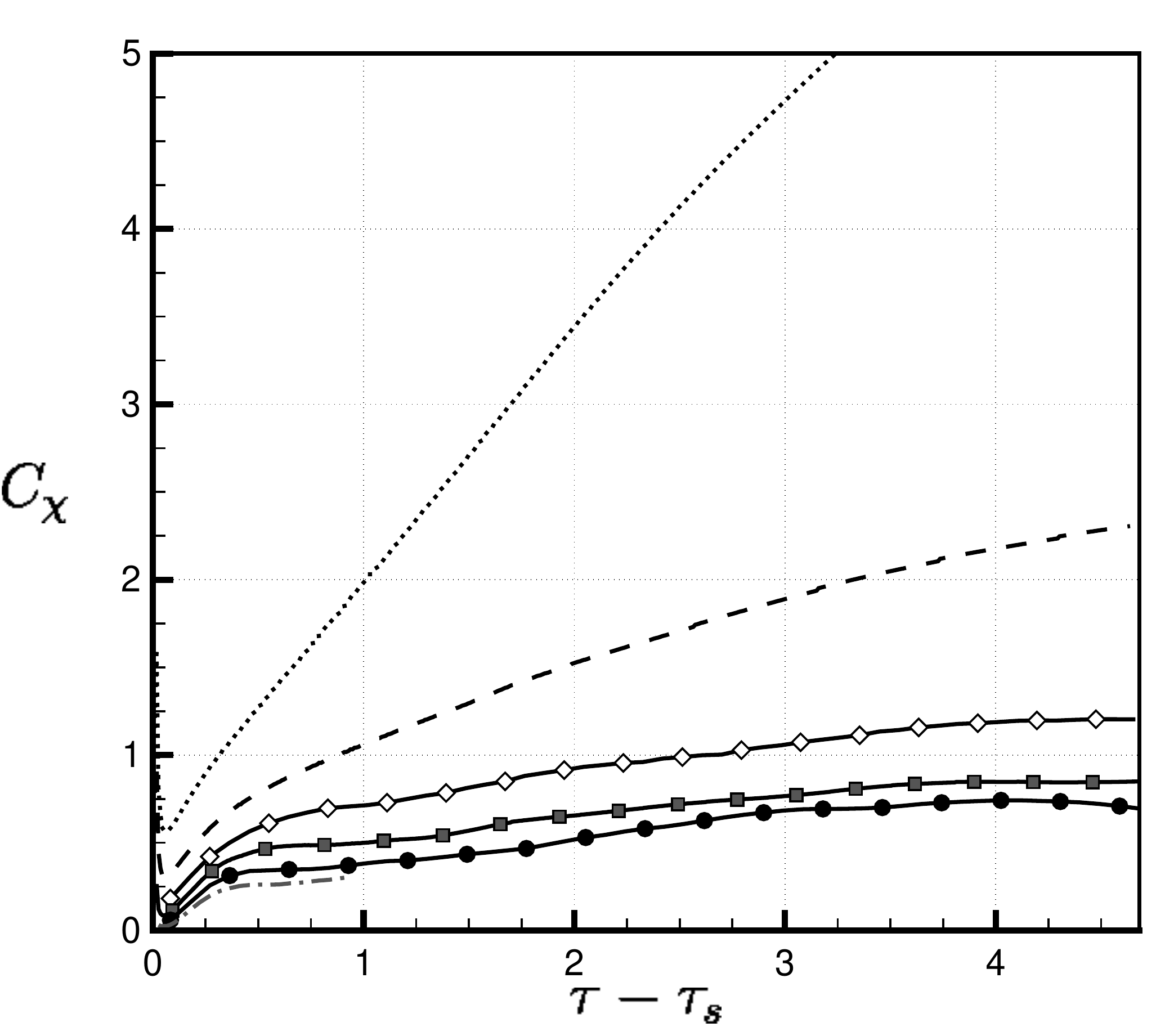}\llap{
		\parbox[b]{0.49\textwidth}{($b$)\\\rule{0ex}{0.40\textwidth}}}
	\caption{Temporal evolution of $(a)$ normalised dissipation rate and $(b)$ normalised scalar dissipation rate at the mixing layer centre plane. Shown are data for $\Rey_0=43$ (dotted black lines), $\Rey_0=86$ (dashed black lines), $\Rey_0=174$ \firstrev{(white diamonds)}, $\Rey_0=348$ \firstrev{(grey squares)}, $\Rey_0=697$ (black circles) and $\Rey_0=1395$ \firstrev{(dash-dot grey lines)}.}
	\label{fig:C_eps-C_chi}
\end{figure}

To conclude this section, the normalised dissipation rate $C_\epsilon$ and normalised scalar dissipation rate $C_\chi$ are examined to assess the degree to which they are independent of the Reynolds number of the flow. Following \citet{Donzis2005}, the normalised dissipation rates are defined as
\begin{equation}
	C_\epsilon = \frac{\langle\epsilon\rangle L_u}{{u^\prime}^3}, \qquad C_\chi = \frac{\langle\chi\rangle L_u}{\langle {\phi^\prime}^2\rangle u^\prime}.
	\label{eqn:C_eps-C_chi}
\end{equation}
Here the dissipation rates $\widetilde{\epsilon^{\prime\prime}}$ and $\widetilde{\chi^{\prime\prime}}$, evaluated at the mixing layer centre plane, are used in place of $\langle\epsilon\rangle$ and $\langle\chi\rangle$. The root mean square velocity $u^\prime$ is calculated from the turbulent kinetic energy at the mixing layer centre as
\begin{equation}
{u^\prime}^2=\frac{2}{3}\widetilde{E_k^{\prime\prime}},
\end{equation}
while the scalar variance $\langle {\phi^\prime}^2\rangle$ is given by that of the heavy fluid mass fraction $\widetilde{Y_1^{\prime\prime^2}}$. Finally the integral length $\Lambda$, calculated using the radial power spectrum of turbulent kinetic energy per unit volume (taken at the mixing layer centre), is used for the characteristic length scale $L_u$. This is calculated as
\begin{equation}
\Lambda=\frac{3\upi}{4}\frac{\displaystyle \int_{0}^{\infty}\frac{E^{(v)}}{k}\:\mathrm{d}k}{\displaystyle \int_{0}^{\infty}E^{(v)}\:\mathrm{d}k}.
\label{eqn:integral-length}
\end{equation}
Note that if the power spectrum of turbulent kinetic energy per unit mass is used instead, the resulting integral length is very similar for this flow \citep{Thornber2016}. Figure \ref{fig:C_eps-C_chi} shows the evolution in time of both $C_\epsilon$ and $C_\chi$ at the mixing layer centre plane. As the current flow under investigation is unsteady, it is not surprising that both quantities are varying with time, especially while the flow is still in the relatively early stages of development. That both quantities are increasing with time is also in agreement with the fact that the outer-scale Reynolds number decreases with time, as shown in \S\ref{subsec:lengthscales}. Of interest is whether this variation becomes independent of Reynolds number at any point in time, a necessary (but not sufficient) criterion for a flow to be classified as fully turbulent. Examining figure \ref{fig:C_eps-C_chi} it can be seen that a high Reynolds number limit is being approached with each increase in $\Rey_0$ for the timescale considered, but has not yet been reached. The data are slightly closer to collapsing for the normalised scalar dissipation rate; at the latest time considered there is a 22\% difference between $\Rey_0=348$ and $\Rey_0=697$ for $C_\chi$ compared to a 32\% difference for $C_\epsilon$. For the higher $\Rey_0$ cases the curves of both $C_\epsilon$ and $C_\chi$ are becoming constant. This behaviour at late time was also observed by \citet{Yoffe2018} for $C_\epsilon$ in simulations of decaying homogeneous isotropic turbulence (HIT), as well as by \citet{Zhou2019a} in DNS of RTI. Note that in the latter study, $C_\epsilon$ is decreasing with time since the Reynolds number is increasing. 

\begin{figure}
	\centering
	\includegraphics[width=0.49\textwidth]{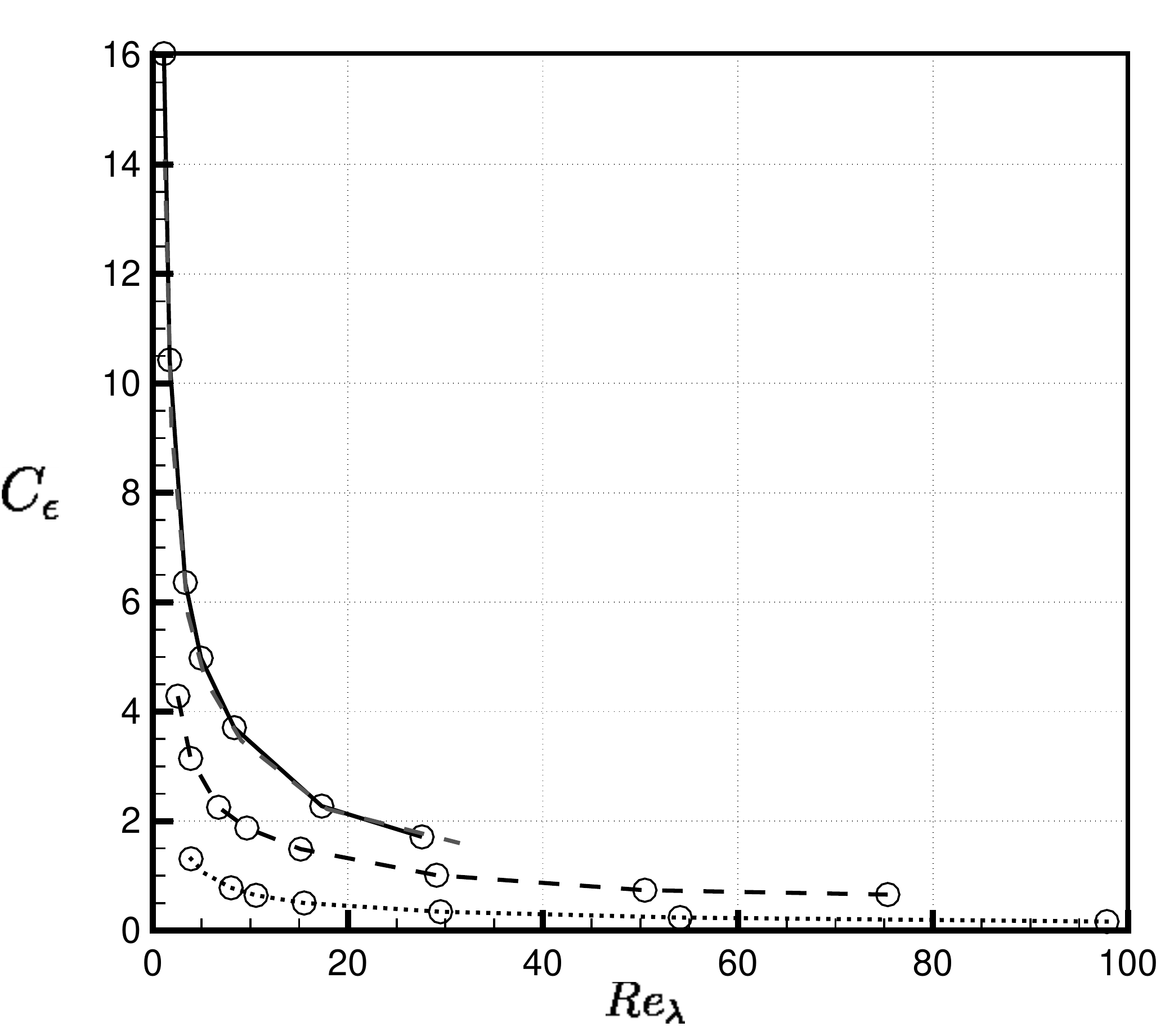}\llap{
		\parbox[b]{0.49\textwidth}{($a$)\\\rule{0ex}{0.40\textwidth}}}
	\includegraphics[width=0.49\textwidth]{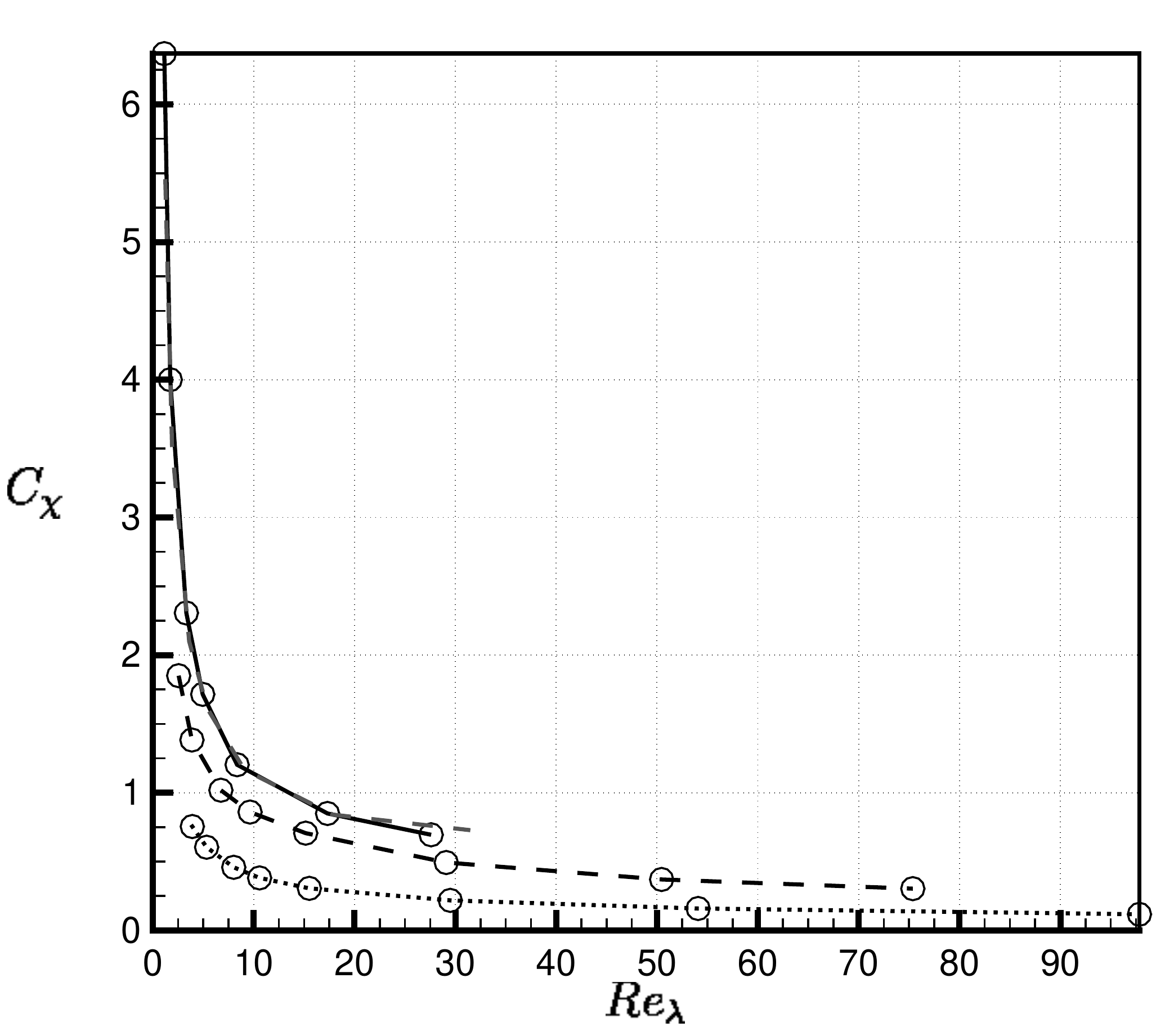}\llap{
		\parbox[b]{0.49\textwidth}{($b$)\\\rule{0ex}{0.40\textwidth}}}
	\caption{\secondrev{Transverse} Taylor microscale Reynolds number vs. $(a)$ normalised dissipation rate and $(b)$ normalised scalar dissipation rate at the mixing layer centre plane. Shown are data for times $\tau=0.187$ (dotted lines), $\tau=0.939$ (dashed lines), $\tau=3.76$ (grey dashed lines) and $\tau=4.70$ (solid lines).}
	\label{fig:C_eps-C_chi-Re}
\end{figure}

The variation with Reynolds number is shown more precisely in figure \ref{fig:C_eps-C_chi-Re}, which plots $C_\epsilon$ and $C_\chi$ against the \secondrev{transverse} Taylor microscale Reynolds number $\Rey_\lambda$ (given by (\ref{eqn:taylor-Re})) for each simulation at four different times. The resulting curves at each time instant follow the same functional form as those produced by isotropic turbulence, but their asymptotic value increases as the simulation progresses. By late time the curves have nearly collapsed, indicating that their high Reynolds number asymptote is also close to becoming independent of time. Nonlinear regression can be used to fit the expected functional form of $C_\epsilon$ and $C_\chi$ to the data and extract this asymptotic value. Following \citet{Donzis2005}, a function is used of the form
\begin{equation}
f=A(1+\sqrt{1+(B/\Rey_\lambda)^2}).
\label{eqn:f1}
\end{equation}
Fitting this function to the $\tau=4.70$ data gives $A=0.77$, $B=27$ for $C_\epsilon$ and $A=0.33$, $B=20$ for $C_\chi$. The lower value of $B$ in the curve fit to the normalised scalar dissipation rate data indicates that the asymptotic value of $C_\chi$ is attained faster than that of $C_\epsilon$. This is in agreement with the observations made in section \ref{subsec:scalar}, as well as those for homogeneous passive scalar turbulence at $\Sc=1$ \citep{Donzis2005}. The asymptotic values of $C_\epsilon$ and $C_\chi$ are equal to $2A$, implying that the high Reynolds number limit of these quantities is $1.54$ and $0.66$ respectively. \firstrev{For the case of the RTI,} \citet{Zhou2019a} split $C_\epsilon$ into normal and transverse components and found values in the range $0.3$-$0.4$ and $0.5$-$0.6$ respectively at the latest time considered, both of which are substantially lower than the estimate of the high Reynolds number limit given here for RMI. This is analogous to the difference in asymptotic value observed between forced and freely decaying HIT \citep{Bos2007}. \citet{Yoffe2018} showed that this difference can be rectified by using values of $C_\epsilon$ at a specified onset time, taken to be either the time of maximum dissipation rate (if it exists) or the time of maximum inertial transfer rate. This onset time typically occurs much earlier than the point at which $C_\epsilon$ becomes time-independent, and estimates of the high Reynolds number limit using this criterion were virtually identical to those obtained in the forced, stationary case. Using the time of maximum dissipation rate in the $\Rey_0=1395$ case as an approximation of the onset time criterion for all cases, the high Reynolds number limit at this time \firstrev{($\tau-\tau_0=0.224$)} is found to be $0.28$, which is a plausible asymptotic value for the normal component of $C_\epsilon$ in RTI. \thirdrev{This may also be compared with the asymptotic value of 0.4 for forced homogeneous turbulence \citep{Donzis2005}.} A more rigorous comparison would involve performing the same split of $C_\epsilon$ (and other key quantities) into normal and transverse components as in \citet{Zhou2019a}, which will be performed in future work.

\subsection{Mixing transition}
\label{subsec:transition}
All of the results presented in the previous sections have been calculated from simulations which are, in reality, at quite modest Reynolds numbers compared to the actual flows observed in experiments or nature. Given this fact it is natural to ask, particularly for those quantities that do not depend predominantly on the large scales, how representative are these results compared to those that would be obtained at higher Reynolds numbers? In particular, it is useful to know when extrapolating results to higher Reynolds numbers whether the amount of turbulence present in the flow is approaching levels that would be considered fully developed in the sense proposed by \citet{Zhou2007}. This is helpful for determining how close the current results are to any high Reynolds number limits that exist for quantities that are known/expected to exhibit universal, asymptotic behaviour once turbulence has fully developed. This section investigates the evolution of various key length scales and Reynolds numbers that are commonly used to characterise turbulent flows. The $\Rey_0=174$, $\Rey_0=348$ and $\Rey_0=697$ cases are used for the analysis in this section, along with the additional $\Rey_0=1395$ case that was run up until time $\tau=0.939$.

\subsubsection{Length scales and Reynolds numbers}
\label{subsec:lengthscales}
In turbulent flows, a number of statistics are used to characterise the typical spatial scales at which energy is generated, transferred and dissipated in the flow. The largest of these is the outer length scale $\delta$, which for RMI and RTI induced flows is identified as the visual width $h$ \firstrev{\citep{Cook2001}}, given by
\begin{equation}
h=x\left(\langle f_1\rangle=0.01\right)-x\left(\langle f_1\rangle=0.99\right) 
\label{eqn:outer-length}.
\end{equation}
This is representative of the largest dynamical motions in the flow. \firstrev{Note that integral definitions have also been presented in the literature \citep{Cook2004}}. Given the definition of $h$, an outer-scale Reynolds number may also be defined,
\begin{equation}
\Rey_h=\frac{h \dot{h}}{\overline{\nu}}
\label{eqn:outer-reynolds},
\end{equation}
where $\dot{h}$ is the time derivative of the outer length scale and $\overline{\nu}$ is the average kinematic viscosity across the layer (from $x\left(\langle f_1\rangle=0.99\right)$ to $x\left(\langle f_1\rangle=0.01\right)$). The next largest length scale to consider is the integral length $\Lambda$, already defined in (\ref{eqn:integral-length}), which characterises the distance over which the fluctuating velocity field is correlated. Related to $\Lambda$ is the Taylor microscale $\lambda$, which is obtained from the curvature of the fluctuating velocity autocorrelation, or equivalently from the variance of fluctuating velocity and its derivatives. \firstrev{This length scale} may be considered to be representative of scales located in some part of the inertial range for fully developed turbulence. To account for anisotropy, directional Taylor microscales may be defined for direction $i$ as
\begin{equation}
	\lambda_i =\left[\frac{\langle u_i^{\prime\prime 2} \rangle}{\langle(\p u_i^{\prime\prime}/\p x_i)^2 \rangle}\right]^{1/2},
	\label{eqn:taylor}
\end{equation}
where plane averages are taken at the mixing layer centre plane. Since isotropy is expected in the transverse directions, a single transverse Taylor microscale is defined as $\lambda_{yz}=(\lambda_y+\lambda_z)/2$ \citep{Cook2001}. Similarly, a transverse Taylor-scale Reynolds number is defined at the mixing layer centre plane as $\Rey_{\lambda_{yz}}=(\Rey_{\lambda_y}+\Rey_{\lambda_z})/2$, where
\begin{equation}
\Rey_{\lambda_i} = \frac{\langle u_i^{\prime\prime 2} \rangle}{\displaystyle \langle \nu \rangle \sqrt{\langle(\p u_i^{\prime\prime}/\p x_i)^2 \rangle}}.
\label{eqn:taylor-Re}
\end{equation}
Finally, the Kolmogorov microscale $\eta$ characterises the scale at which motions in the flow \firstrev{are dominated} by viscosity and is given by
\begin{equation}
	\eta =\left(\frac{\langle \nu \rangle ^3}{\langle\epsilon^{\prime\prime}\rangle}\right)^{1/4}.
	\label{eqn:kolmogorov}
\end{equation}

\begin{figure}
	\centering
	\includegraphics[width=0.33\textwidth]{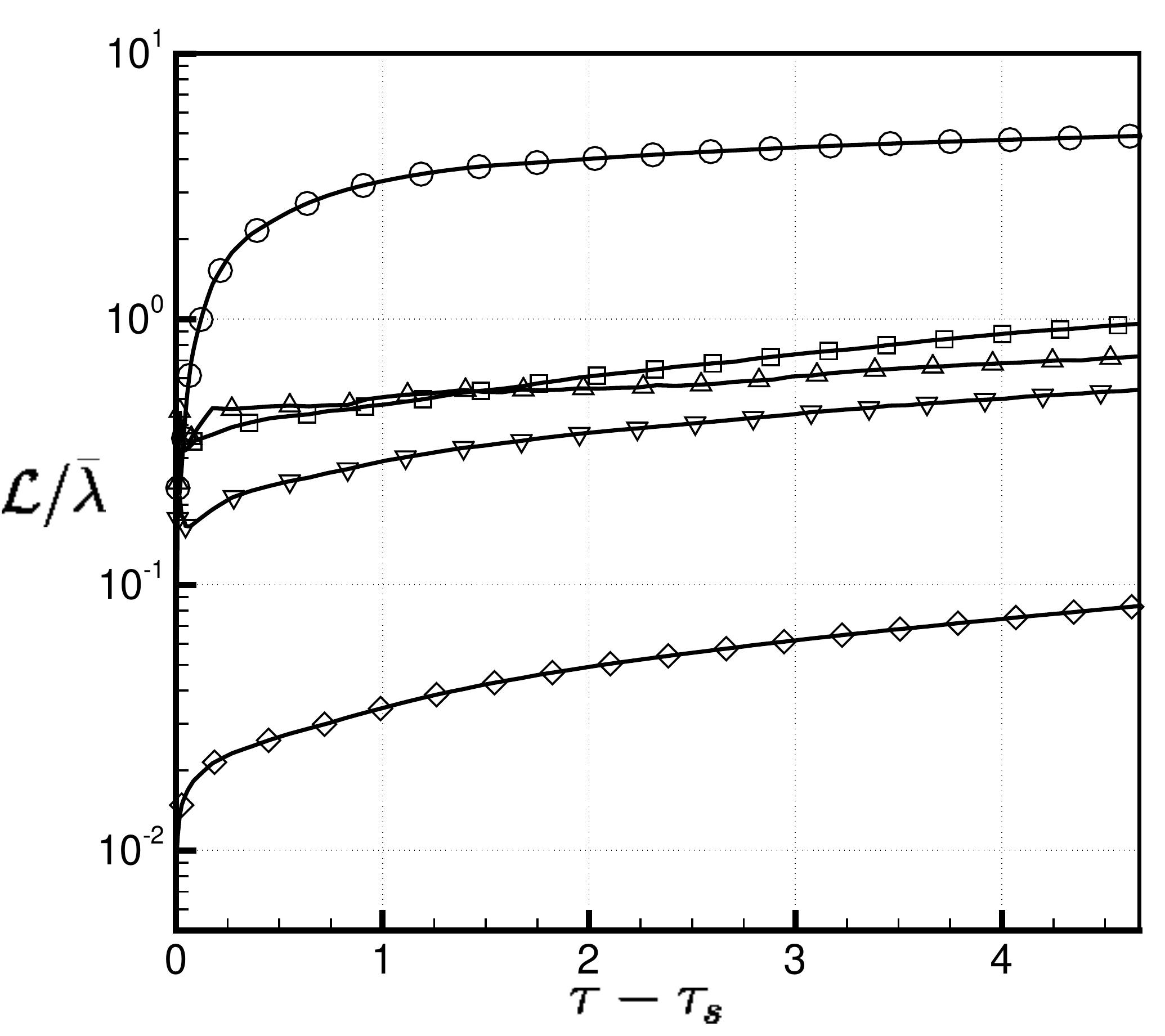}\llap{
		\parbox[b]{0.33\textwidth}{($a$)\\\rule{0ex}{0.267\textwidth}}}
	\includegraphics[width=0.33\textwidth]{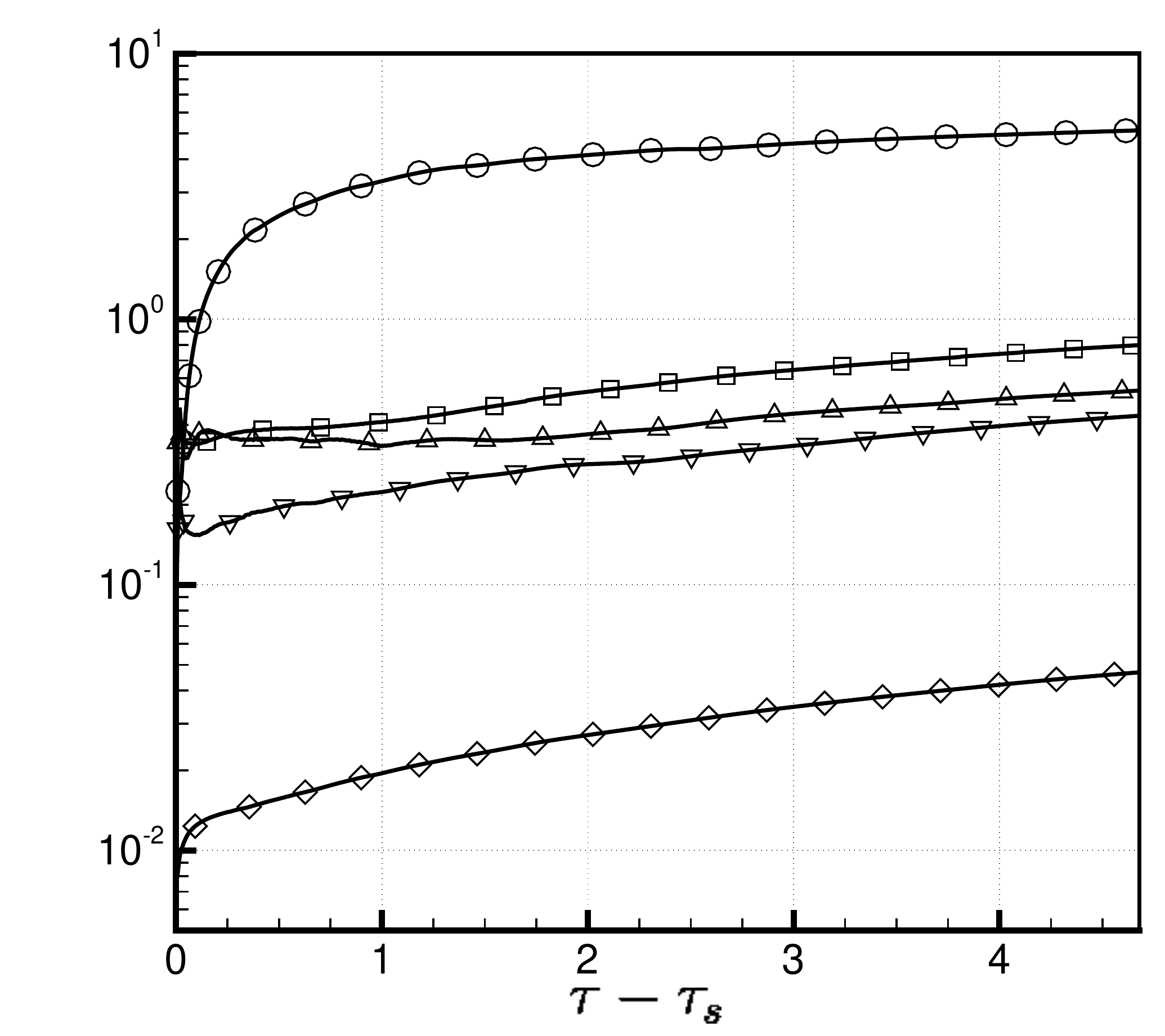}\llap{
		\parbox[b]{0.33\textwidth}{($b$)\\\rule{0ex}{0.267\textwidth}}}
	\includegraphics[width=0.33\textwidth]{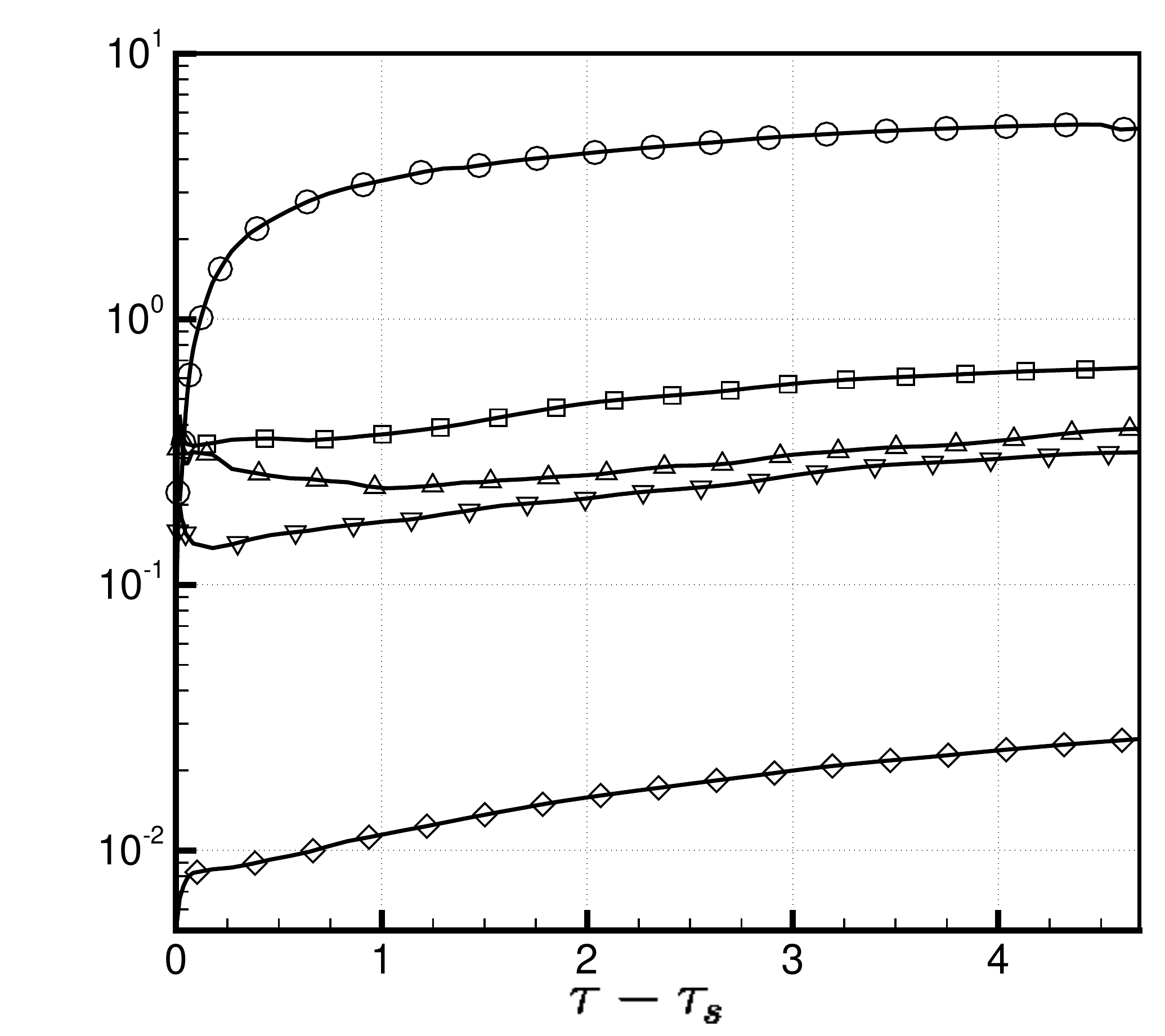}\llap{
		\parbox[b]{0.33\textwidth}{($c$)\\\rule{0ex}{0.267\textwidth}}}
	\caption{Evolution of length scales for $\Rey_0=174$ $(a)$, $\Rey_0=348$ $(b)$ and $\Rey_0=697$ $(c)$. Plotted are the outer scale $h$ (circles), the integral length $\Lambda$ (squares), \secondrev{the Taylor microscales $\lambda_{x}$ (upward triangles) and $\lambda_{yz}$ (downward triangles)} and the Kolmogorov microscale $\eta$ (diamonds).}
	\label{fig:lengthscales-tau}
\end{figure}

The temporal evolution of the visual width, integral length and Taylor and Kolmogorov microscales is shown in figure \ref{fig:lengthscales-tau}, from which a clear trend of increasing scale separation with increasing $\Rey_0$ can be observed. Comparing results across the three simulations, there is only a small difference in the outer-scale (mostly at late time), whereas the integral length, Taylor microscales and Kolmogorov microscale all decrease uniformly in time with increasing $\Rey_0$. In addition to this observed decrease in each of these individual length scales, the relative distance between each length scale for a given $\Rey_0$ is also increasing. This is consistent with the notion that the mixing layer is becoming progressively more turbulent as the damping of fine-scale motions due to viscosity is reduced, as observed in figure \ref{fig:isosurface}. \citet{Tritschler2014pre} also observed a similar increase in the separation of scales but by varying the initial Mach number $M_0$ of the problem rather than the initial Reynolds number. This is because, for a fixed initial perturbation, decreasing the viscosity $\nu$ and increasing the interface velocity jump $\Delta u$ (through increasing $M_0$) have approximately the same effect on the level of turbulence that subsequently develops. In \citet{Groom2019} it was shown that the highest Mach number case of \citet{Tritschler2014pre} corresponds to an initial Reynolds number $\Rey_0=739$ (ignoring any correction factor for the initial diffuse interface).

\begin{figure}
	\centering
	\includegraphics[width=0.33\textwidth]{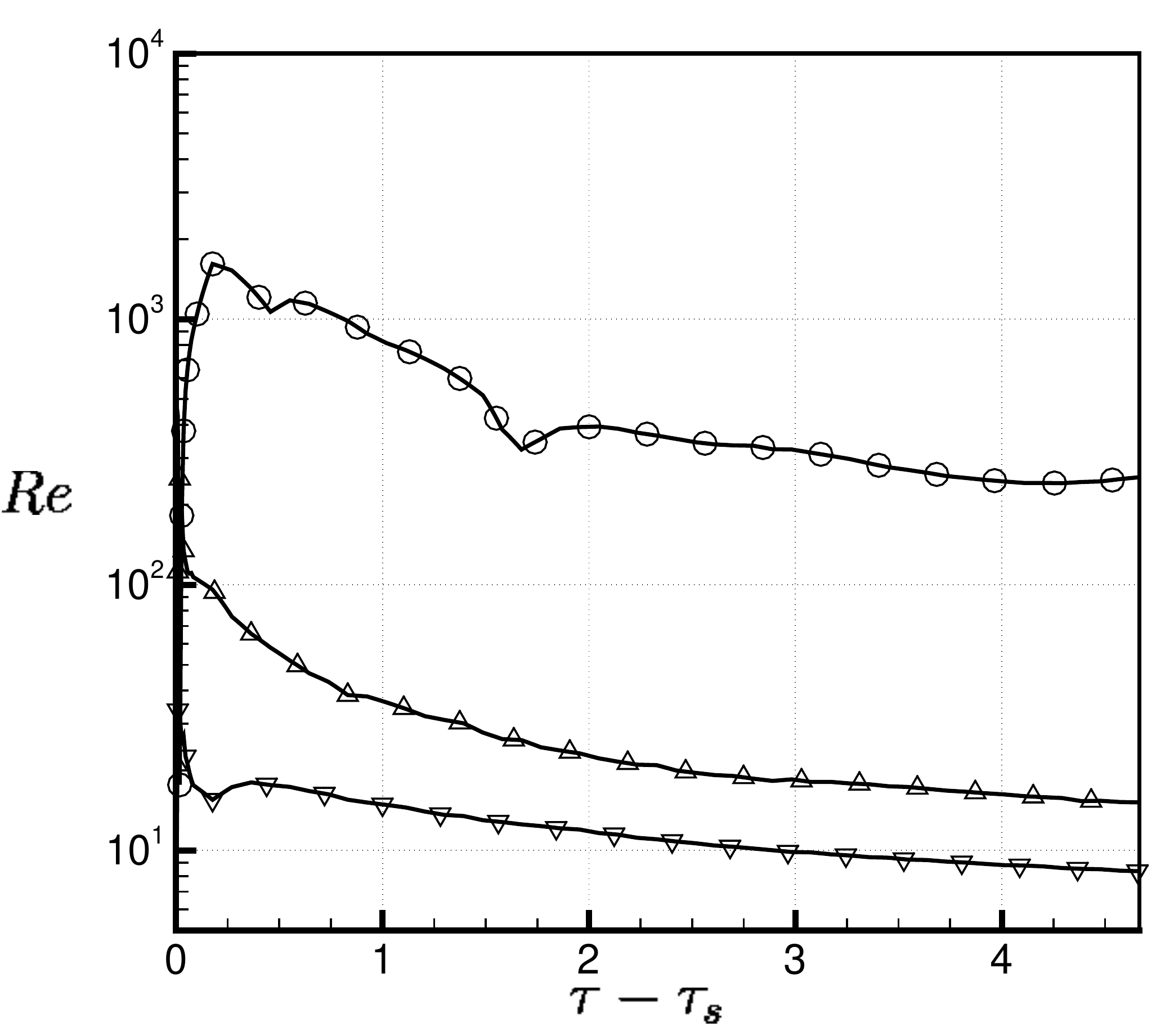}\llap{
		\parbox[b]{0.33\textwidth}{($a$)\\\rule{0ex}{0.267\textwidth}}}
	\includegraphics[width=0.33\textwidth]{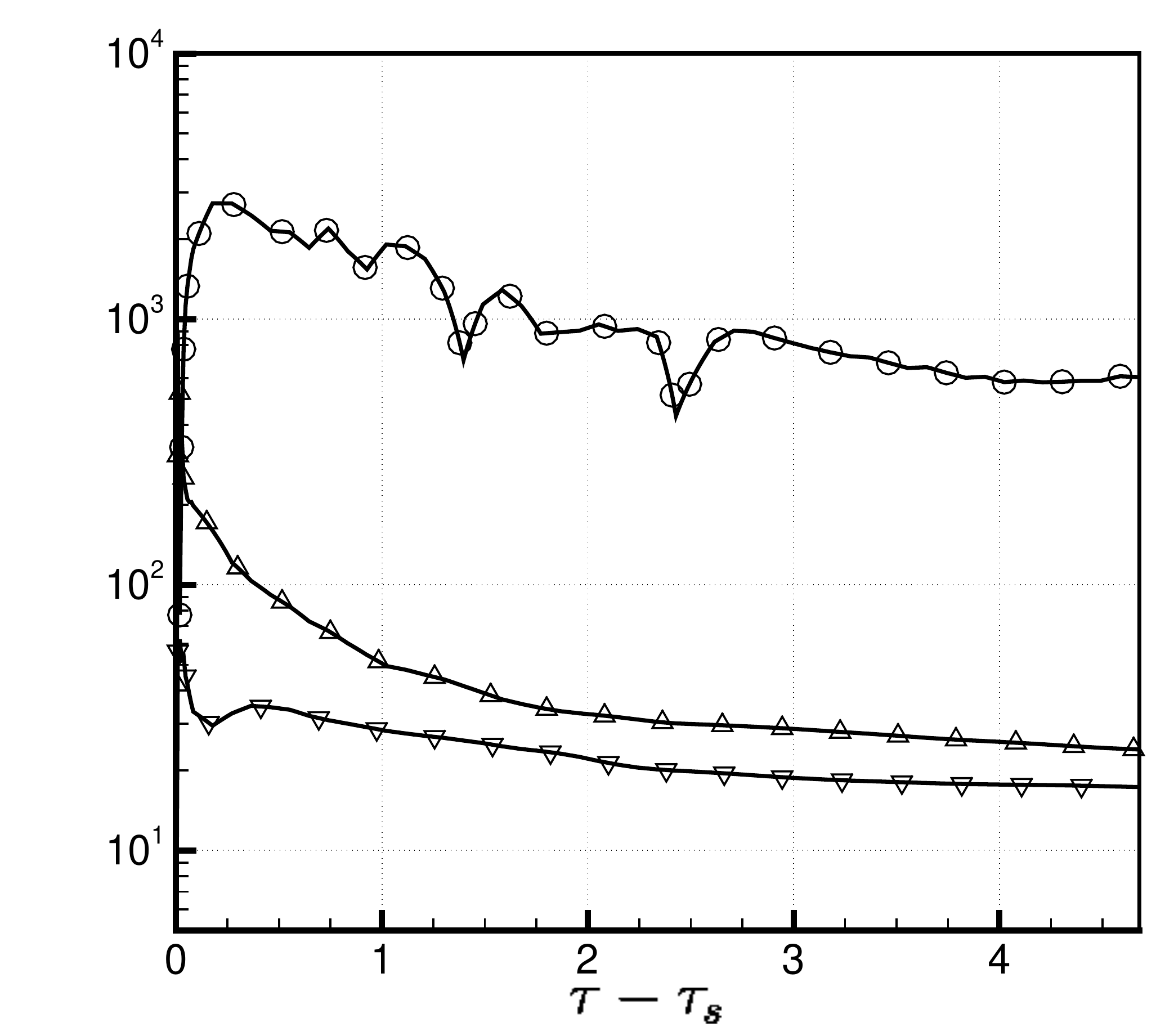}\llap{
		\parbox[b]{0.33\textwidth}{($b$)\\\rule{0ex}{0.267\textwidth}}}
	\includegraphics[width=0.33\textwidth]{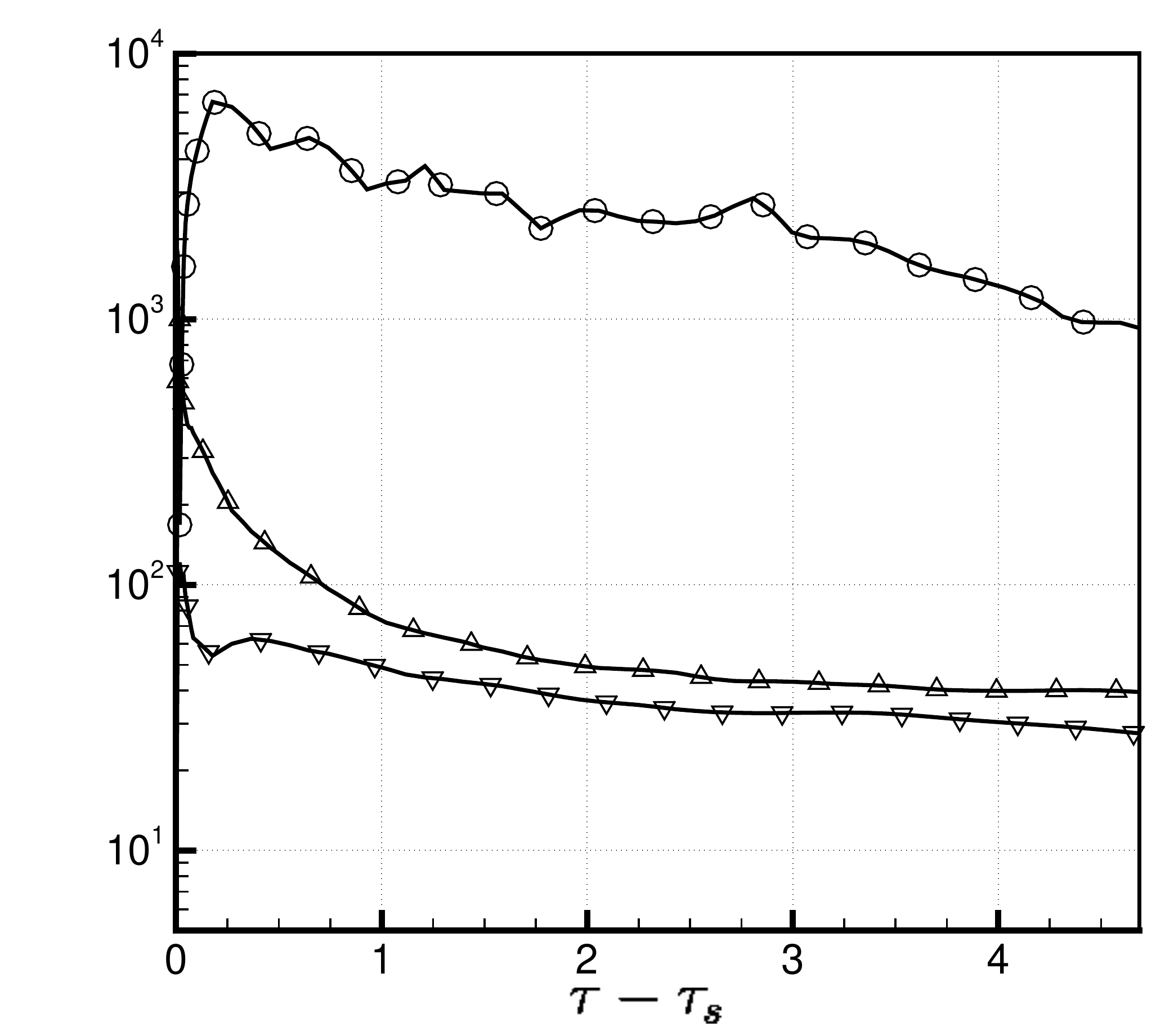}\llap{
		\parbox[b]{0.33\textwidth}{($c$)\\\rule{0ex}{0.267\textwidth}}}
	\caption{Evolution of Reynolds numbers for $\Rey_0=174$ $(a)$, $\Rey_0=348$ $(b)$ and $\Rey_0=697$ $(c)$. Plotted are the outer-scale Reynolds number $\Rey_h$ (circles) \secondrev{and the Taylor-scale Reynolds numbers $\Rey_{\lambda_{x}}$ (upward triangles) and $\Rey_{\lambda_{yz}}$ (downward triangles)}.}
	\label{fig:reynolds-tau}
\end{figure}

Figure \ref{fig:reynolds-tau} shows the temporal evolution of the outer-scale and Taylor-scale Reynolds numbers. \citet{Dimotakis2000} proposed, based on experimental evidence, that fully developed \secondrev{stationary} turbulent flow requires $\Rey_\delta\geq$ 1--2$\times 10^4$, or equivalently $\Rey_\lambda\geq$100--140, in order for it to be sustained. Hence both Reynolds numbers are important parameters for assessing the transition to fully developed turbulence. From figure \ref{fig:reynolds-tau} it can be seen that for the $\Rey_0=697$ case a peak outer-scale Reynolds number of $\Rey_h=6.57\times10^3$ is obtained shortly after shock passage, before decaying to a value of $\Rey_h=926$ at the latest time. It is worth noting that, for a compressible simulation, the visual width $h$ is easily contaminated by small acoustic waves and imperfect boundary conditions. Hence when calculating  $\Rey_h$, which requires the derivative of $h$, these small fluctuations get amplified and result in a rather noisy signal. For the \secondrev{transverse} Taylor-scale Reynolds number, a peak value of $\Rey_\lambda=121$ is observed in the $\Rey_0=697$ case, at an earlier time than the peak in $\Rey_h$. This is very similar to the value of $\Rey_\lambda$ that \citet{Tritschler2014pre} observed shortly after shock passage for their $M_0=1.5$ case. By the end of the simulation $\Rey_\lambda$ has decayed to a value of 27.6 in the $\Rey_0=913$ case, which is very close to the value of 26 obtained by \citet{Tritschler2014pre} at the end of the estimated period of uncoupled length scales for the $M_0=1.5$ case. It is important to note that the drop in Taylor microscale Reynolds numbers occurs very rapidly across all three cases, at around the same time that the peak in outer-scale Reynolds number occurs. Thus the peak value of $\Rey_\lambda$ is not sustained for very long, but conversely the subsequent decay is quite gradual. \secondrev{The ratio of Taylor microscale Reynolds numbers also indicates significant anisotropy is present in the velocity field, which has been documented previously in \citet{GroomAFMC}. This anisotropy is persistent at the latest time considered and also appears to be decreasing with increasing $\Rey_0$.}

\begin{figure}
	\centering
	\includegraphics[width=0.49\textwidth]{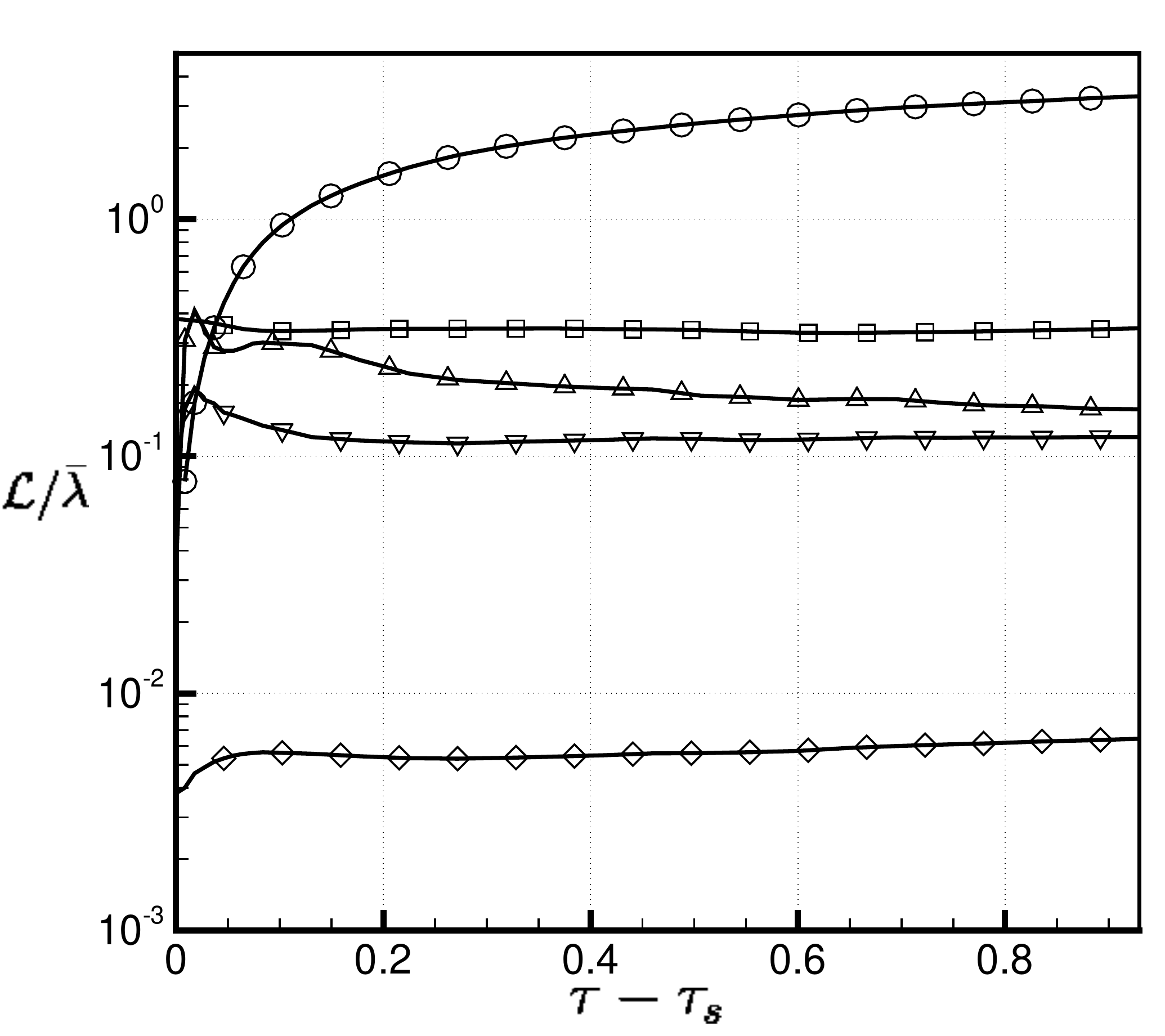}\llap{
		\parbox[b]{0.49\textwidth}{($a$)\\\rule{0ex}{0.40\textwidth}}}
	\includegraphics[width=0.49\textwidth]{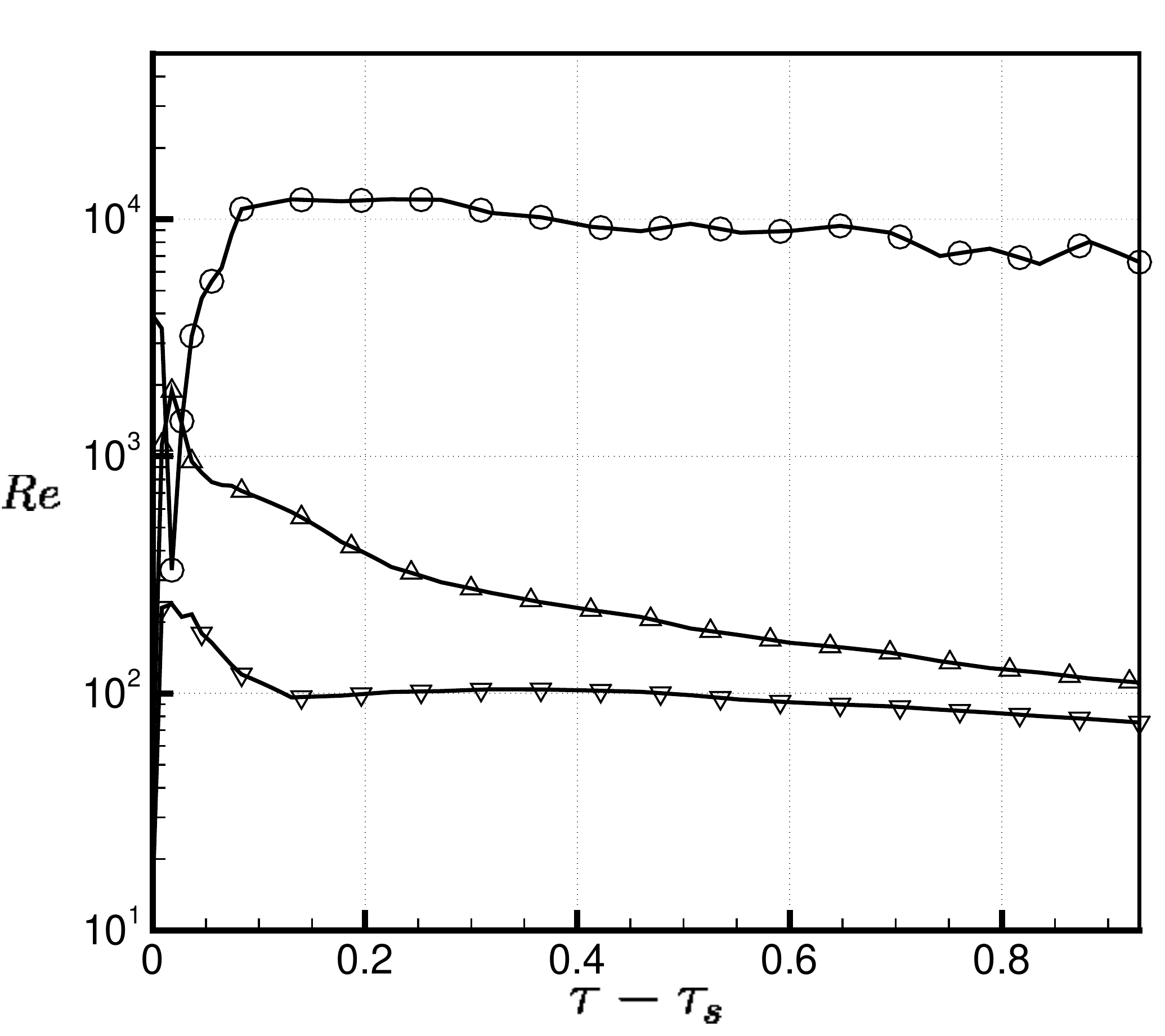}\llap{
		\parbox[b]{0.49\textwidth}{($b$)\\\rule{0ex}{0.40\textwidth}}}
	\caption{Evolution of length scales $(a)$ and Reynolds numbers $(b)$ for $\Rey_0=1385$. Plotted are the outer scale $h$ and associated Reynolds number $\Rey_h$ (circles), the integral length $\Lambda$ (squares), \secondrev{the Taylor microscales $\lambda_{x}$ (upward triangles) and $\lambda_{yz}$ (downward triangles) and associated Reynolds numbers $\Rey_{\lambda_{x}}$ and $\Rey_{\lambda_{yz}}$} and the Kolmogorov microscale $\eta$ (diamonds).}
	\label{fig:lengthscales-reynolds-Re1826}
\end{figure}

Therefore in the $\Rey_0=697$ case, the requirement of $\Rey_\delta\geq$ 1--2$\times 10^4$ for fully developed \secondrev{stationary} turbulence is not met at any point of the simulation, while the equivalent requirement that $\Rey_\lambda\geq$100--140 is met only very briefly. In the lower $\Rey_0$ cases, neither requirements are met at any point. However, given that there is little change in $h$ between simulations, the additional $\Rey_0=1395$ case should achieve $\Rey_h\ge10^4$ for at least a small fraction of time. This is confirmed in figure \ref{fig:lengthscales-reynolds-Re1826}, which shows the evolution of both the length scales and Reynolds numbers up until a time of $\tau=0.939$ for this additional case. Between approximately \firstrev{$\tau-\tau_s=0.1$} and \firstrev{$\tau-\tau_s=0.4$} the outer scale Reynolds number for this case is greater than $1\times10^4$, while the \secondrev{transverse} Taylor-scale Reynolds number is also greater than 100 up until a time of about \firstrev{$\tau-\tau_s=0.5$}. This indicates that there may be significant levels of turbulence within the mixing layer at this time, even if it is not yet fully developed. 

\subsubsection{Mixing transition criterion}
\begin{figure}
	\centering
	\includegraphics[width=0.5\textwidth]{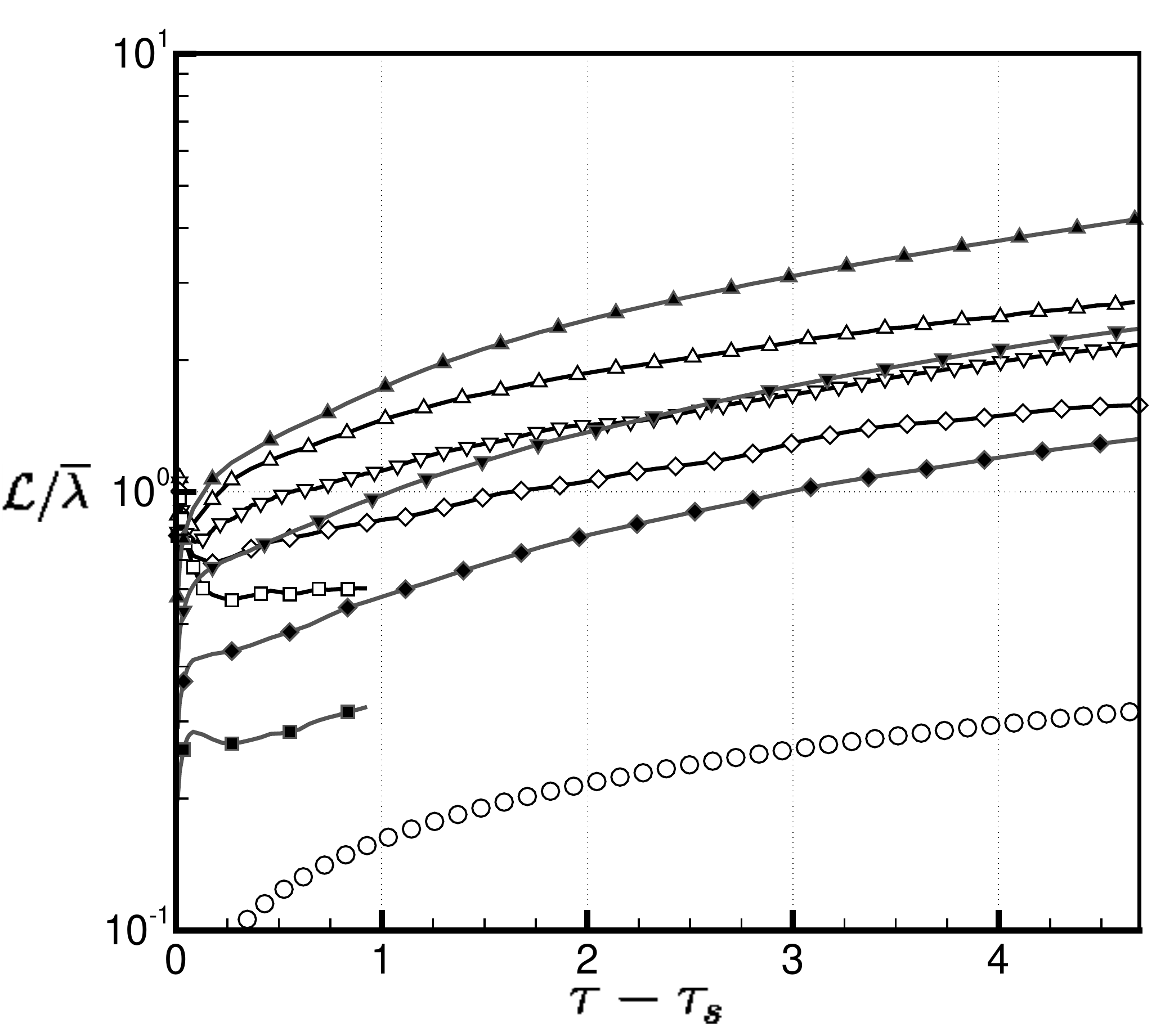}
	\caption{\firstrev{Liepmann--Taylor (black lines, white symbols) and inner-viscous (grey lines, black symbols) length scales vs. time. Results are shown for $\Rey_0=174$ (upward triangles), $\Rey_0=348$ (downward triangles), $\Rey_0=697$ (diamonds) and $\Rey_0=1395$ (squares). Also shown is the estimated diffusion layer length scale (white circles) for the $\Rey_0=1395$ case}.}
	\label{fig:lambda-tau}
\end{figure}

Qualitatively the mixing transition criterion for unsteady flows, presented in \S\ref{sec:intro}, may be expressed as saying that an additional amount of time is required in the presence of a sufficient Reynolds number in order to generate the range of scales that produces a mixing transition. In particular, the hypothesis is that uncoupled fluctuations develop within laminar boundary layers created by viscous diffusion at locations of significant shear. Therefore transition to turbulence occurs once these viscous layers grow for a long enough time such that their extent exceeds the inner-viscous scale.
\secondrev{It is important to note that (\ref{eqn:diffusion-scale}) only describes the late-time behaviour of the diffusion layer scale; the virtual time origin has been neglected \citep{Zhou2019b}, which implies $\lambda_D=0$ at $t=0$. This may be rectified by providing an estimate for the virtual time origin, or equivalently the initial momentum thickness of the shear layer. Here the post-shock integral width $W_0^+$ is used as an estimate for the initial momentum thickness, which gives 
\begin{equation}
\lambda_D=C_{lam}(\nu \bar{t})^{1/2}+W_0^+,
\label{eqn:diffusion-scale2}
\end{equation}
where $\bar{t}=t-t_s$ ($t_s$ being the shock arrival time).} The temporal evolution of the Liepmann--Taylor and inner-viscous scales at the mixing layer centre plane for the four highest $\Rey_0$ cases is shown in figure \ref{fig:lambda-tau}, along with the diffusion layer length scale for the $\Rey_0=1395$ case. Since $\lambda_D$ is trivial to calculate, it has been plotted up until the end time of $\tau=4.70$ for the rest of the simulations. The Liepmann--Taylor scale is almost independent of $\Rey_0$ at early time, due to each case having the same amount of kinetic energy imparted by the shock. Subsequent differences in $\lambda_L$ are due to the fact that as $\Rey_0$ is increased, the fluctuating velocity gradients increase at a faster rate than the turbulent kinetic energy. Meanwhile for each successive doubling of $\Rey_0$, the inner-viscous scale is reduced by a factor of close to 1.4 at early time. In the high Reynolds number limit this factor is expected to approach $2^{3/4}\approx1.7$ according to (\ref{eqn:kolmogorov}), assuming that the dissipation rate becomes independent of $\nu$.

\begin{figure}
	\centering
	\includegraphics[width=0.49\textwidth]{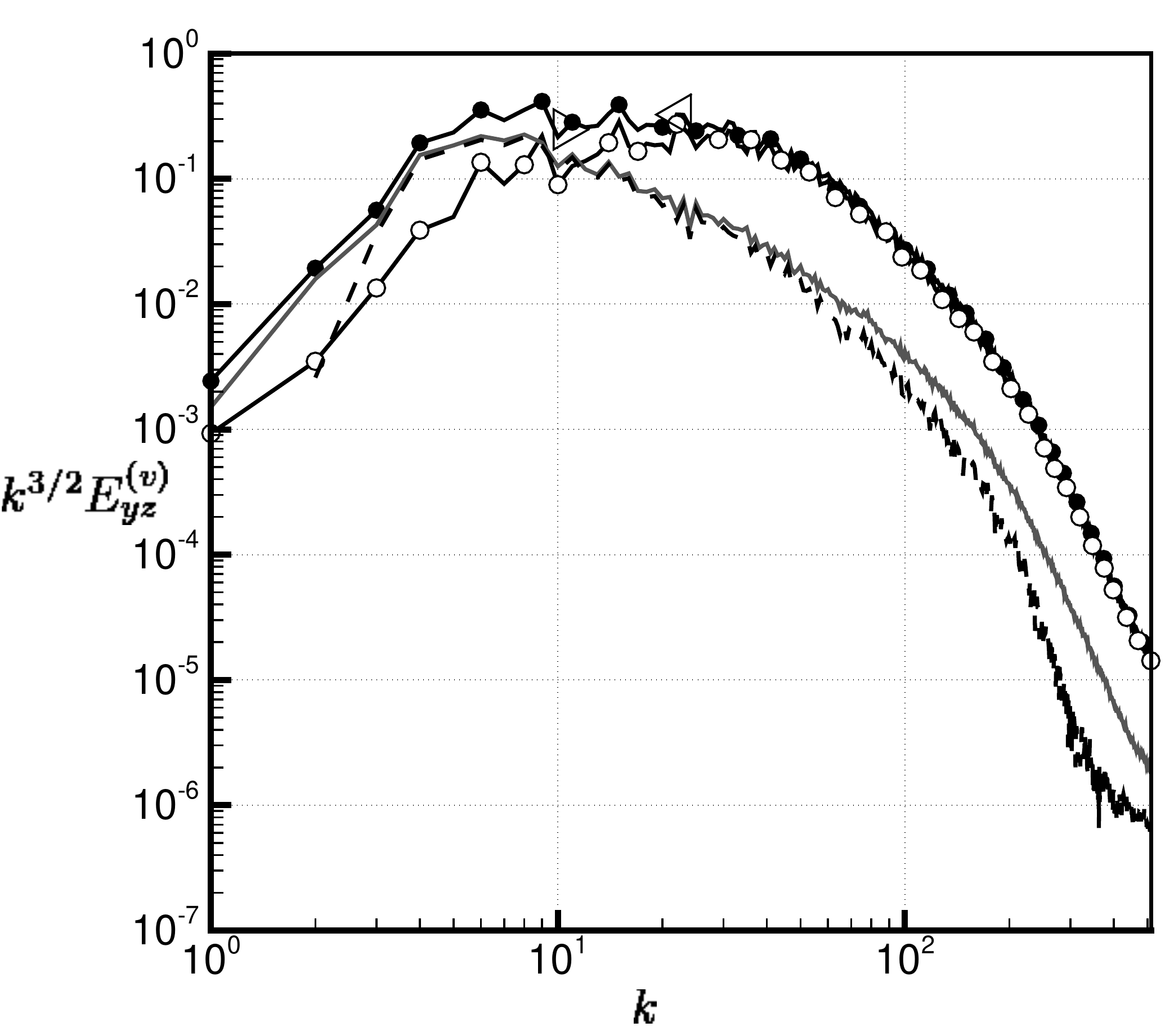}\llap{\parbox[b]{0.49\textwidth}{($a$)\\\rule{0ex}{0.40\textwidth}}}
	\includegraphics[width=0.49\textwidth]{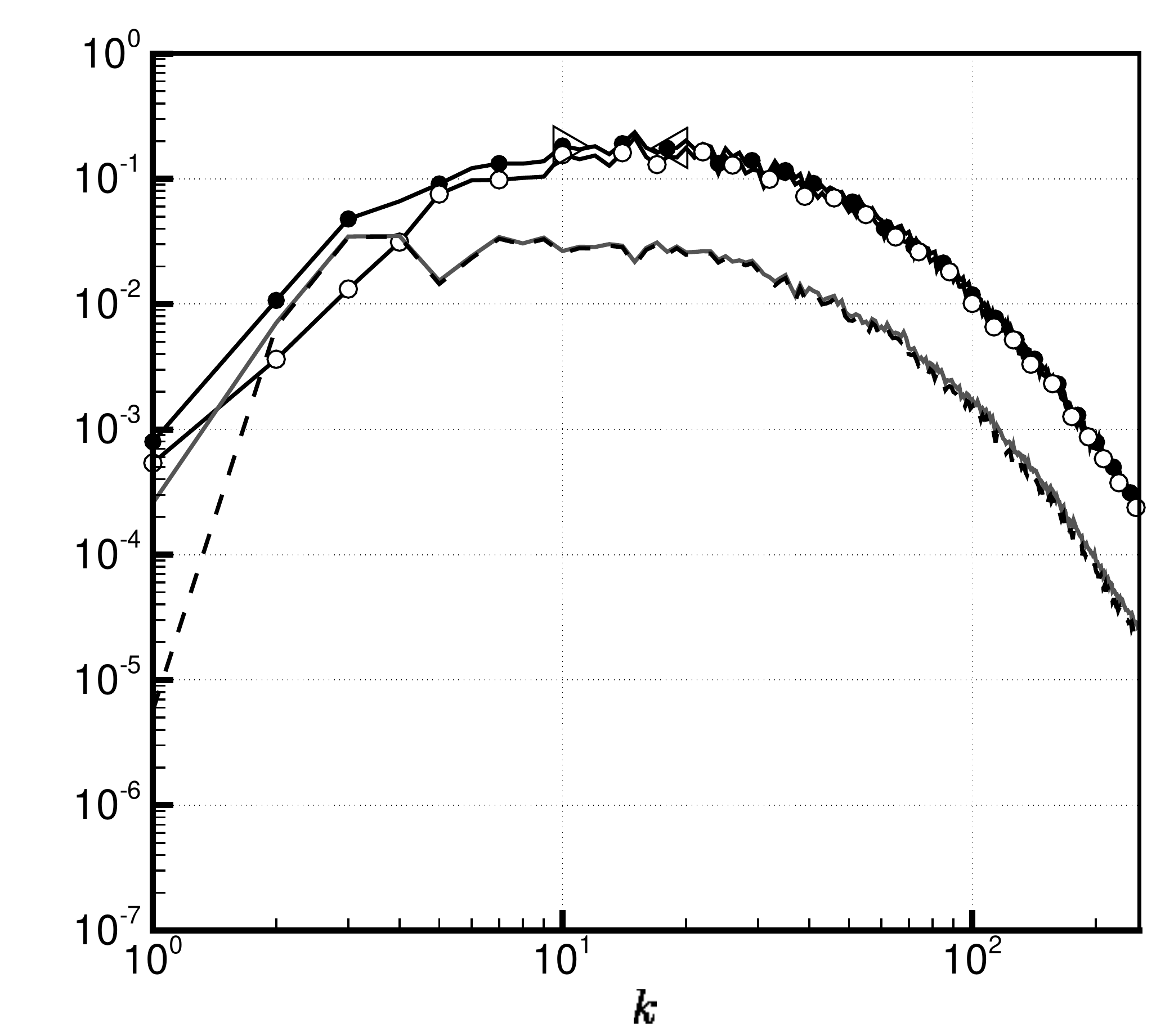}\llap{\parbox[b]{0.49\textwidth}{($b$)\\\rule{0ex}{0.40\textwidth}}}
	\caption{Decomposition of the \thirdrev{compensated} average transverse turbulent kinetic energy per unit volume (solid black lines) in the $\Rey_0=1395$ case into solenoidal \firstrev{(white circles)}, total dilatational \firstrev{(solid grey lines)} and compressible \firstrev{(dashed black lines)} components for times ($a$) $\tau=0.187$ and ($b$) $\tau=0.939$. Also shown are the wavenumbers corresponding to the Liepmann--Taylor (right triangles) and inner-viscous (left triangles) length scales.}
	\label{fig:KEV-Re1826}
\end{figure}

Figure \ref{fig:lambda-tau} shows that for $\Rey_0=348$ there exists a period of time from the beginning of the simulation to approximately \firstrev{$\tau-\tau_s=2.5$} during which $\lambda_L\ge\lambda_V$, due to the observations given above. For the $\Rey_0=697$ case this period has extended to the end time of the simulation and the separation of scales has increased, while for the $\Rey_0=1395$ case the two scales have separated even further. However, it can also be seen that $\lambda_D<\lambda_V$ for the entirety of the simulation, from which it can be concluded that the turbulence in the flow has not yet passed the mixing transition. Figure \ref{fig:KEV-Re1826} shows the turbulent kinetic energy spectra, \thirdrevtwo{compensated by a factor of $k^{3/2}$}, for the $\Rey_0=1395$ case at times $\tau=0.187$ and $\tau=0.939$, annotated with the wavenumbers corresponding to the Liepmann--Taylor and inner-viscous length scales. These scales are intended to represent the smallest of the energy containing scales and the largest of the dissipative scales respectively, and qualitatively this appears to be true when examining the spectrum. The slope of the narrow inertial range that is formed between these two scales is also measured; at $\tau=0.187$ the \thirdrev{(uncompensated)} turbulent kinetic energy scales as $k^{-1.59}$ while at $\tau=0.939$ the scaling is $k^{-1.47}$. As was the case for the lower Reynolds number cases, care must be taken when interpreting the early-time spectra, which contain a significant acoustic component that influences the slope. If the slope is measured purely from the solenoidal component, the resulting scaling is $k^{-0.93}$ instead. At the later time of $\tau=0.939$, the contribution from the acoustic component to the overall kinetic energy has substantially diminished; the scaling measured from the solenoidal spectra is $k^{-1.44}$. This scaling of the inertial range is very close to the $k^{-3/2}$ scaling that has been observed in ILES computations of this case \citep{Groom2019} and which is predicted by the theoretical analysis of \citet{Zhou2001}, rather than the $k^{-5/3}$ scaling for canonical turbulence. Admittedly the slope has only been measured over a small number of data points, therefore higher Reynolds number cases that have passed the mixing transition would be needed to verify these findings.

\begin{figure}
	\centering
	\includegraphics[width=0.49\textwidth]{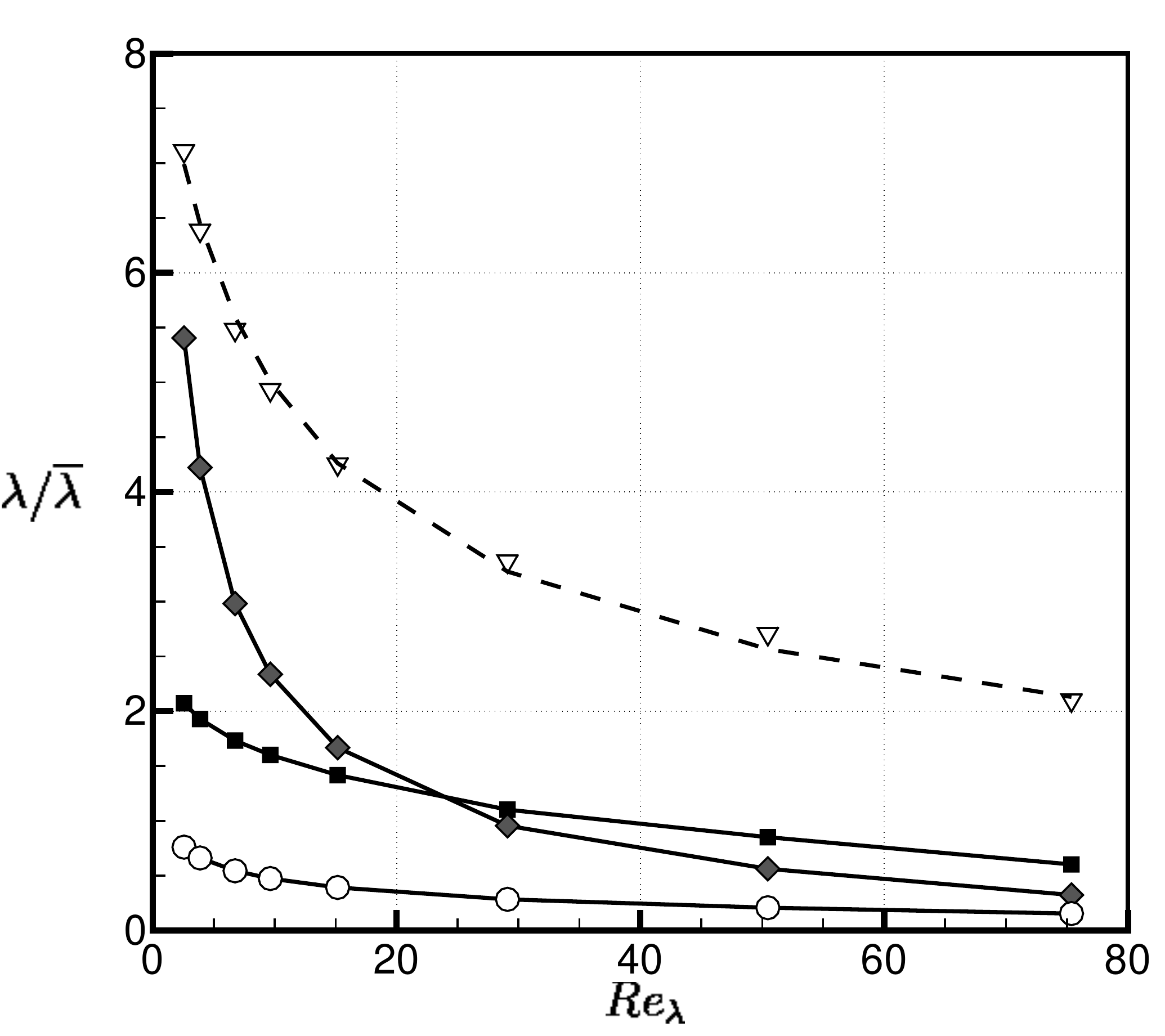}\llap{\parbox[b]{0.49\textwidth}{($a$)\\\rule{0ex}{0.40\textwidth}}}
	\includegraphics[width=0.49\textwidth]{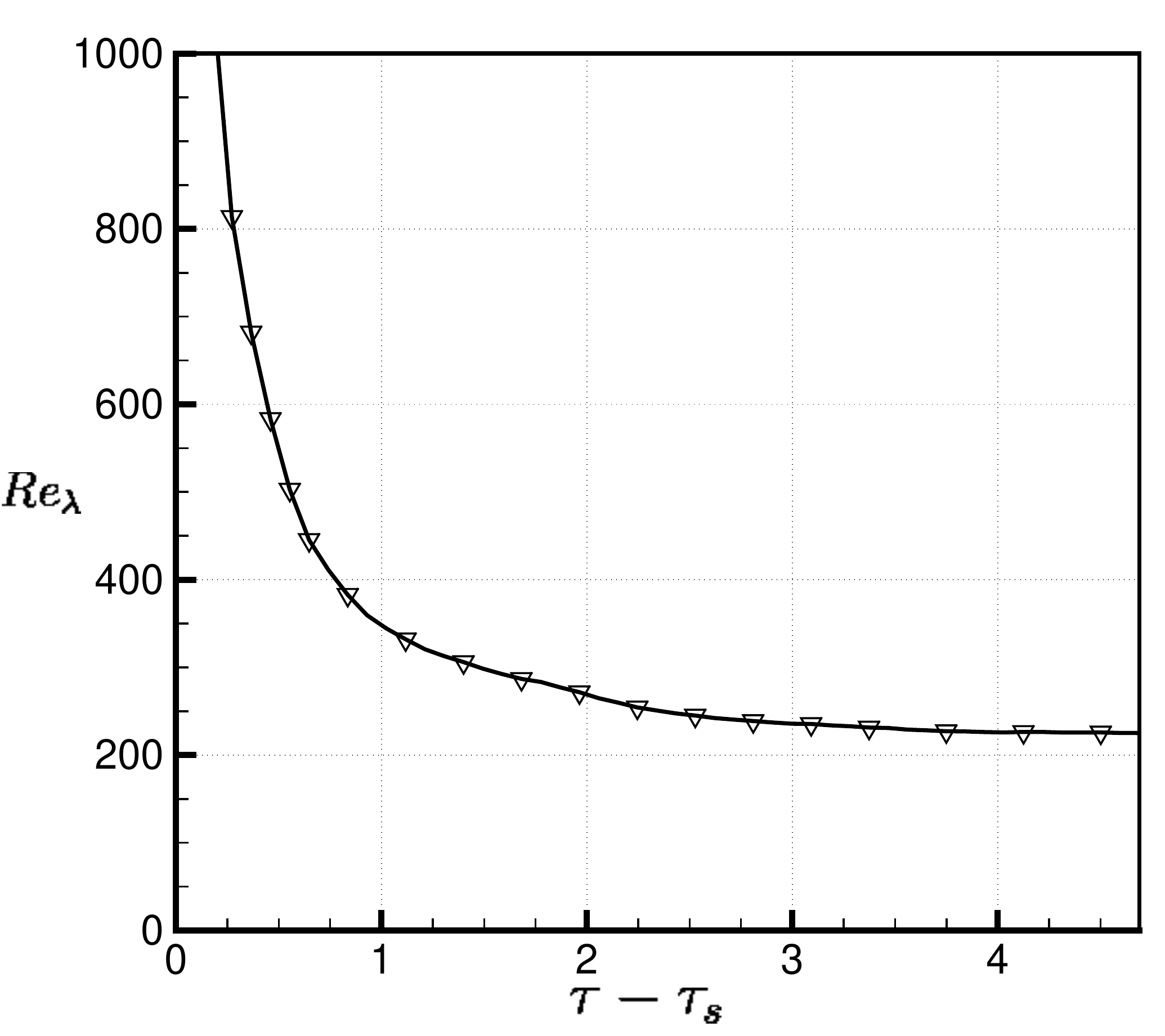}\llap{\parbox[b]{0.495\textwidth}{($b$)\\\rule{0ex}{0.40\textwidth}}}
	\caption{($a$) Liepmann--Taylor \firstrev{(black squares)}, inner-viscous \firstrev{(grey diamonds)} and diffusion layer (white circles) length scales \firstrev{vs. Taylor-scale Reynolds number} at the mixing layer centre plane for time $\tau=0.939$ \firstrev{($\tau-\tau_s=0.929$)}. Also shown is the ratio of the inner-viscous to diffusion layer length scales vs. Taylor-scale Reynolds number \firstrev{(triangles)} including the line of best fit (dashed lines). ($b$) Critical Taylor-scale Reynolds number vs. time.}
	\label{fig:lambda-re}
\end{figure}

While the present set of simulations do not allow for an explicit evaluation of the Reynolds number at which the mixing transition will occur, they can still be used to infer this by considering the variation in length scales with $\Rey_\lambda$ at a given time. This is shown in figure \ref{fig:lambda-re} for a time of $\tau=0.939$, from which it can be seen that the inner-viscous scale is smaller than the Liepmann--Taylor scale for $\Rey_\lambda\gtrsim25$ and is approaching the diffusion layer length scale as $\Rey_\lambda$ is further increased. The ratio of $\lambda_V/\lambda_D$ is also shown in figure \ref{fig:lambda-re} at time $\tau=0.939$. By using a curve fitting procedure similar to the one performed for $C_\epsilon$ and $C_\chi$, the Reynolds number at which this ratio equals one can be estimated. The choice of an appropriate functional form is guided by considering how $\lambda_V$ and $\lambda_D$ vary with Reynolds number. Both $\lambda_V$ and $\lambda_D$ can be related to the outer-scale Reynolds number by
\begin{subeqnarray}
	\lambda_V & \approx & 50\eta \propto \Rey_\delta^{-3/4}, \\
	\lambda_D & = & C_{lam}(\nu \bar{t})^{1/2} \propto \Rey_\delta^{-1/2},
	\label{eqn:eta-lambda}
\end{subeqnarray}
which can be combined with the relation $\Rey_\delta=3/20\Rey_\lambda^2$ for isotropic turbulence to derive that $\lambda_V/\lambda_D\propto\Rey_\lambda^{-1/2}$. Thus the curve that is fit to the data is chosen to be of the form
\begin{equation}
f=\sqrt{\frac{B}{\Rey_\lambda+C}}.
\label{eqn:f2}
\end{equation}
The curve of best fit is also shown in figure \ref{fig:lambda-re}, for which the parameters are $B=364$, $C=4.90$. Therefore the critical Taylor-scale Reynolds number at which $\lambda_V/\lambda_D=1$ is estimated to be $\Rey_\lambda=359$ at time $\tau=0.939$.

The curve-fitting procedure is repeated for a range of times between $\tau=0.187$ and $\tau=4.70$, with the estimated critical Taylor-scale Reynolds number at each time also plotted in figure \ref{fig:lambda-re}. 
It can be seen that at very early times the required Taylor-scale Reynolds number \secondrev{that satisfies the mixing transition} is very high, for example at $\tau=0.187$ it is estimated to be $\Rey_\lambda=1068$. A caveat must be made here; it is expected that $\Rey_\lambda\rightarrow \infty$ for some time $\tau>\tau_s$ since a finite amount of time is required for the initial energy injected at the driving scales to pass down to smaller and smaller scales via nonlinear transfer and form an inertial range. This process is not explicitly represented in (\ref{eqn:unsteady-mixing-transition}), therefore the estimated critical Taylor-scale Reynolds number may not be accurate as $\tau\rightarrow\tau_s$. Nonetheless it will still be extremely large, which is sufficient for the purposes of this study. Of much greater interest is the behaviour at late time; at the latest time considered it is estimated that flows with $\Rey_\lambda=225$ or greater will \secondrev{pass the mixing transition}.  This is significantly greater than the estimate of $35\le\Rey_\lambda\le80$ given by \citet{Tritschler2014pre}. The corresponding critical outer-scale Reynolds number is approximately $8\times$ that of the $\Rey_0=697$ case at this time \firstrev{(derived using the relation $\Rey_\delta=3/20\Rey_\lambda^2$)}, which suggests that a case with $\Rey_0=5576$ would begin to pass the mixing transition \secondrev{(but not necessarily obtain the minimum state)}. Such a case is also likely currently achievable using a substantial portion of the computational resources on one of the world's top supercomputers. The critical Taylor-scale Reynolds number curve also has an approximate $t^{-1}$ dependence (based on a curve fit to the data) and asymptotically approaches of value of $\Rey_\lambda=174$ as $t\rightarrow\infty$, which is approaching the $\Rey_\lambda\geq$100--140 requirement for stationary flows. Furthermore, as was observed in figure \ref{fig:reynolds-tau}, the outer-scale and Taylor-scale Reynolds numbers also decrease in time, beyond some initial peak shortly after shock passage, implying that the mixing transition criterion may only be satisfied temporarily. Such a phenomenon does not occur in turbulence induced by the Rayleigh--Taylor instability, for which the Reynolds number increases as $\Rey_h\propto t^3$, and reflects a fundamental difficulty in attaining sufficiently high Reynolds numbers for universal behaviour to be observed in experiments or simulations of the Richtmyer--Meshkov instability.


\section{Conclusions}
\label{sec:conclusion}
This paper has investigated the effects of initial Reynolds number $\Rey_0$ on a turbulent mixing layer induced by Richtmyer--Meshkov instability evolving from narrowband initial conditions using a series of direct numerical simulations. After the initial shock passage the turbulence in the layer is freely decaying, with the outer-scale Reynolds number obtaining its maximum value at very early time, after which it continually decreases for each of the simulations. An analysis of various mixing measures showed that there was little variation in the integral width for the range of $\Rey_0$ considered here, while lower $\Rey_0$ cases have more mixed mass at early times. At later times the amount of mixed mass in these cases is overtaken by that of higher $\Rey_0$ cases due to a larger interfacial surface area and steeper gradients. A clear trend of increasing growth rate exponent $\theta$ was also observed with increasing $\Rey_0$, although the overall variation was only 25\%. The molecular mixing fraction showed a dependence on $\Rey_0$, in particular the point of minimum mix which was estimated to be $\Theta_{min}=0.10$.  The late time asymptotic value also varied with $\Rey_0$, with the data observed to be approaching a high Reynolds number limit of $0.765$.

A detailed analysis of the Reynolds number dependence of various statistics of the velocity and scalar fields was also presented. The decay rates of turbulent kinetic energy and its dissipation rate were shown to decrease with increasing $\Rey_0$. The spatial distribution of both of these quantities was also shown to be biased towards the spike side of the layer. An analysis of the turbulent kinetic energy spectra showed that the distribution of energy at the largest scales was extremely similar across all cases, while substantially more energy is contained in the small scales as $\Rey_0$ is increased. The spectra were also decomposed into solenoidal, total dilatational and purely compressible components, which showed that at early time the energy at low to intermediate wavenumbers is dominated by compressible modes. At later times the solenoidal component begins to dominate the overall energy spectrum, indicating that the mixing layer is approaching incompressible flow. For the scalar variance and scalar dissipation rate, similar trends to the turbulent kinetic energy and dissipation rate were observed. There was found to be less variation with $\Rey_0$ however, particularly for the scalar dissipation rate at later times.

The Reynolds number dependence of the normalised dissipation rate $C_\epsilon$ and scalar dissipation rate $C_\chi$ was assessed, showing that a high Reynolds number limit is being approached. At early times the asymptotic values of $C_\epsilon$ and $C_\chi$ vs. Reynolds number vary significantly as the flow continues to develop, while at late times the curves have collapsed. Fitting an appropriate functional form to the data showed that the asymptotic value of $C_\chi$ is attained faster than that of $C_\epsilon$, in agreement with similar observations made for homogeneous passive scalar turbulence at $\Sc=1$. Finally, an evaluation of the mixing transition was performed, showing that although the highest $\Rey_0$ case satisfies the criteria of \citet{Dimotakis2000} for fully developed stationary turbulence, it does not meet the additional requirement of \citet{Zhou2003b} for unsteady flows and therefore cannot be considered fully turbulent. By considering the ratio between the inner-viscous and diffusion layer length scales, the critical Reynolds number at which the mixing transition criterion is satisfied was able to be estimated, which translates to an initial Reynolds number around $4\times$ larger than the current highest $\Rey_0$ case. This case also exhibited a narrow inertial range in the turbulent kinetic energy spectra with a scaling close to $k^{-3/2}$ (as predicted by \cite{Zhou2001}), which shows that such an observation is insufficient for assessing whether the turbulence is fully developed.

As was mentioned in the introduction, if the layer width grows as $\sim t^\theta$ then the outer-scale Reynolds number grows/decays as $\sim t^{2\theta-1}$. For narrowband initial conditions this means that the Reynolds number decreases in time, however if $\theta>0.5$ then the Reynolds number will increase in time. For broadband perturbations with an initial power spectrum $P(k)\propto k^{m}$ and $m<-1$, the results of \citet{Groom2020} show that this will indeed be the case, at least while the layer is growing in the self-similar regime. These perturbations are also more representative of RMI flows encountered in reality, thus future work will involve performing DNS of RMI-induced turbulence evolving from broadband perturbations while the layer is growing self-similarly. Another area for extending the current work is to evaluate the Reynolds number dependence of various quantities such as spectra, length scales and normalised dissipation rates at different planes in the mixing layer. Presently these are only evaluated at the mixing layer centre plane, however the spatial distributions of many of the quantities analysed in this paper show that this is not necessarily the location of peak turbulent activity. Finally, it would be useful to extend the current set of simulations to include a much larger parameter sweep of different Schmidt, Atwood and Mach numbers. This would allow for interaction effects between different parameters to be captured that have not been explored in the present study.

\section{Acknowledgements}
The authors would like to acknowledge the computational resources at the National Computational Infrastructure provided through the National Computational Merit Allocation Scheme, as well as the Sydney Informatics Hub and the University of Sydney's high performance computing cluster Artemis, which were employed for all cases presented here. The authors also wish to acknowledge Dr. Ye Zhou for helpful discussions regarding the unsteady mixing transition criterion. 

Declaration of Interests: The authors report no conflict of interest.

\appendix
\section{}\label{app:B} 
\begin{figure}
	\centering
	\includegraphics[width=0.33\textwidth]{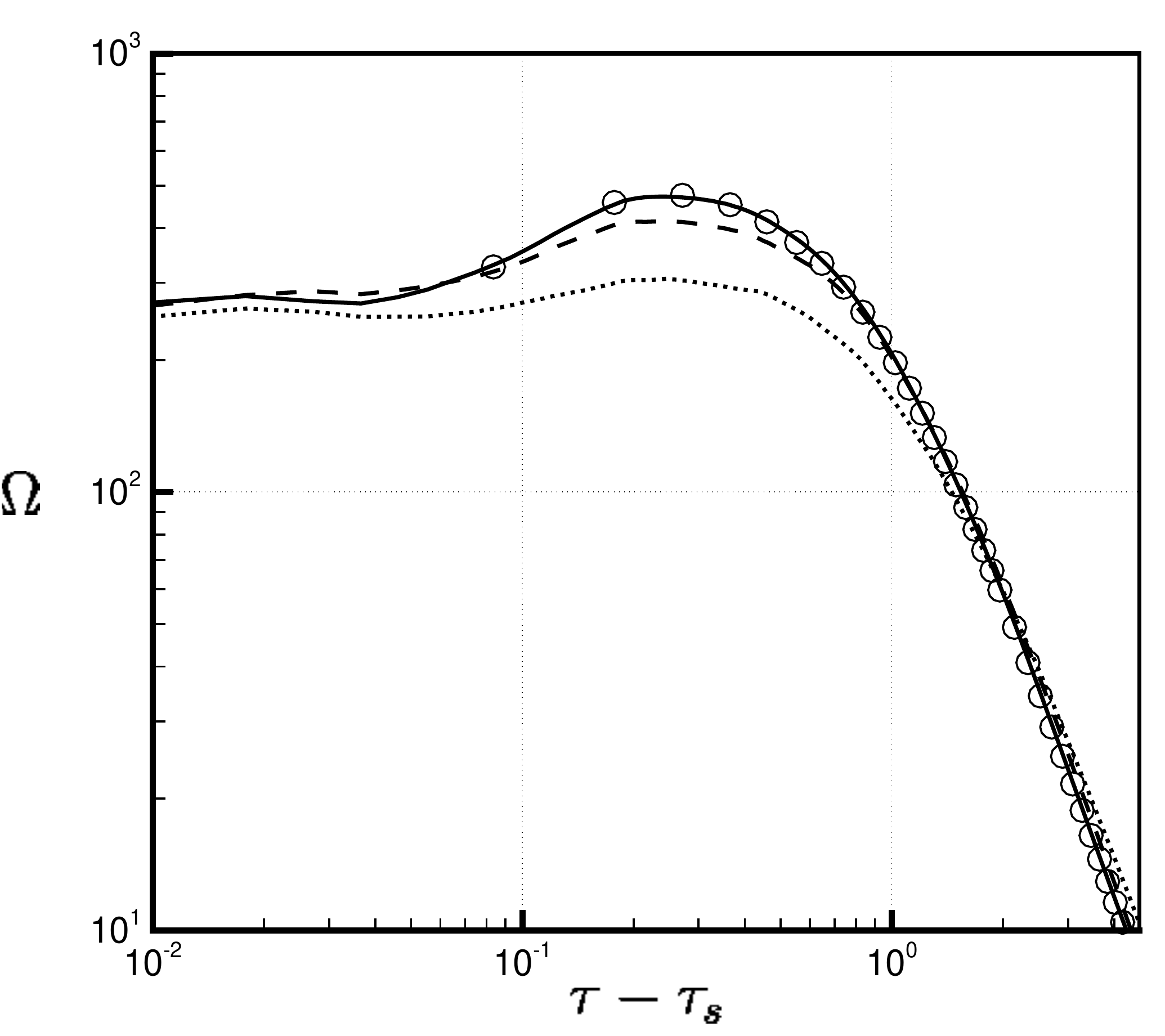}
	\includegraphics[width=0.33\textwidth]{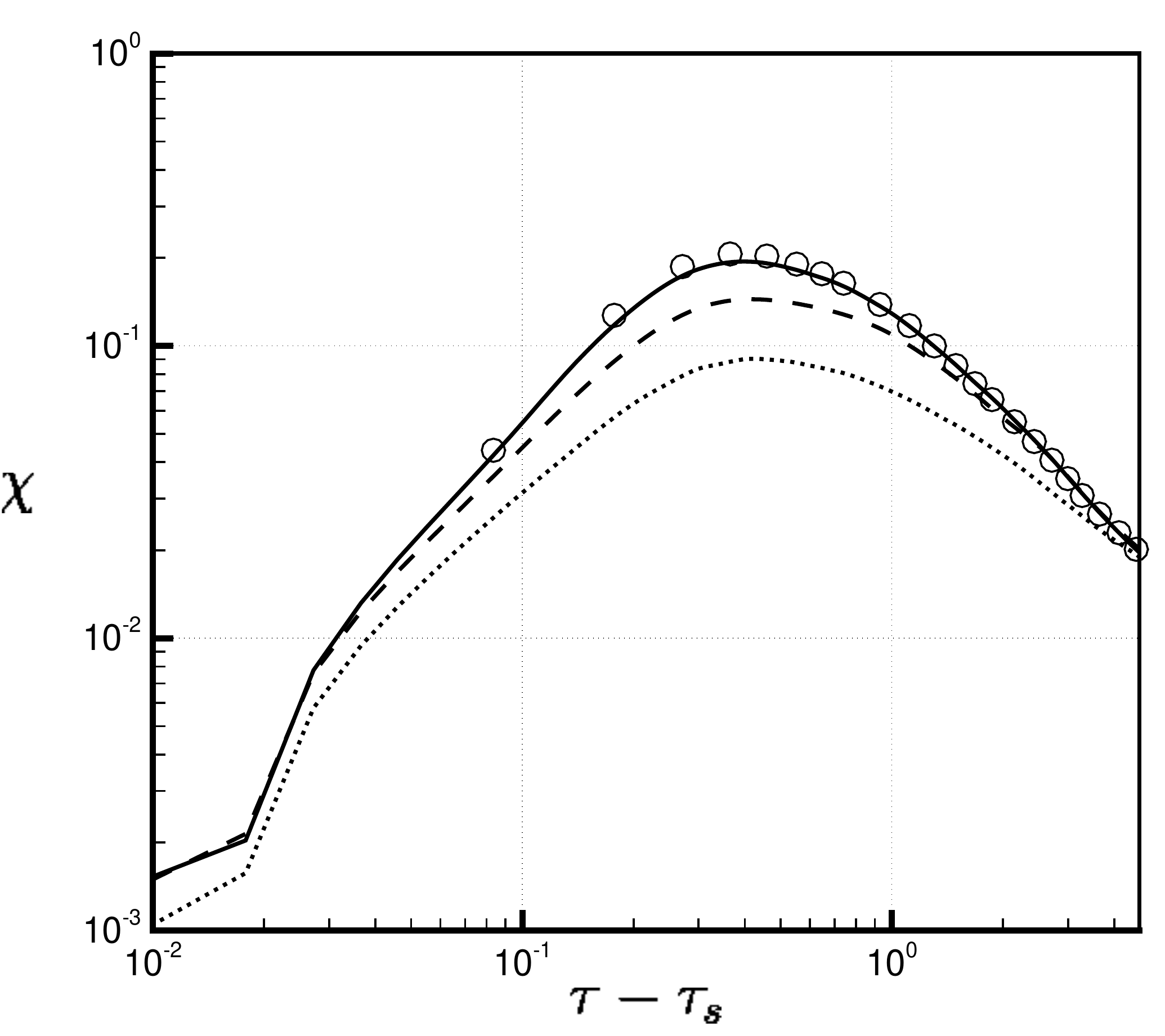}
	\caption{Temporal evolution of enstrophy $(a)$ and scalar dissipation rate $(b)$ for the $\Rey_0=697$ case. Results are shown for grid resolutions of $180\times128^2$ (dotted lines), $360\times256^2$ (dashed lines), $720\times512^2$ (solid lines) and $1440\times1024^2$ (circles).}
	\label{fig:omega-chi-time}
\end{figure}

This appendix summarises how grid convergence is assessed in the current set of direct numerical simulations. Full details can be found in \citet{Groom2019}. Following \citet{Olson2014}, the instantaneous enstrophy $\Omega$ and scalar dissipation rate $\chi$ are computed and compared for successively increasing grid resolutions, as these quantities are dependent on the small scales and therefore are more difficult to demonstrate convergence for than statistics such as integral width or turbulent kinetic energy. Domain integrated values of $\Omega$ and $\chi$ are calculated as
\begin{subeqnarray}
	\Omega & = & \int\rho\norm{\boldsymbol{\omega}}^2\:\mathrm{d}x\:\mathrm{d}y\:\mathrm{d}z, \\
	\chi & = & \int D\bnabla Y_1\bcdot\bnabla Y_1\:\mathrm{d}x\:\mathrm{d}y\:\mathrm{d}z,
	\label{eqn:omega-chi}
\end{subeqnarray}
where $\boldsymbol{\omega}=\bnabla\times\boldsymbol{u}$ is the vorticity. Radial power spectra are also calculated, in an analogous manner to (\ref{eqn:E2D}). Figure \ref{fig:omega-chi-time} shows the temporal evolution of domain integrated enstrophy and scalar dissipation rate for the $\Rey_0=697$ case computed on various grid resolutions. The solutions for both of these quantities are clearly converging with each successive doubling of the grid resolution, with a sufficiently small difference observed between the $720\times512^2$ and $1440\times1024^2$ grids. The differences between the solutions obtained on the two finest grids are greatest at early time when the layer is being thinned and stretched, resulting in large gradients across the interface \citep{Groom2019}.

\begin{figure}
	\centering
	\includegraphics[width=0.33\textwidth]{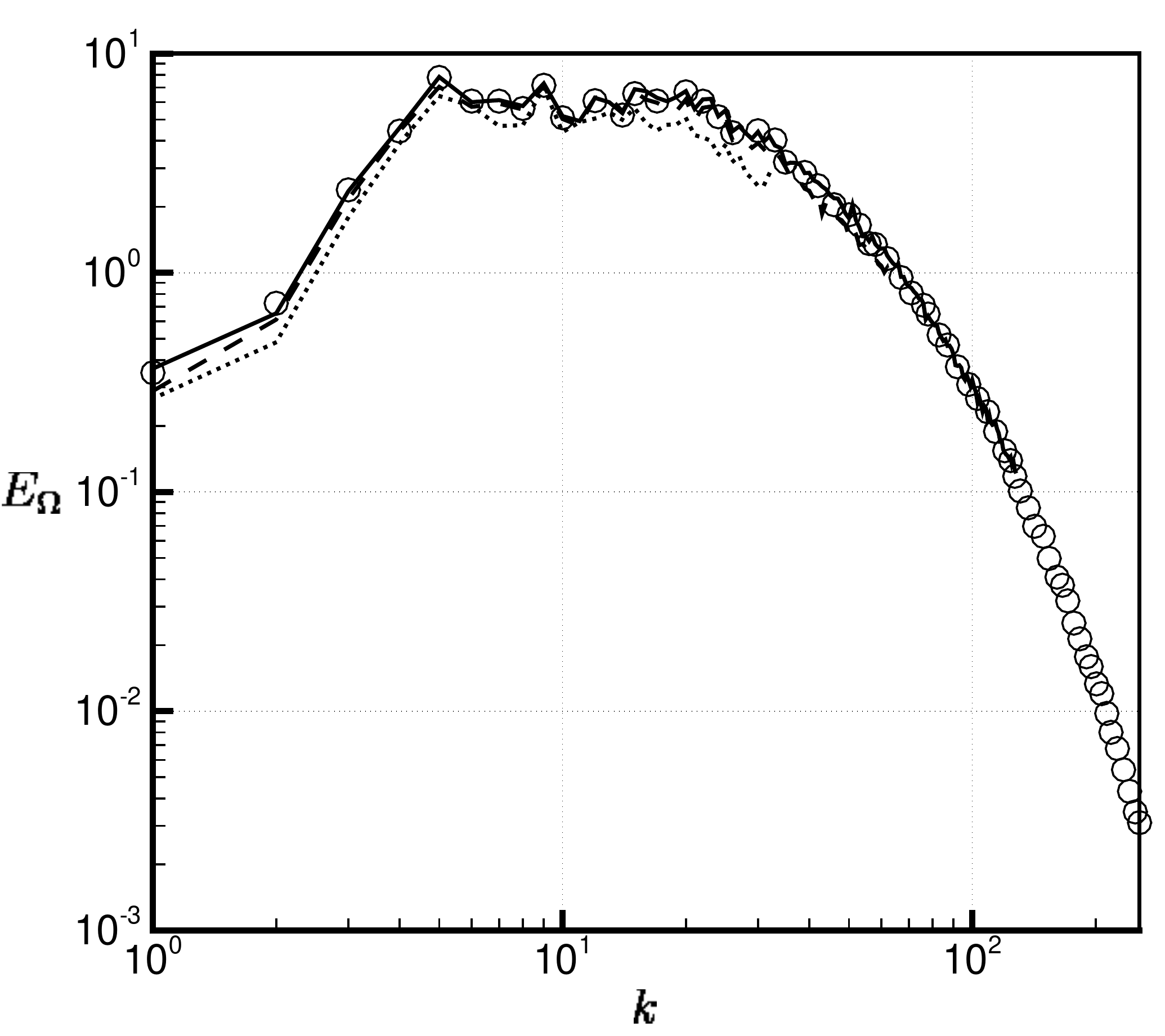}
	\includegraphics[width=0.33\textwidth]{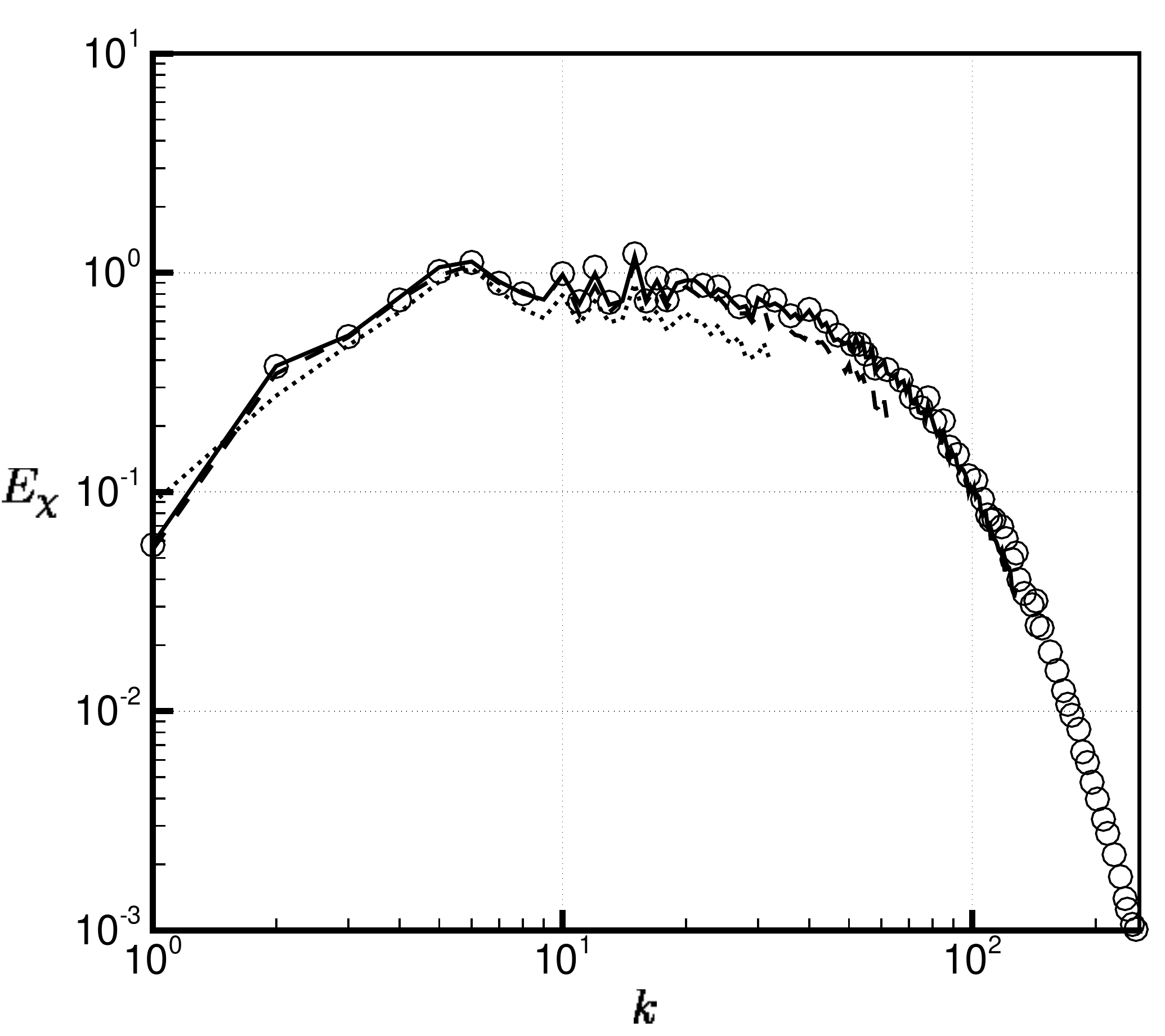}
	\caption{Power spectra of enstrophy $(a)$ and scalar dissipation rate $(b)$ at time $\tau=0.470$. Results are shown for grid resolutions of $180\times128^2$ (dotted lines), $360\times256^2$ (dashed lines), $720\times512^2$ (solid lines) and $1440\times1024^2$ (circles)}
	\label{fig:omega-chi-spectra}
\end{figure}

Power spectra of $\Omega$ and $\chi$ are presented in figure \ref{fig:omega-chi-spectra}, taken at time $\tau=0.470$ corresponding to the peak scalar dissipation rate. There is excellent agreement across all grid resolutions for the low wavenumber end of the spectra ($k\le 10$), while for the two highest grid resolutions the results are converged up to at least $k\le 128$ for the enstrophy spectra and $k\le100$ for the scalar dissipation rate spectra. This represents the least converged region of the entire solution; for later times the scalar dissipation rate spectra are converged up to at least $k\le 128$ also. These results should be compared with similar ones presented in \citet{Olson2014} and \citet{Tritschler2014pre} for DNS of Richtmyer--Meshkov flows. 


\bibliographystyle{jfm}
\bibliography{bibliography} 
\end{document}